\documentclass[twocolumn,superscriptaddress,nofootinbib,aps,prd]{revtex4-2}

\usepackage{graphicx}
\usepackage{float}
\usepackage{dcolumn}
\usepackage{bm}
\usepackage{amsmath}
\usepackage{amssymb}
\usepackage{siunitx}
\usepackage{upgreek}
\usepackage{xcolor}
\usepackage{hyperref}
\usepackage{ulem}

\begin{document}

\title{Dark-field Search for Quantum Vacuum Signals in the Low-Energy Regime}

\newcommand{\HZDR}{\affiliation{Helmholtz-Zentrum Dresden-Rossendorf, Bautzner Landstra\ss e 400, 01328 Dresden, Germany}}
\newcommand{\HIJ}{\affiliation{Helmholtz-Institut Jena, Fr\"obelstieg 3, 07743 Jena, Germany}}
\newcommand{\GSI}{\affiliation{GSI Helmholtzzentrum f\"ur Schwerionenforschung, Planckstra\ss e 1, 64291 Darmstadt, Germany}}
\newcommand{\FSU}{\affiliation{Physikalisch-Astronomische Fakultät, Friedrich-Schiller-Universität Jena, 07743 Jena, Germany}}
\newcommand{\XFEL}{\affiliation{European XFEL, Holzkoppel 4, 22869 Schenefeld, Germany}}
\newcommand{\IOP}{\affiliation{Department of Radiation and Chemical Physics, Institute of Physics, Czech Academy of Sciences, Na Slovance 2, 182 00 Prague 8, Czech Republic}}
\newcommand{\CUNI}{\affiliation{Department of Surface and Plasma Science, Faculty of Mathematics and Physics, Charles University in Prague, V Hole\v{s}ovi\v{c}k\'ach 2, 182 00 Prague 8, Czech Republic}}
\newcommand{\CTU}{\affiliation{Department of Nuclear Chemistry, Faculty of Nuclear Science and Physical Engineering, Czech Technical University in Prague, B\v{r}ehov\'a 7, 115 19 Prague 1, Czech Republic}}
\newcommand{\IPP}{\affiliation{Laser Plasma Department, Institute of Plasma Physics, Czech Academy of Sciences, U Slovanky 2525/1a, 182 00 Prague 8, Czech Republic}}
\newcommand{\DESY}{\affiliation{Deutsches Elektronen-Synchrotron DESY, Notkestra\ss e 85, 22607 Hamburg, Germany}}
\newcommand{\TUD}{\affiliation{Technische Universit\"at Dresden, 01062 Dresden, Germany}}
\newcommand{\CASUS}{\affiliation{Center for Advanced Systems Understanding, Untermarkt 20, 02826 G\"orlitz, Germany}}
\newcommand{\PLY}{\affiliation{Centre for Mathematical Sciences, University of Plymouth, Plymouth, PL4 8AA, United Kingdom}}
\newcommand{\ELI}{\affiliation{ELI Beamlines Facility, The Extreme Light Infrastructure ERIC, Doln\'i B\v{r}e\v{z}any, Czech Republic}}
\newcommand{\ESRF}{\affiliation{European Synchrotron Radiation Facility, CS 40220, F-38043 Grenoble Cedex 9, France}}
\newcommand{\PALM}{\affiliation{PALM Scientific LLC, Naperville, IL, USA}}

\author{Aim\'e~Matheron$^{\,\parallel}$}\email{a.matheron@hi-jena.gsi.de}\HIJ\GSI\FSU
\author{Michal~\v{S}m\'id$^{\,\parallel}$}\email{m.smid@hzdr.de}\HZDR
\author{Felix~Karbstein}\email{f.karbstein@hi-jena.gsi.de}\HIJ\GSI\FSU
\author{Sebastian~G\"ode}\XFEL
\author{Willi~Hippler}\HIJ\GSI\FSU
\author{Robert~L\"otzsch}\HIJ\GSI\FSU

\author{Michal~Andrzejewski}\XFEL
\author{Sergey~Antipov}\PALM
\author{Carsten~Baehtz}\XFEL
\author{Victorien~Bouffetier}\HZDR
\author{Erik~Brambrink}\XFEL
\author{Jakub~Buli\v{c}ka}\IOP\CUNI
\author{Samuele~Di~Dio~Cafiso}\HZDR
\author{Rafael~Celestre}\ESRF
\author{Jarom\'ir~Chalupsk\'y}\IOP
\author{Thomas~E.~Cowan}\HZDR\TUD
\author{J\"org~Grenzer}\HZDR
\author{Travis~Griffin}\XFEL
\author{V\v{e}ra~H\'ajkov\'a}\IOP
\author{Jens~Hauser}\HZDR
\author{Peter~Hilz}\HIJ\GSI\FSU
\author{Hauke~H\"oppner}\HZDR
\author{Al\v{z}b\v{e}ta~Horynov\'a}\IOP\CTU
\author{Lingen~Huang}\HZDR
\author{Oliver~Humphries}\XFEL
\author{\v{S}imon~Jel\'inek}\IOP\IPP\CUNI
\author{Libor~Juha}\IOP
\author{Pooyan~Khademi}\HIJ\GSI\FSU
\author{Alejandro~Laso-Garc\'ia}\HZDR
\author{Masruri~Masruri}\HZDR
\author{Nikita~Medvedev}\IOP\IPP
\author{Motoaki~Nakatsutsumi}\XFEL
\author{Gerhard~G.~Paulus}\HIJ\GSI\FSU
\author{Jennifer~Popp}\XFEL
\author{Thomas~R.~Preston}\XFEL
\author{Isabella~Probst}\HIJ\GSI\FSU 
\author{Sripati~V.~Rahul}\XFEL
\author{Thomas~Roth}\ESRF
\author{Felipe~C.~Salgado}\HIJ\GSI\FSU 
\author{Roland~Sauerbrey}\HZDR\TUD
\author{Alexander~S\"avert}\XFEL
\author{Hans-Peter~Schlenvoigt}\HZDR
\author{Jan~Patrick~Schwinkendorf}\HZDR\XFEL
\author{Frank~Seiboth}\DESY
\author{Georgiy~Shoulga}\HZDR
\author{Minxue~Tang}\XFEL
\author{Abha~Thazhathe~Peedika}\HIJ\GSI\FSU
\author{Monika~Toncian}\HZDR
\author{Toma~Toncian}\HZDR
\author{Vojt\v{e}ch~Vozda}\IOP
\author{Edgar~Weckert}\DESY
\author{Colin~Wessel}\HIJ\GSI\FSU
\author{Jan~Wild}\CUNI
\author{Ulf~Zastrau}\XFEL

\author{Matt~Zepf}\email{m.zepf@hi-jena.gsi.de}\HIJ\GSI\FSU

\collaboration{BIREF@HIBEF Collaboration}\noaffiliation

\date{\today}

\begin{abstract}
 We report on the experimental details of a dark-field search for light-by-light scattering signals performed at the High Energy Density (HED) instrument of the European XFEL (EuXFEL) in a pump-probe type experiment colliding the EuXFEL beam with the 300-TW-class near-infrared Relativistic Laser at XFEL (ReLaX) at a center-of-mass photon energy of $\omega_\star = 116\,\mathrm{eV}$. Especially in combination with polarization filtering, the dark-field concept allows the signal-to-background ratio to be significantly enhanced and both polarization channels to be recorded. This work elaborates on the experimental setup and the diagnostics used, from the per-shot spatiotemporal-overlap diagnostics, via the polarizing beam splitter, to the characterization of the residual background. Moreover, it provides details on the start-to-end simulations of the setup used for analysis. Accumulating 2751 pump-probe collisions, no excess above background is observed, and an upper limit on the light-by-light scattering cross section of $7\times10^{4}$ times the QED prediction at a 95\% confidence level is derived. This is the most stringent bound on low-energy light-by-light scattering achieved to date, representing an improvement by ten orders of magnitude upon previous experimental limits.
\end{abstract}

\maketitle

\begingroup
\renewcommand\thefootnote{\(\parallel\)}
\footnotetext[1]{These authors contributed equally to this work.}
\endgroup

\section{Introduction}
\label{sec:intro}

\begin{figure}[hbt]
	\centering
	\includegraphics[width=0.8\columnwidth]{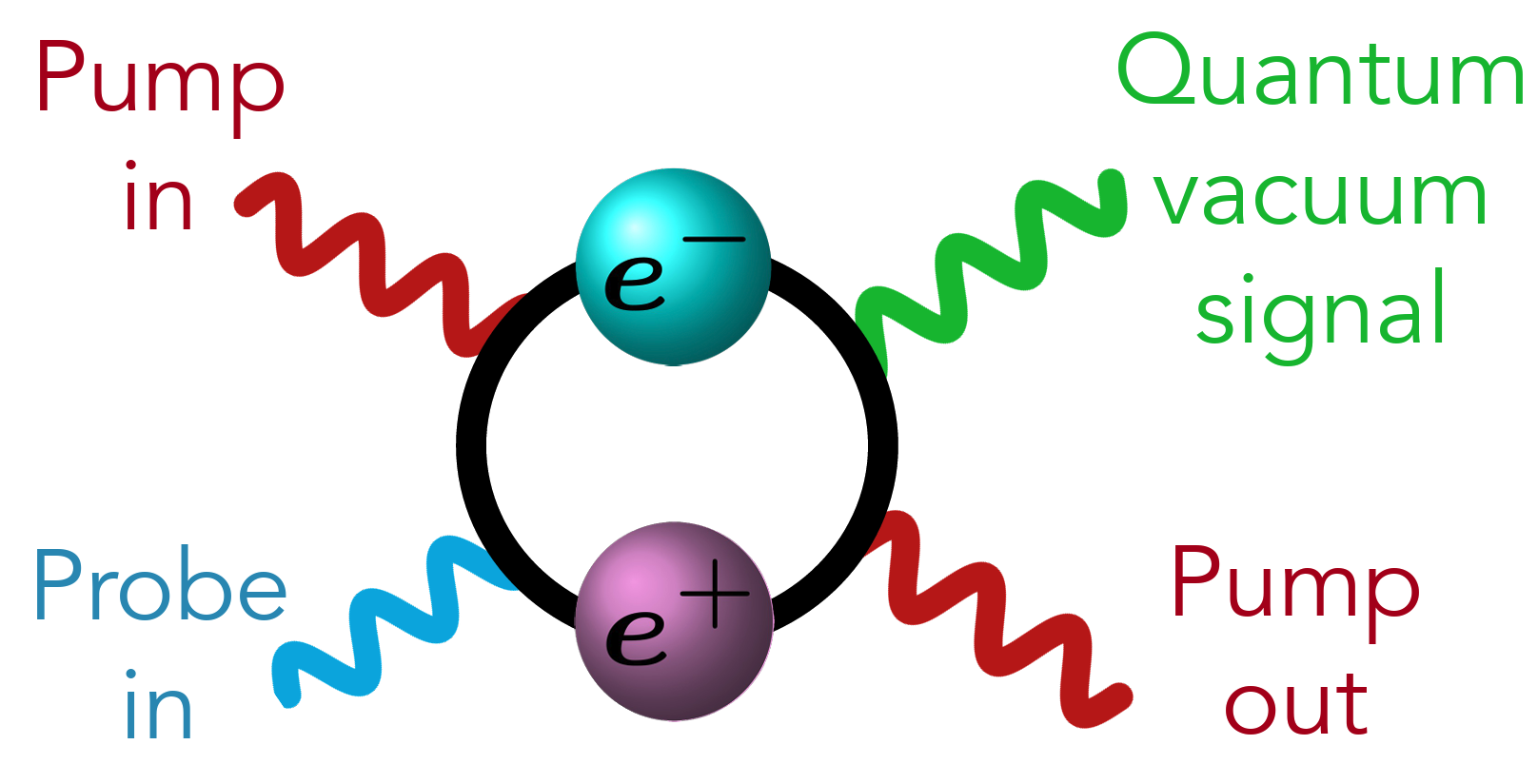}
	\caption{Photon--photon scattering. Probe and pump photons interact through a virtual electron--positron pair loop represented by the black line. The scattered probe photon, whose momentum and, possibly, polarization have changed, constitutes the quantum vacuum signal searched for in the present experiment.}
	\label{fig:box_diag}
\end{figure}

We report in detail on an experimental campaign carried out at the High Energy Density (HED) instrument of the European XFEL, in which femtosecond X-ray pulses of \SI{8.766}{keV} were collided with the focus of the \SI{300}{TW}-class optical laser ReLaX, in a search for elastic photon--photon (light-by-light) scattering. Predicted by QED ninety years ago as a consequence of the coupling of photons through virtual electron--positron pairs (Fig.~\ref{fig:box_diag}) \cite{Euler:1935zz,Euler:1935qgl,Heisenberg:1936nmg,Karplus:1950zza,Karplus:1950zz}, light-by-light scattering has been evidenced at high center-of-mass energies in ultraperipheral heavy-ion collisions at the LHC \cite{ATLAS:2017fur,CMS:2018erd,ATLAS:2019azn}, but never directly observed in the low-energy regime accessible to colliding photon beams (see \cite{Ahmadiniaz:2024xob} and references therein), where past searches with optical lasers and X-ray beams have set bounds far above the QED light-by-light scattering cross section \cite{Moulin:1996vv,Bernard:2000ovj,Yamaji:2016xws,Watt:2024brh}.

In such a pump--probe collision, the interaction induces a quantum vacuum signal: the probe angular distribution changes, and a fraction $\sim 10 \%$ of the signal photons has a polarization flipped by $90^{\circ}$ relative to the probe.
Resolving the polarization of the signal is what makes the low-energy regime (i.e. a probe photon energy below the electron rest mass energy $h \nu \lesssim m_e c^2$) particularly attractive: the two low energy constants $a$ and $b$ governing the nonlinear self-interaction of the electromagnetic field at leading order in a weak-field expansion in QED \cite{Heisenberg:1936nmg,Ritus:1975pcc,Dittrich:1985yb,Gies:2016yaa} can be extracted from the concurrent measurement of $N_\parallel$ and $N_\perp$ \cite{Karbstein:2022uwf} when the relative pump-probe polarisation angle is 45°, as: 
\begin{equation}
	a \propto \sqrt{N_\parallel} - \sqrt{N_\perp},
	\qquad
	b \propto \sqrt{N_\parallel} + \sqrt{N_\perp},
	\label{eq:ab_meas}
\end{equation}
with the numerical values of $a=4$ and $b=7$ at leading loop order in QED.

\begin{figure*}[ht!]
	\centering
	\includegraphics[width=\textwidth]{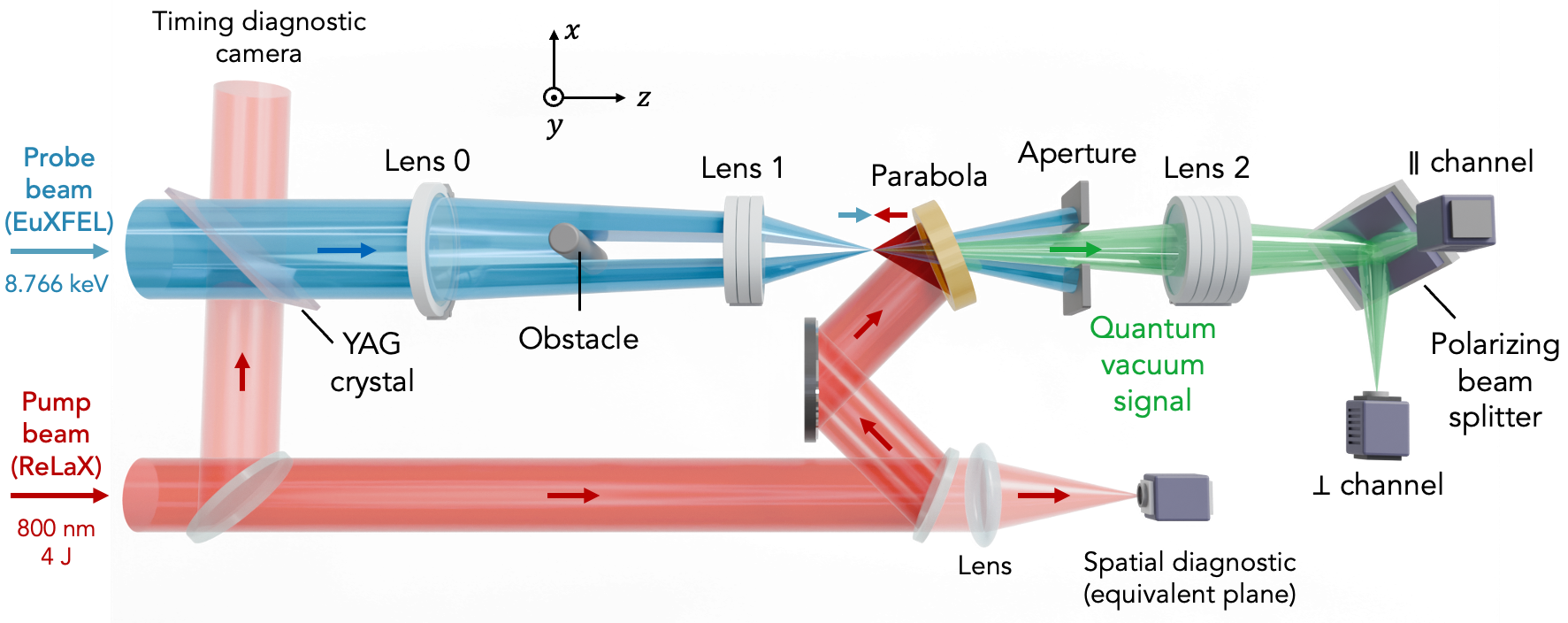}
	\caption{Experimental setup. The X-ray probe (blue) is pre-focused by Lens~0, partially blocked by the obstacle, and focused by Lens~1 to collide at its focus with the optical pump (red), which is focused by an $f/1.3$ off-axis parabola. Probe photons scattered by the pump and the quantum vacuum signal (green), pass the aperture, are refocused by Lens~2, and are split by the polarizing beam splitter into the parallel ($\parallel$) and perpendicular ($\perp$) channels, each recorded by its own detector. The unscattered probe is blocked by the aperture and the polarizing beam splitter. Also shown are the two per-shot proxy diagnostics described in Sec.~\ref{sec:overlap}: the timing diagnostic, based on a YAG crystal intercepting both beams upstream of the dark-field setup, and the spatial diagnostic, which images the pump equivalent plane by refocusing the leakage of the pump through its last turning mirror.}
	\label{fig:setup}
\end{figure*}

The experimental challenge is one of background suppression: every shot typically delivers $\lesssim 1.1\times10^{11}$ probe photons, of which only a small fraction is expected to scatter off the interaction with the pump. A number of schemes have been considered to separate the signal from the probe \cite{Ahmadiniaz:2024xob}. We note that schemes based on polarization filtering alone rely on X-ray polarimeters of purity $\sim 10^{-12}$ \cite{Aleksandrov:1985,Marx:2013,Schulze:2022,Heinzl:2006xc,DiPiazza:2006pr,Schlenvoigt:2016,Mosman:2021vua} and give access only to the flipped photon signal $N_\perp\sim(a-b)^2$ and preclude the constants $a$ and $b$ from being determined separately.
The dark-field scheme \cite{Karbstein:2020gzg,Karbstein:2022uwf,Smid:2025wmj} adopted here rejects the probe primarily through the change of the angular distribution: an obstacle imprints a shadow onto the angular distribution of the probe, an aperture downstream of the interaction transmits only this shadow region, and a polarizing beam splitter finally separates the transmitted light into the $\parallel$ and $\perp$ channels. From 2751 pump--probe collisions at a center-of-mass photon energy of $\omega_\star = 116\,\mathrm{eV}$, we derive an upper limit on the light-by-light scattering cross section of $7\times10^{4}$ times the QED prediction at 95\% confidence level, improving on previous low-energy experiments by approximately ten orders of magnitude \cite{Bernard:2000ovj}. These results are reported in a companion Letter \cite{Matheron:2026}; the present article provides the details of the experimental setup and the simulations performed to arrive at these results.

The present article is organized as follows. Section~\ref{sec:concept} introduces the dark-field concept, the experimental setup, and its working principle. Section~\ref{sec:beams} characterizes the probe and pump beams. Section~\ref{sec:overlap} describes how the spatiotemporal overlap of the two beams was established and tracked for every shot. Section~\ref{sec:polarimetry} is devoted to the polarizing beam splitter. Section~\ref{sec:data} presents the photon counts measured in the two polarization channels. Section~\ref{sec:simulations} details the start-to-end simulations of the setup performed within the diffractive beam propagation code ``Vacuum Induced Birefringence Explorer'' (VIBE) \cite{Matheron:2025owt}. Section~\ref{sec:scattering} discusses the dominant sources of background, and Sec.~\ref{sec:limit} derives the resulting upper bound on the QED light-by-light scattering cross section.

\section{The dark-field concept and its implementation}
\label{sec:concept}

\subsection{Concept}
\label{sec:df_concept}
The dark-field scheme \cite{Karbstein:2020gzg,Karbstein:2022uwf,Smid:2025wmj} exploits the fact that light-by-light scattering modifies two observables of the scattered probe photons at once: their angular distribution and, for part of them, their polarization. The idea is to prepare a probe beam with a ``hole'' in its angular spectrum, by blocking its central part with an obstacle before focusing, and to detect scattered photons precisely inside that hole. In the plane where the obstacle is imaged, after the interaction, the unscattered probe is concentrated in two bright lobes on either side of the obstacle image and can be blocked by an aperture, whereas the modified angular distribution of the signal photons can partially pass through. The transmitted light is then split into the $\parallel$ and $\perp$ channels by an analyzer crystal operated in the Baronova configuration \cite{Baronova:2003,Wallace:2020,Presura:2021}, hereafter referred to as the polarizing beam splitter, completing the two-observable separation.

Throughout this paper, we use the following terminology. The quantum vacuum signal refers to the probe photons that underwent light-by-light scattering off the pump at the interaction point; they pass the aperture by virtue of the change in angular distribution, and a fraction of them additionally undergoes the $90^{\circ}$ polarization flip that directs them into the $\perp$ channel. The background refers to the probe photons that did not scatter off the pump but nevertheless reach the detectors, because they were imperfectly rejected by the dark-field optics (obstacle and aperture) and/or by the polarizing beam splitter. At the detector, background photons are indistinguishable from the quantum vacuum signal on a photon-by-photon basis; their suppression and characterization are therefore central to the experiment (Sec.~\ref{sec:scattering}).

\subsection{Implementation at the European XFEL}
\label{sec:setup}

The experiment was conducted at the High Energy Density (HED) scientific instrument \cite{Zastrau:2021} of the European XFEL \cite{Decking:2020} and is sketched in Fig.~\ref{fig:setup}. The probe is an X-ray seeded beam of pulse energy $\sim$\SI{150}{\micro J} and horizontally polarized. The pump is the ReLaX optical laser (\SI{800}{nm} central wavelength, about \SI{4}{J} on target; Sec.~\ref{sec:pump}), linearly polarized at $45^{\circ}$ with respect to the probe polarization and colliding with the probe in a counter-propagating geometry.

Following the beam path, the probe is first pre-focused by Lens~0, a polished-diamond compound refractive lens (CRL) \cite{Snigirev1996,Tomie1997, Celestre:2020}, and its central part is then blocked by the obstacle, a vertical polished tungsten wire of \SI{250}{\micro m} diameter. The resulting beam is focused by Lens~1 (a second polished-diamond CRL) and collides at its focus with the pump, which is focused by an $f/1.3$ off-axis parabola facing the probe. After the interaction point the probe passes through a central hole drilled in the parabola and reaches the aperture, placed in the plane where Lens~1 images the obstacle. There, the unscattered probe is blocked while the quantum vacuum signal passes through the central opening, as described above. Lens~2 (a third polished-diamond CRL) refocuses the transmitted light through a secondary spatial filter, the detector aperture (not represented in Fig.~\ref{fig:setup} for simplicity), onto the polarizing beam splitter, a germanium crystal in the Baronova configuration \cite{Baronova:2003,Wallace:2020,Presura:2021} (Sec.~\ref{sec:polarimetry}) that routes each polarization to its own detector: photons retaining the initial horizontal probe polarization go to the $\parallel$ channel, photons flipped by $90^{\circ}$ to the $\perp$ channel. Each channel is recorded by a Jungfrau 500k detector \cite{Sikorski:2023} with \SI{75}{\micro m} pixels and quantum efficiency approaching unity. The exact positions, focal lengths, and further parameters of all elements are given together with their simulation model in Sec.~\ref{sec:vibe_aligned}.

The probe was operated at \SI{10}{Hz} and the pump at \SI{0.5}{Hz}, so that one probe shot in twenty is a pump--probe collision (``pump-on'' shot) while the remaining nineteen are probe-only (``pump-off'') shots. This interleaving is a major asset of the experiment: every pump-on shot is accompanied by several contemporaneous pump-off shots that sample the background under identical beamline conditions, especially the shots directly before and after the pump-on shot, enabling the powerful per-shot background subtraction of Sec.~\ref{sec:limit}. In total, the data set comprises 2751 pump-on and 52441 pump-off shots.

\section{Probe and pump characterization}
\label{sec:beams}

\subsection{Probe characterization}
\label{sec:probe}
\begin{figure}[hbt]
	\centering
	\includegraphics[width=0.95\columnwidth]{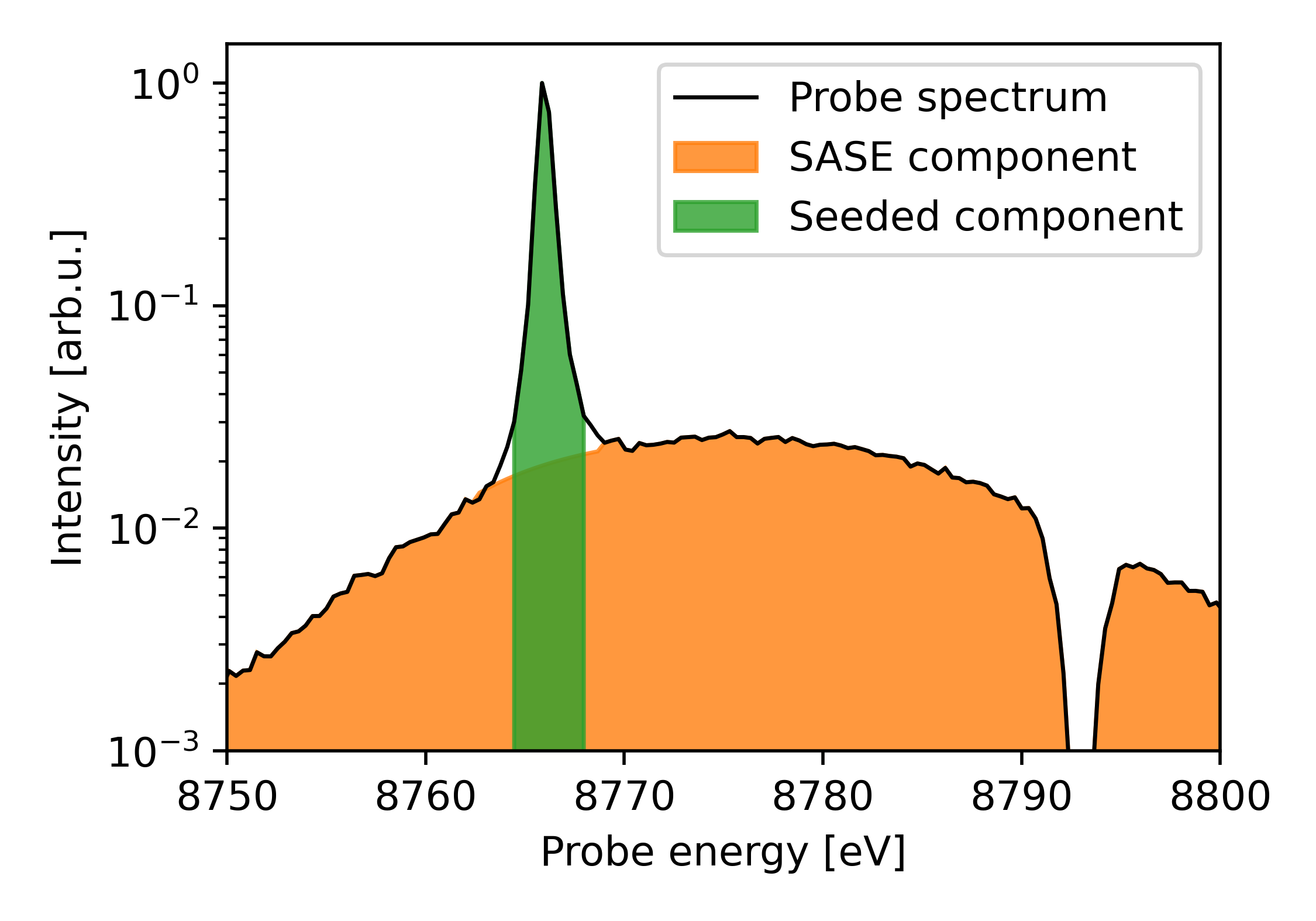}	
	\caption{Probe spectrum. The measured spectrum (black line) is decomposed into its broadband SASE component (orange) and its seeded component (green), the latter having a FWHM bandwidth of \SI{0.7}{eV} centered at about \SI{8766}{eV} and carrying 61\% of the integrated intensity of the pulse.}
	\label{fig:Probe_spec}
\end{figure}

\begin{figure}[hbt]
	\centering
	\includegraphics[width=1.15\columnwidth]{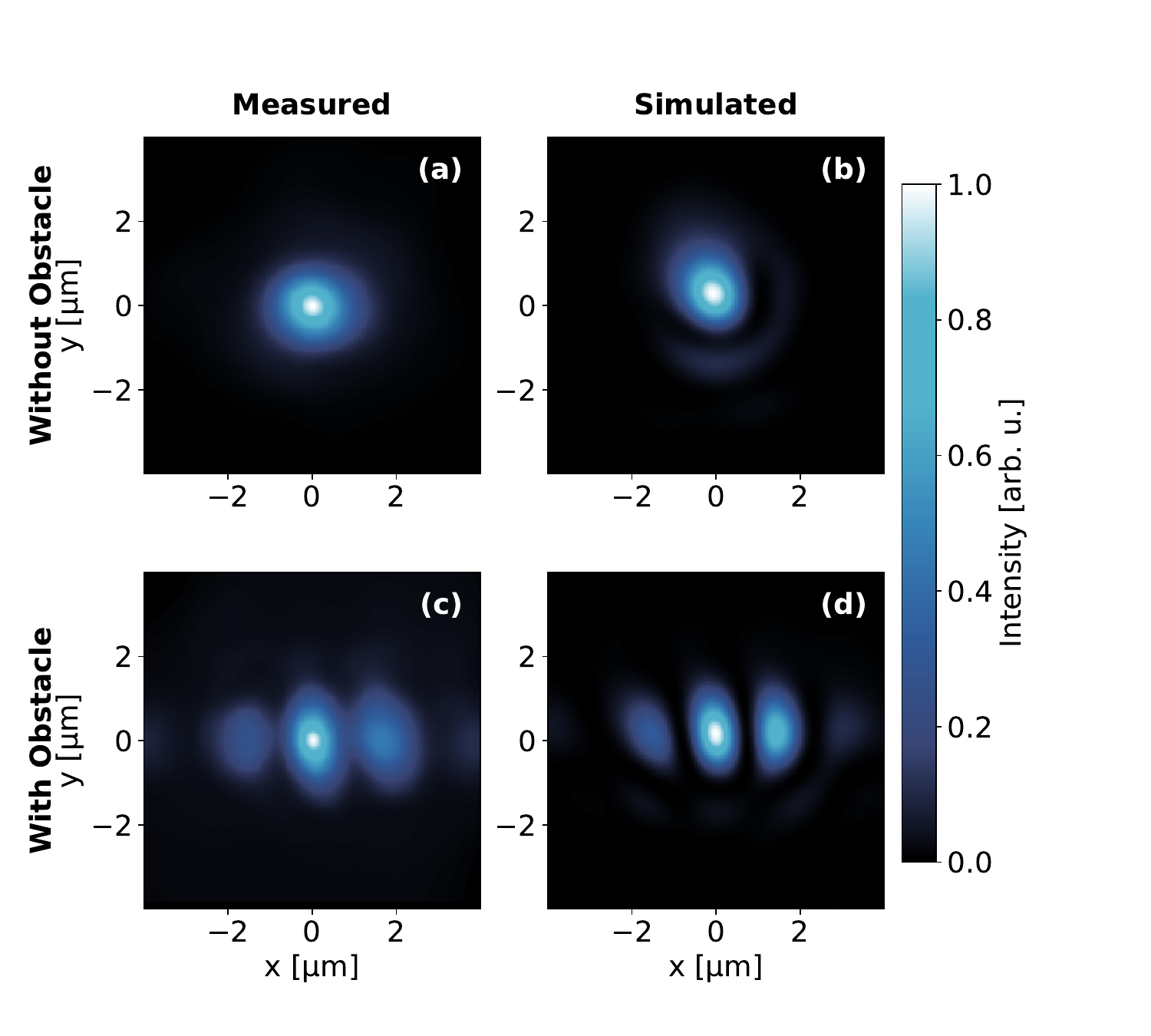}
	\caption{Probe spatial intensity distribution at focus. Measured (left column) and simulated (right column) profiles without the wire obstacle [(a) and (b)] and with the obstacle inserted [(c) and (d)]. All profiles are normalized to their maximum intensity.}
	\label{fig:imprints}
\end{figure}

The probe beam is generated by the European XFEL via a self-seeding process requiring two undulator sections, one triggering the self-amplified spontaneous emission (SASE) and one performing the seeded amplification, which produces a quasi-monoenergetic, intense X-ray beam of typical pulse energy \SI{150}{\micro\joule} on target. The probe spectrum is shown in Fig.~\ref{fig:Probe_spec}. While the SASE component has a large energy bandwidth, the seeded peak has a full width half-maximum (FWHM) bandwidth of only \SI{0.7}{eV}, centered at about \SI{8766}{eV}, and carries 61\% of the integrated intensity of the entire pulse.
The generated beam is then guided by two sets of CRLs positioned at about \SI{274}{m} and \SI{902}{m} from the undulator exit. They deliver an essentially collimated beam with a size of about \SI{500}{\micro m} at the entrance of the dark-field setup (Lens~0), located about \SI{1007}{m} downstream of the undulator exit. A measurement of the probe spatial profile at the entrance of the dark-field
setup (before Lens~0) is shown in Fig.~\ref{fig:VIBE1}(b); it is also used as
input to the simulations of Sec.~\ref{sec:vibe_aligned}. The beam is then
brought to a focus at the interaction point by the combination of Lens~0 and
Lens~1. The probe profile at focus was both measured and simulated, with and
without the wire obstacle, as illustrated in Fig.~\ref{fig:imprints}. Without
the obstacle [Fig.~\ref{fig:imprints}(a)], the focal spot has a FWHM size of
\SI{1.3}{\micro m}. Inserting the obstacle produces the characteristic
diffraction pattern shown in Fig.~\ref{fig:imprints}(c), with a fringe
spacing of \SI{1.7}{\micro m} and a central-spot FWHM of \SI{0.9}{\micro m}.
The simulations [Figs.~\ref{fig:imprints}(b) and \ref{fig:imprints}(d)]
reproduce the measured focal distributions in both configurations. The focused beam profiles were characterized using the ablation imprint method \cite{Chalupsk2010,Chalupsk2023}. Ablation imprints were produced in a thin layer of lead iodide deposited on a diamond substrate (PbI2/CVD) and imaged using scanning electron microscopy. The threshold boundaries of the imprints correspond to iso-fluence contours of the beam at a fixed threshold fluence. By varying the pulse energy, and consequently the peak fluence, a set of contours was obtained, enabling reconstruction of the beam fluence profile. Figs.~\ref{fig:imprints}(a) and (c) show the reconstructed fluence profiles, obtained from the ablation contours and interpolated using the thin-plate spline method \cite{Duchon:1976}.

\subsection{Pump characterization}
\label{sec:pump}

\begin{figure*}[hbt]
	\centering
	\includegraphics[width=2.0\columnwidth]{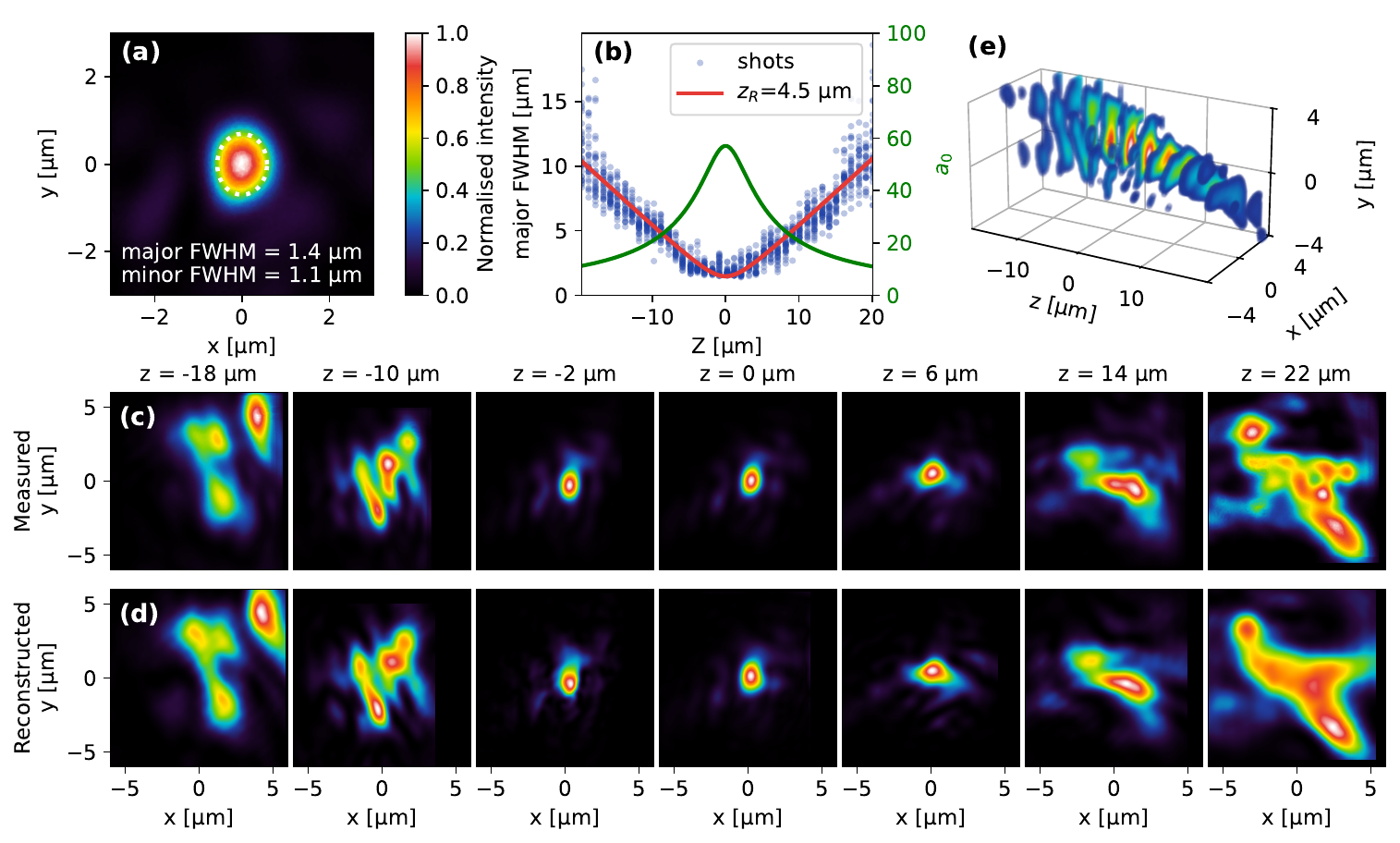}
	\caption{Pump laser characterization. (a) Measured focal spot at best focus. The dotted yellow ellipse indicates the FWHM of the Gaussian fit. (b) Major-axis FWHM of the beam through the focus scan (blue dots: direct shot measurements; red line: beam caustic fit) and associated inferred $a_0$ (green line, right axis). (c) Measured and (d) reconstructed transverse intensity profiles at several positions $z$ along the caustic. (e) Rendering of the three-dimensional intensity distribution of the reconstructed field.}
	\label{fig:relax_focus}
\end{figure*}

The pump beam is delivered by the ReLaX laser \cite{LasoGarcia:2021} operated by HED, a \SI{300}{TW}-class titanium:sapphire system with a central wavelength of \SI{800}{nm}, a pulse duration of \SI{23}{fs} (intensity FWHM), and an energy of \SI{6.8}{J} before the compressor, of which 65\% is transmitted to the interaction point.
It is focused by an off-axis parabola with a \SI{15}{cm} focal length and an effective diameter of \SI{11.5}{cm}, limited by the near-field beam size (compared with the nominal \SI{15}{cm} diameter reported in \cite{LasoGarcia:2021}). This corresponds to an effective f-number of $f/1.3$, rather than the nominal $f/1$.

The ReLaX pump beam was systematically characterized just before the data-taking measurements. Figure~\ref{fig:relax_focus}(a) shows a typical focal spot at the interaction plane (later fed into the simulation of the interaction, see Sec.~\ref{sec:vibe_aligned}). The transverse profile is fitted with a rotated two-dimensional Gaussian,
\begin{equation}
\begin{aligned}
A \exp\!\Bigl[-\bigl(&\alpha(x-x_0)^2 + 2\beta (x-x_0)(y-y_0) \\
&+ \gamma (y-y_0)^2\bigr)\Bigr],
\end{aligned}
\end{equation}
where $x$ and $y$ are the horizontal and vertical coordinates, $A$ is a normalization constant, and the coefficients
$\alpha = \frac{\cos^2\theta}{2\sigma_x^2} + \frac{\sin^2\theta}{2\sigma_y^2}$,
$\beta  = \frac{\sin 2\theta}{4\sigma_x^2} - \frac{\sin 2\theta}{4\sigma_y^2}$, and
$\gamma = \frac{\sin^2\theta}{2\sigma_x^2} + \frac{\cos^2\theta}{2\sigma_y^2}$
depend on the widths $\sigma_{x,y}$ and the rotation angle $\theta$ of the elliptical spot. At best focus the spot measures \SI{1.4}{\micro m} (major axis) by \SI{1.1}{\micro m} (minor axis) FWHM, with $\theta \simeq 5^{\circ}$, i.e., the major axis is essentially vertical. The typical laser spot quality factor, defined as the encircled-energy fraction inside the fitted FWHM ellipse, is measured to an average of $q = 0.24$. For comparison, $q$ reaches 0.5 for a perfect gaussian and $0.47$ for an Airy spot. A longitudinal scan of the measured focal spot was also performed around the best focus position and is shown in Fig.~\ref{fig:relax_focus}(b), where the major-axis FWHM as a function of $z$ is fitted with the beam caustic
\begin{equation}
	w_{\mathrm{FWHM}}(z) = w_{0,\mathrm{FWHM}} \sqrt{1+\left(\frac{z-z_0}{z_R}\right)^{2}},
\end{equation}
where $z_0$ is the focus position and the fitted Rayleigh range is $z_R = \SI{4.5}{\micro m}$. Combining the spot size with the pulse energy and duration, and assuming a Gaussian temporal profile, yields the normalized vector potential along the caustic shown in green in Fig.~\ref{fig:relax_focus}(b), peaking at $a_0 \sim 60$, corresponding to a peak intensity of \SI{7.7e21}{W/cm^2}.

The full complex laser field, including its phase, is reconstructed at every $z$ position, including planes where no data were taken, with a Gerchberg--Saxton-type phase-retrieval algorithm \cite{GerchbergSaxton:1972, Allen_2001_Phase_retrieval_OpC, Chalupsky_2015_Imprinting_PRA, Jelinek_2026_HPLSE}: the field is propagated numerically back and forth between the measured planes, and at each plane its amplitude is replaced by the square root of the measured intensity while the accumulated phase is retained; iterating this procedure converges to a phase consistent with all measured intensity profiles. The reconstruction is validated by comparing the measured intensities [Fig.~\ref{fig:relax_focus}(c)] with those of the reconstructed field, $|\vec{E}(x,y,z)|^2$ [Fig.~\ref{fig:relax_focus}(d)]; a rendering of the resulting three-dimensional intensity distribution is shown in Fig.~\ref{fig:relax_focus}(e). Small features differ between measurement and reconstruction because the algorithm was fed with shot-averaged profiles: the focus scan spans many shots subject to fluctuations, and the profiles at each $z$ were recentered before averaging to avoid artificial broadening. The averaged profiles are thus representative of a typical shot, but they cannot correspond exactly to a single coherent field, which limits the achievable convergence.

\section{Spatiotemporal overlap of the probe and pump beams}
\label{sec:overlap}

\begin{figure*}[hbt]
	\centering
	\includegraphics[width=2\columnwidth]{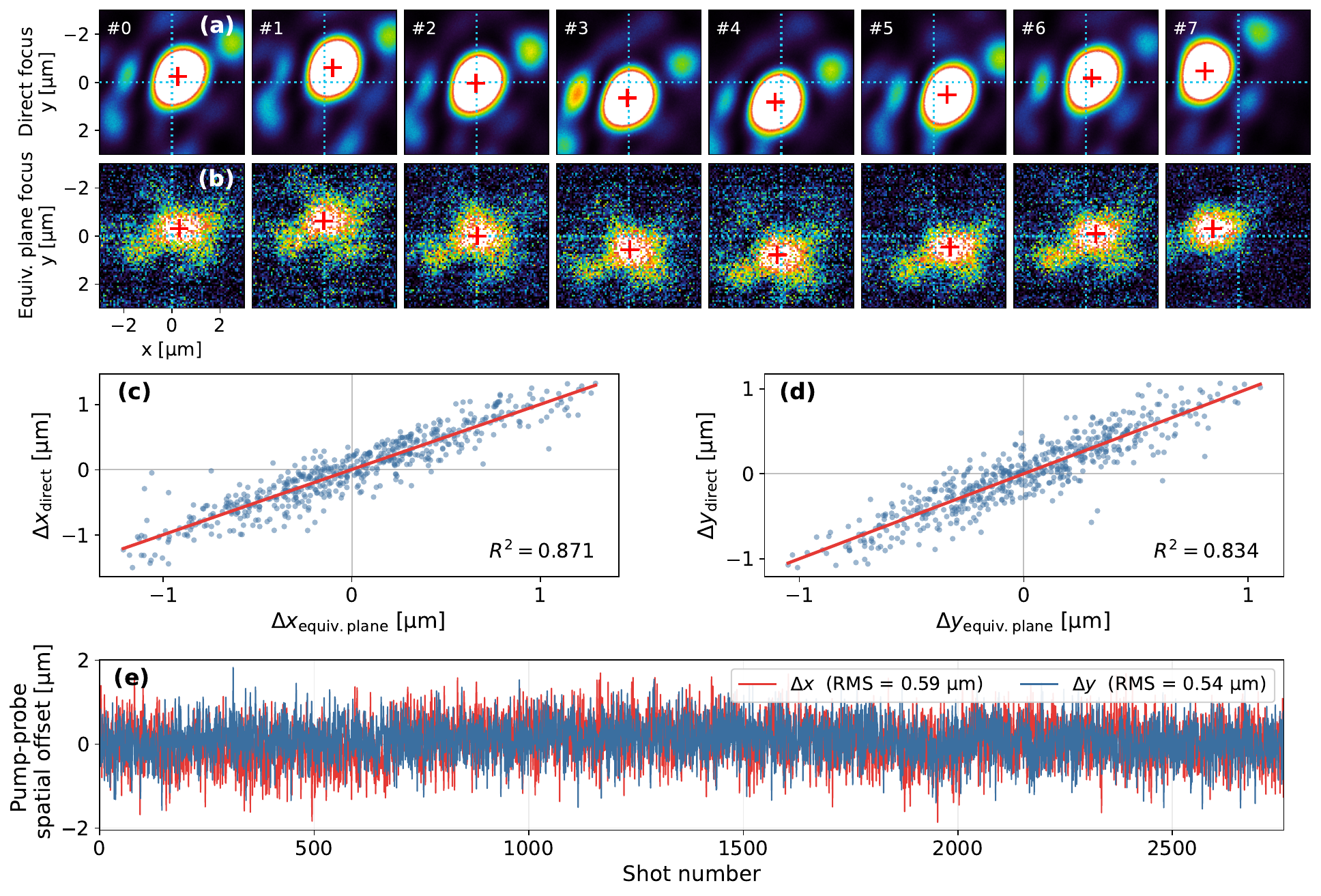}
	\caption{Pump--probe spatial alignment. Consecutive shots of the pump beam imaged by the microscope objective at the interaction point (a) and on the equivalent plane proxy diagnostic (b). The red cross marks the beam center computed via an rms calculation on the profile clipped at 20\% of the maximum. The blue dotted lines cross at $(x,y)=(0,0)$, defining the X-ray axis, assumed fixed in space. (c) and (d) Correlations between the interaction-point diagnostic and the equivalent plane diagnostic. The blue points are individual shot measurements taken from the same data set as the shots shown in (a)--(b), and the red line is the least-squares linear fit with its correlation coefficient $R^2$. (e) Shot-to-shot misalignment in $x$ (red) and $y$ (blue) over the full 2751 shots of the in-vacuum data taking.}
	\label{fig:spatial_alignment}
\end{figure*}

A decisive requirement of the experiment is to establish the spatial and temporal overlap of the pump and probe at the interaction point, and to track the overlap at every shot. Unlike experiments that can optimize alignment on their own signal, for instance when colliding an electron beam with laser and producing Compton scattering \cite{Matheron:2024gev}, the quantum vacuum signal is far too weak to serve as an alignment proxy. We therefore adopted a two-step strategy. 
First, the spatial and temporal overlap was established at the interaction point. Second, the spatial and temporal jitter was correlated with two proxy diagnostics that remain available during data taking: (i) for space, a camera imaging the equivalent plane of the pump, sensitive to the pointing jitter of the pump focus; and (ii) for time, the HED photon arrival monitor (PAM) \cite{Zastrau:2021} located upstream of the dark-field setup, based on a YAG crystal.

To align the spatial and temporal overlap with sufficient accuracy, a dedicated procedure and hardware were designed. 
First, a 50x Mitutoyo objective is inserted and used to image and optimize the focus of the pump beam with high resolution. Second, a timing sample (\SI{2}{\micro m} thick TiO$_2$ layer deposited on a \SI{20}{\micro m} thick diamond screen) is inserted, backlit by pump in flatfield mode, and its position along $z$ is adjusted into the focal plane of the 50x objective, therefore into pump focus. Now, the 50x lens can be exchanged for a 20x lens with a through hole for the probe beam. This allows observation of both beams on the same TiO$_2$ layer diagnostics, therefore aligning their spatial overlap in $x$ and $y$ direction. Furthermore, the sample material (TiO$_2$) is chosen so that the X-ray pulse upon its arrival creates a transient population of free carriers in its conduction and valence bands, which decreases its optical transmission. Therefore a dark spot appears in the imaged pump beam. The onset of this absorption is about 6 fs after irradiation, therefore enabling precise time zero setup.

\subsection{Spatial alignment}
\label{sec:spatial}

The pump equivalent plane camera exploits the leakage of the pump through the last turning mirror before the focusing parabola, focusing this leak with a lens of \SI{4}{m} focal length. To calibrate it, the attenuated pump beam was imaged with a 50$\times$ microscope objective, so that its focus was observed simultaneously at the interaction point [Fig.~\ref{fig:spatial_alignment}(a)] and on the equivalent plane camera [Fig.~\ref{fig:spatial_alignment}(b)]. The intensity centroid (red cross) was tracked on both diagnostics for each shot, and the transverse coordinates $(x,y)$ of the centroid were correlated [Figs.~\ref{fig:spatial_alignment}(c) and \ref{fig:spatial_alignment}(d)], yielding correlation coefficients of $R^2 = 0.87$ in $x$ and $R^2 = 0.83$ in $y$. This confirms that, once calibrated, the equivalent plane camera measures the pump focus position at the interaction point even after the screen and microscope objective are driven out of the beam path for in-vacuum data taking.

The calibrated equivalent plane camera then provided the transverse offset $(\Delta x, \Delta y)$ of the pump focus for each of the 2751 pump-on in-vacuum data-taking shots [Fig.~\ref{fig:spatial_alignment}(e)], with rms jitter of \SI{0.59}{\micro m} in $x$ and \SI{0.54}{\micro m} in $y$. The pointing jitter of the probe focus at the interaction point was measured to be smaller and is therefore neglected in the following.

\subsection{Temporal alignment}
\label{sec:timing}

\begin{figure}[hbt]
	\centering
	\includegraphics[width=\columnwidth]{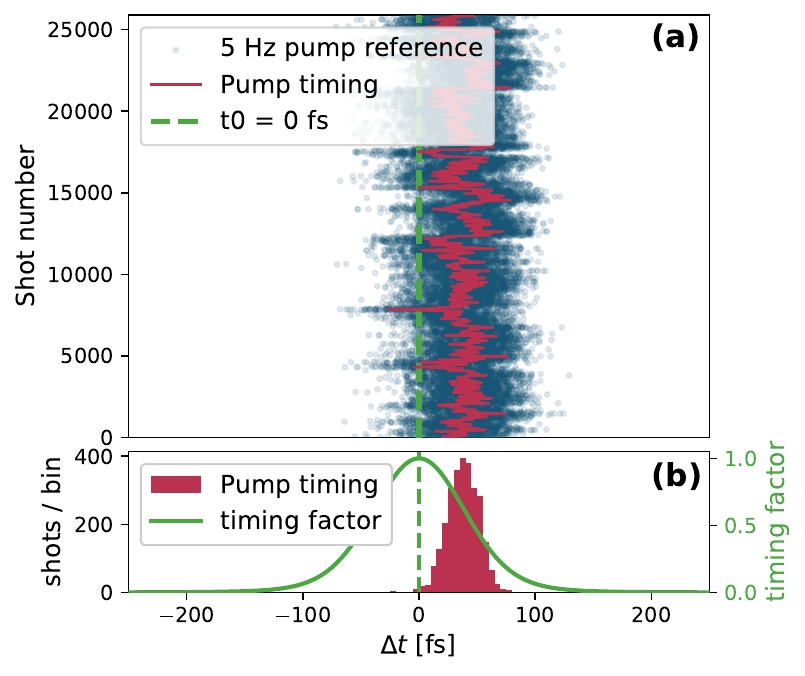}
	\caption{Pump--probe timing alignment. (a) Waterfall view of the timing difference $\Delta t$ between the pump and probe arrival times at the interaction point, as a function of shot number. The timing of the full-power pump shots is shown in red, while the diagnostic also allows tracking the timing of the laser oscillator operating at \SI{5}{Hz}, shown as blue dots. The green dashed line corresponds to a perfectly timed pump--probe collision, $\Delta t = 0$. (b) Histogram of the data shown in (a); the green curve shows the timing factor defined in Eq.~(\ref{eq:timing_factor}).}
	\label{fig:timing}
\end{figure}

Temporal alignment followed the same calibrate-and-remove strategy: a timing diagnostic at the interaction point was correlated with a permanent one upstream of the dark-field setup \cite{Zastrau:2021}, after which the former was removed. The upstream diagnostic is the HED photon arrival monitor (PAM) using a \SI{10}{\micro m} YAG screen intercepting both beams. After cross-calibration, the interaction-point diagnostic was removed, and the upstream monitor tracked the per-shot timing offset $\Delta t = t_{\mathrm{collision}} - t_0$, where $t_{\mathrm{collision}}$ is the measured arrival-time difference of the two pulses and $t_0$ is the optimum for which both pulses meet at the focus. Figure~\ref{fig:timing}(a) shows the timing measured during the data run. The upstream diagnostic operates on the pump oscillator at \SI{5}{Hz} (blue points), providing higher statistics than the \SI{0.5}{Hz} full-power shots (red line). The histogram of the full-power timing [Fig.~\ref{fig:timing}(b)] is centered about \SI{30}{fs} away from $t_0$.

A timing offset reduces the number of quantum vacuum signal photons by the timing factor. For pump and probe lasers that are well-characterized by Gaussian temporal pulse shapes, an approximate integral expression is available for this timing factor \cite{Karbstein:2018omb} (see also \cite{Matheron:2025owt}). It is given by 
\begin{equation}
	\mathcal{T}(\Delta t) = \tilde{F}\Bigl(\frac{2z_\mathrm{R}}{cT}, \frac{\Delta t}{T}, \frac{t_\mathrm{probe}}{T}\Bigr)\,/\,
	\tilde{F}\Bigl(\frac{2 z_\mathrm{R}}{cT}, 0, \frac{t_\mathrm{probe}}{T}\Bigr)\,,
	\label{eq:timing_factor}
\end{equation}
with
\begin{multline}
	\tilde{F}(\tilde\chi,\tilde\chi_0,\rho)
	= \sqrt{\frac{\pi}{2}}\frac{\ln^{3}2}{1+2\rho^2}\,\tilde\chi^2
	\int_{-\infty}^\infty {\rm d}K\, {\rm e}^{-2\ln^{2}2\,K^2}\\
	\times
	\biggl|
	\sum_{s=\pm1}
	{\rm erfcx}\Bigl(
	\frac{\sqrt{2}\ln 2}{\sqrt{1+2\rho^2}}
	\bigl[
	s (\rho K -{\rm i}\tilde\chi_0)
	+ \tilde\chi
	\bigr]
	\Bigr)
	\biggr|^2 .
	\label{eq:tildeF}
\end{multline}
Here, $z_\mathrm{R}$ is the Rayleigh range of the pump measured in Fig.~\ref{fig:relax_focus}(b), the time scale $T$ is related to the pump FWHM pulse duration $\tau_{\rm pump}$ via $T=\sqrt{\ln2/2}\,\tau_{\rm pump}$, $t_{\mathrm{probe}}$ is the probe FWHM pulse duration, and ${\rm erfcx}(x)$ denotes the scaled complementary error function. The timing factor $\mathcal{T}$ is shown in green in Fig.~\ref{fig:timing}(b). For the data presented here, the timing factor at the peak of the timing histogram is about $\mathcal{T}\simeq 0.5$ and fluctuates for every single shot.

\subsection{Per-shot spatiotemporal alignment}
\label{sec:pershot}

\begin{figure}[hbt]
	\centering
	\includegraphics[width=1\columnwidth]{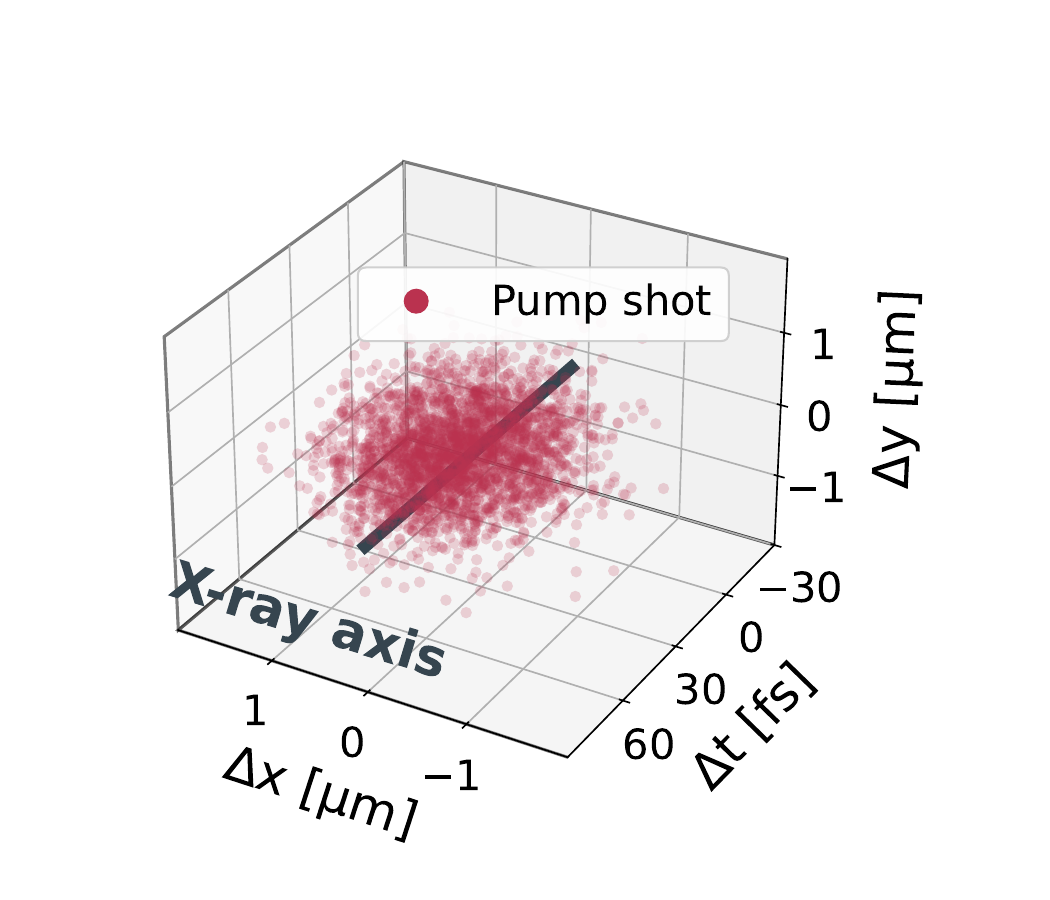}
	\caption{Spatiotemporal misalignment of each pump--probe collision shot with respect to the X-ray axis. Each red dot represents one of the 2751 pump laser shots. The dark line represents the X-ray axis.}
	\label{fig:3Dscatter}
\end{figure}

Combining the two proxy diagnostics yields, for every pump-on shot, the misalignment in the two transverse directions and in time, $(\Delta x, \Delta y, \Delta t)$. Figure~\ref{fig:3Dscatter} shows where each of the 2751 pump shots landed relative to the (fixed) probe axis. This per-shot information is the key input that allows the simulations of Sec.~\ref{sec:simulations} to predict the expected quantum vacuum signal of every individual shot.

\section{Polarimetry measurements}
\label{sec:polarimetry}

The polarizing beam splitter is based on a germanium crystal used in the Baronova configuration \cite{Baronova:2003,Wallace:2020,Presura:2021}: the incident beam is Bragg-diffracted such that the polarization component parallel to the incident probe polarization is routed into the $\parallel$ channel (vertical in our setup), while the orthogonal component is routed into the $\perp$ channel (horizontal in our setup), allowing both channels to be recorded simultaneously.

\begin{figure}[hbt]
	\centering
	\includegraphics[width=\columnwidth]{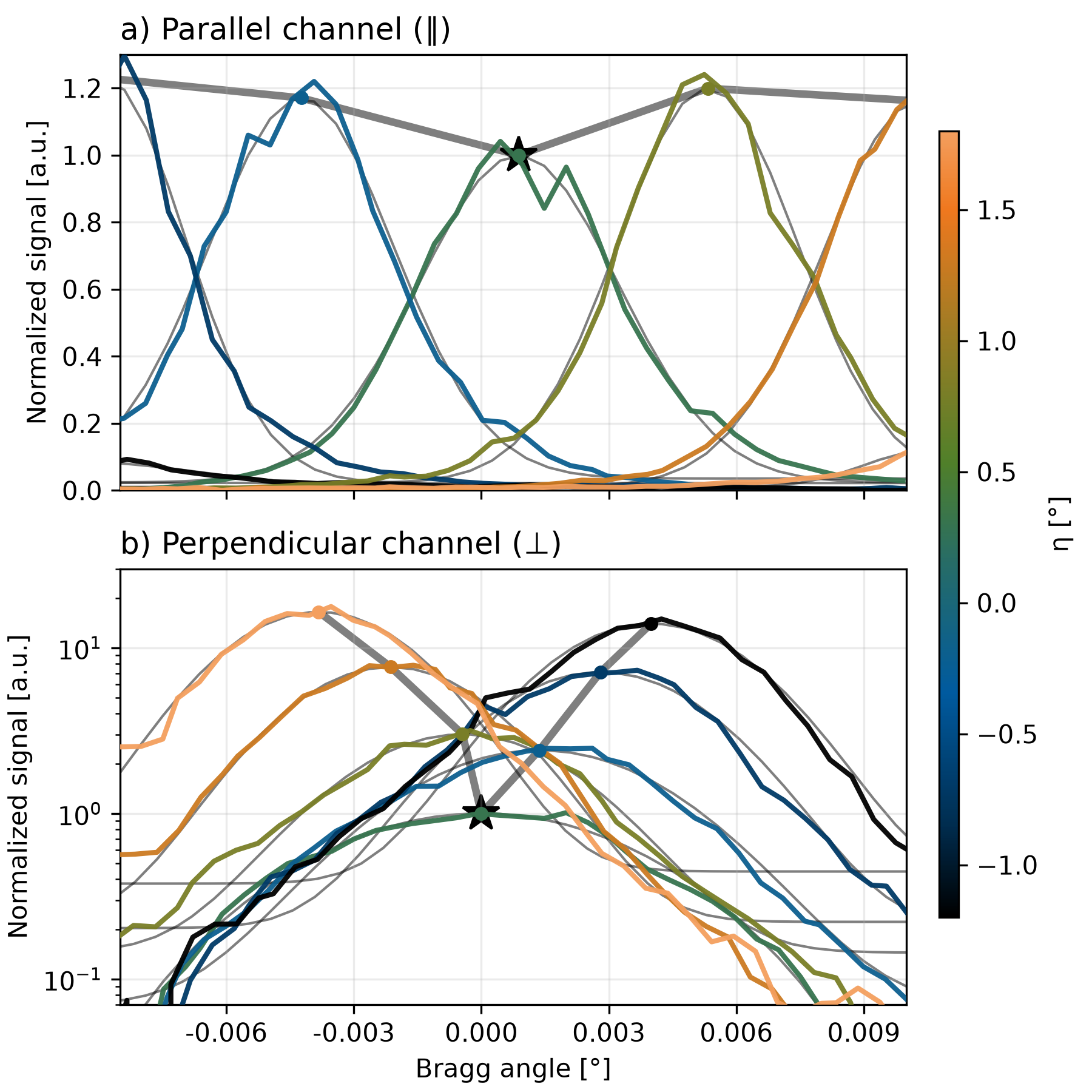}
	\caption{Rocking curves of the polarizing beam splitter for various polarization angles $\eta$, measured with all lenses inserted. $\parallel$ channel (a) and  $\perp$ channel on a logarithmic scale (b). Each rocking curve is fitted with a Gaussian (gray lines), and their maxima are connected by thick gray lines. The minimum of these maxima in the $\perp$ channel (star) defines the optimal crystal configuration.}
	\label{fig:rocking}
\end{figure}

\begin{figure}[hbt]
	\centering
	\includegraphics[width=\columnwidth]{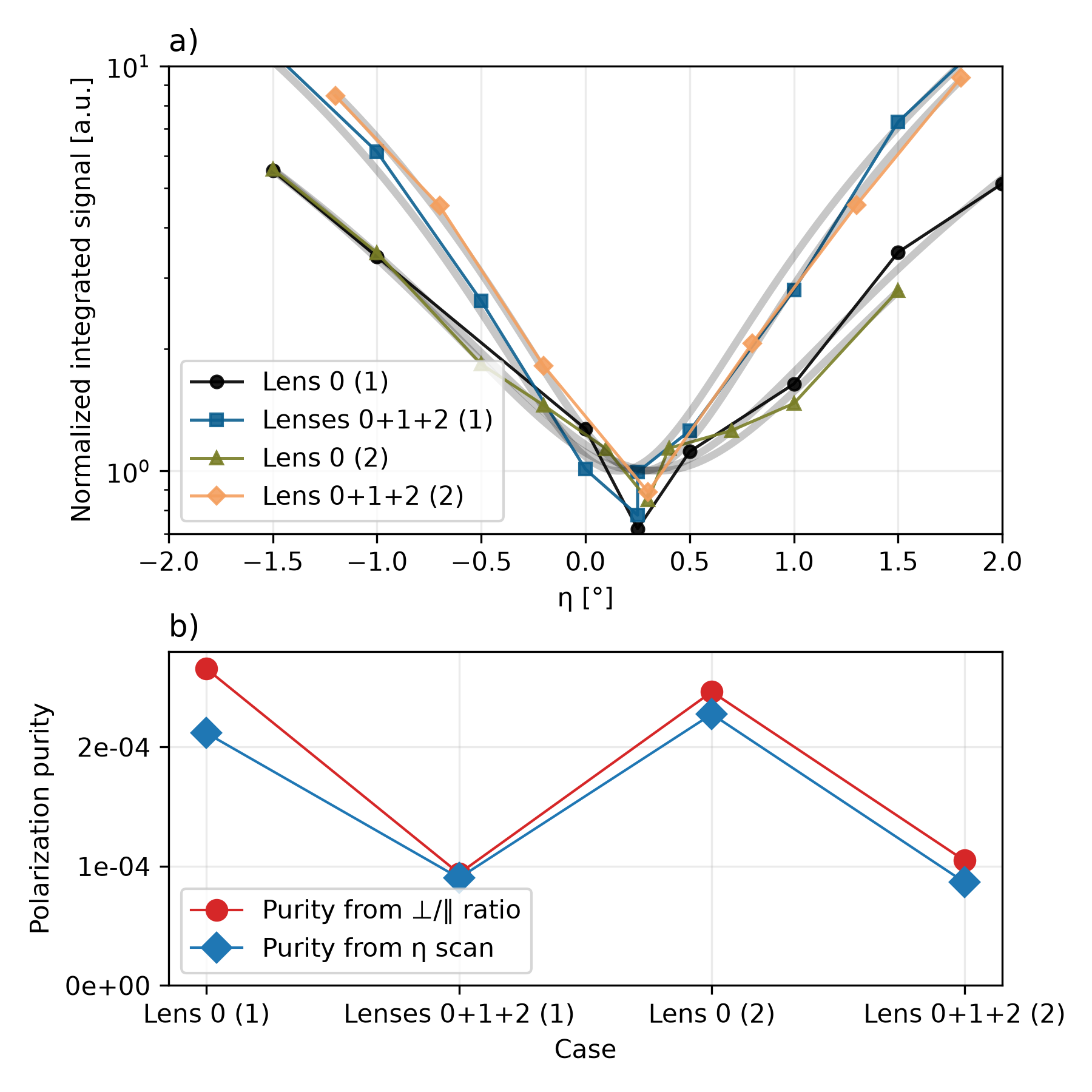}
	\caption{$\eta$ scans for various lens configurations, each fitted with the Malus law in gray lines(a); Measured purities for the four setup cases, fitted from the $\eta$ scans and obtained from the ratio of the $\perp$ to $\parallel$ channel signals (b). The number in parenthesis indicate the measurement day.}
	\label{fig:eta_scans}
\end{figure}

In order to set it up properly, all three rotation angles have to be precisely adjusted. For a given azimuthal angle $\phi$, a set of rocking-curve scans over the Bragg angle $\theta$ was performed for various polarization angles $\eta$. These are shown in Fig.~\ref{fig:rocking}. All rocking curves are fitted with Gaussian functions, whose maxima are connected with a line. The minimum of all these maxima in the $\perp$ channel is marked with a star and defines the best setup configuration. If the maximum of the rocking curve for the corresponding $\eta$ (marked with a star in the upper panel) lies at the same Bragg angle, the angle $\phi$ is set up properly, otherwise it has yet to be adjusted.

The $\eta$ scans are calculated as integrals of the rocking curves over the Bragg angle. They are shown in Fig.~\ref{fig:eta_scans}(a) for several lens configurations. All are fitted with the Malus law, and the fitted polarization purity $P$ is plotted as the blue line in Fig.~\ref{fig:eta_scans}(b). The red line there shows the purity inferred as the ratio of the signal in the $\perp$ channel to that in the $\parallel$ channel. The agreement of both curves indicates that the crystal was sufficiently well aligned.
 
The four measurements were performed on different days, as indicated by the numbers in brackets in the figure. This confirms that the observed change in purity is a consequence of the lens configuration rather than a drift of the probe beam. The improvement in purity with the addition of further lenses can possibly be explained by their limited aperture, which may crop the outer regions of the probe beam where the polarization is less pure. Moreover, a full simulation of the spectrally and angularly resolved beam on the analyser crystal shows that a purity down to \num{1e-5} could in principle be reached with the current configuration.

\section{Measured photons in the \texorpdfstring{$\parallel$}\ \ and \texorpdfstring{$\perp$}\ \ channels}
\label{sec:data}

\begin{figure*}[hbt]
	\centering
	\includegraphics[width=2\columnwidth]{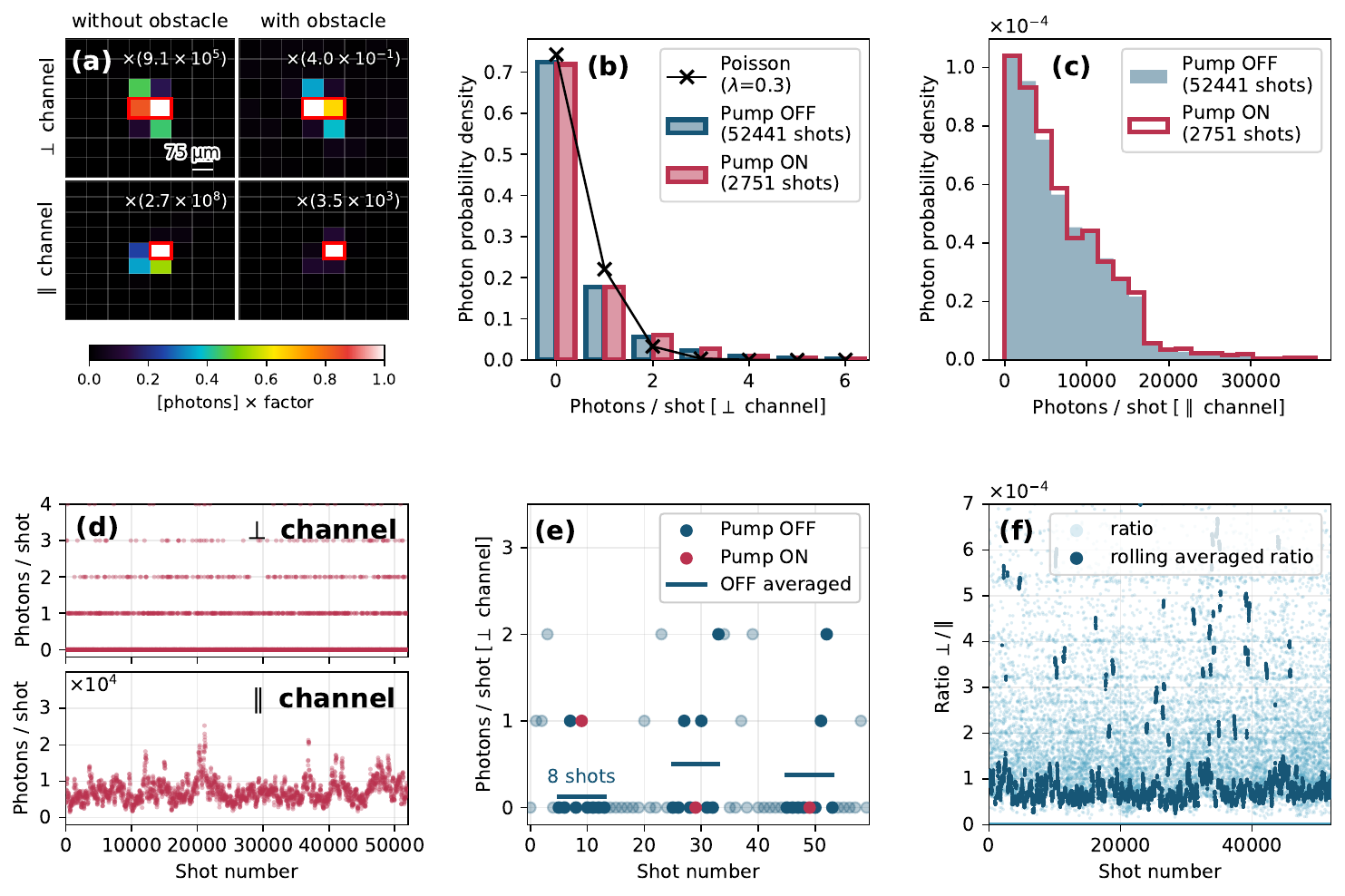}
	\caption{Photons measured in the $\perp$ and $\parallel$ channels during dark-field data taking. (a) Detector images of the $\perp$ (top row) and $\parallel$ (bottom row) channels, without the obstacle (left column) and with the obstacle inserted (right column). The images are averaged over 500 probe shots. The counts in each map are normalized to the maximum count of the image, indicated by the multiplicative factor on the image. The red rectangles show the pixels chosen for photon integration during data shots. (b) and (c) Photon probability distribution histograms (normalized to unit integral) for the $\perp$ channel (b) and the $\parallel$ channel (c). Pump-off shots are shown as blue filled bars and pump-on shots as red-framed distributions. The black curve in (b) is the Poisson fit obtained via $\chi^2$ minimization. (d) Time evolution of the photon counts in the $\perp$ (top) and $\parallel$ (bottom) channels for pump-on (red) and pump-off (blue) shots. (e) Zoom of the $\perp$-channel counts over a 60-shot range: pump-on shots in red, pump-off shots in blue; the 8 pump-off shots used for background subtraction around each pump-on shot are shown in dark blue, together with a horizontal line representing their average value. (f) Ratio of the photon counts detected in the $\perp$ channel to those in the $\parallel$ channel as a function of shot number: light blue dots show individual shots, dark blue dots the 8-shot rolling average.}
	\label{fig:perp_parall}
\end{figure*}

After the polarizing beam splitter, the photons of each shot are split between the $\parallel$ and $\perp$ channels and recorded on two dedicated Jungfrau 500k detectors \cite{Sikorski:2023}, with \SI{75}{\micro m} pixels and quantum efficiency approaching unity. Figure~\ref{fig:perp_parall}(a) shows the measured intensity distributions on the two detector planes ($\perp$ channel top row, $\parallel$ channel bottom row). To determine on which pixels the quantum vacuum signal is expected, a reference measurement was first taken with the obstacle removed: the probe then passes through the aperture and follows exactly the path expected for the quantum vacuum signal [left column of Fig.~\ref{fig:perp_parall}(a)]. This measurement shows that the signal is essentially concentrated on two pixels in the $\perp$ channel and on a single pixel in the $\parallel$ channel. The analysis therefore integrates only these pixels (framed in red), since including further pixels would only add background without appreciable signal. With the obstacle reinserted (dark-field configuration, right column), the count rate drops dramatically; the remaining counts are background leaking through the dark-field optics and the polarizing beam splitter.

The suppression of this background is the central experimental challenge. Although the aperture blocks the vast majority of the unscattered probe, photons deflected by scattering in the lenses, for instance (Sec.~\ref{sec:scattering}), can pass the aperture. The polarizing beam splitter then removes most of these, since they retain the incident probe polarization and are routed to the $\parallel$ channel; only the small fraction transmitted through the finite polarization purity contaminates the $\perp$ channel.

Figures~\ref{fig:perp_parall}(b) and \ref{fig:perp_parall}(c) show the photon probability density distributions in the two channels, separated into 2751 pump-on shots (containing quantum vacuum signal plus background) and 52441 pump-off shots (background only); Fig.~\ref{fig:perp_parall}(d) shows their evolution in time. In the $\perp$ channel most shots contain between 0 and 1 photons, with mean values of $0.4649$ (pump-on) and $0.4635$ (pump-off), an rms of $0.9570$, and standard errors of the mean of $0.0180$ and $0.0043$, respectively. In the $\parallel$ channel the counts range up to $\sim$\,20\,000 photons per shot, with means of $7264$ (pump-on) and $7193$ (pump-off), an rms of $6724$, and standard errors of the mean of $128$ and $29$. Transmission filters were placed in front of the $\parallel$-channel detector to protect it from excessive photon flux. The reported values are corrected for the filter transmission rather than being the raw detector counts. In both channels the pump-on and pump-off means agree within uncertainties: no quantum vacuum signal is observed at this level of sensitivity.

The width of the $\parallel$-channel distribution is dominated by the intrinsic fluctuations of the XFEL parameters, like pointing, divergence, and central energy. In the $\perp$ channel, by contrast, the combined spatial filtering (dark-field optics) and polarization filtering (polarizing beam splitter) suppress the transmitted intensity to the level where the distribution approaches a Poisson distribution of mean $\lambda = 0.3$ with a Fano factor~\cite{Cowan:1998ji} $F=1.97$ [Fig.~\ref{fig:perp_parall}(b)].

The high repetition rate of the XFEL (\SI{10}{Hz}) compared to the pump ReLaX laser (\SI{0.5}{Hz}) is an advantage for performing an efficient background subtraction. Figure~\ref{fig:perp_parall}(e) shows a zoom of the photon counts observed in the $\perp$ channel over 60 shots. For the background subtraction, an average over the 8 contemporaneous pump-off shots around a given pump-on shot is calculated. This average removes the long-term ($>\SI{1}{s}$) drifts of the XFEL machine while not being too time-averaged, thereby retaining the local information about the background level at that moment in time. Section~\ref{sec:limit_bkg} gives a detailed explanation of how the background subtraction is used to estimate the upper limit.

An instructive quantity is the shot-to-shot ratio of $\perp$ to $\parallel$ counts [Fig.~\ref{fig:perp_parall}(f)]. While it fluctuates from shot to shot with the XFEL parameters, its 8-shot rolling average is steady at $\sim 3 \times 10^{-4}$. In the present background-dominated regime this ratio simply measures the polarization purity achieved by the polarizing beam splitter (Sec.~\ref{sec:polarimetry}). As future experiments approach the signal-dominated regime, the same ratio should instead converge to the QED prediction $N_\perp / N_\parallel \simeq 9/121$ at leading loop order, giving direct access to the low-energy constants $a$ and $b$ through Eq.~\eqref{eq:ab_meas} \cite{Karbstein:2020gzg,Karbstein:2022uwf}.

\section{Simulations of the dark-field setup}
\label{sec:simulations}

Realistic simulations are a central element of the analysis: they must describe not only the pump--probe interaction itself, but the propagation of both the quantum vacuum signal and the background through the complete setup, from the initial probe to the detectors. The VIBE code \cite{Matheron:2025owt}  implementing a vacuum-emission module in the diffractive beam propagation toolkit LightPipes \cite{LightPipes:1997, LightPipes:2.1.5} was developed for exactly this purpose. It computes the dominant quantum vacuum signal in the X-ray domain induced in the interaction of the colliding laser fields, and propagates both the probe and the signal fields through every optical element of the beamline. To this end, it accepts measured beam profiles and measured optics imperfections as inputs, allowing for a faithful, start-to-end modeling of the experiment.

\subsection{Simulation of a perfectly aligned interaction}
\label{sec:vibe_aligned}

\begin{figure*}[hbt]
	\centering
	\includegraphics[width=2\columnwidth]{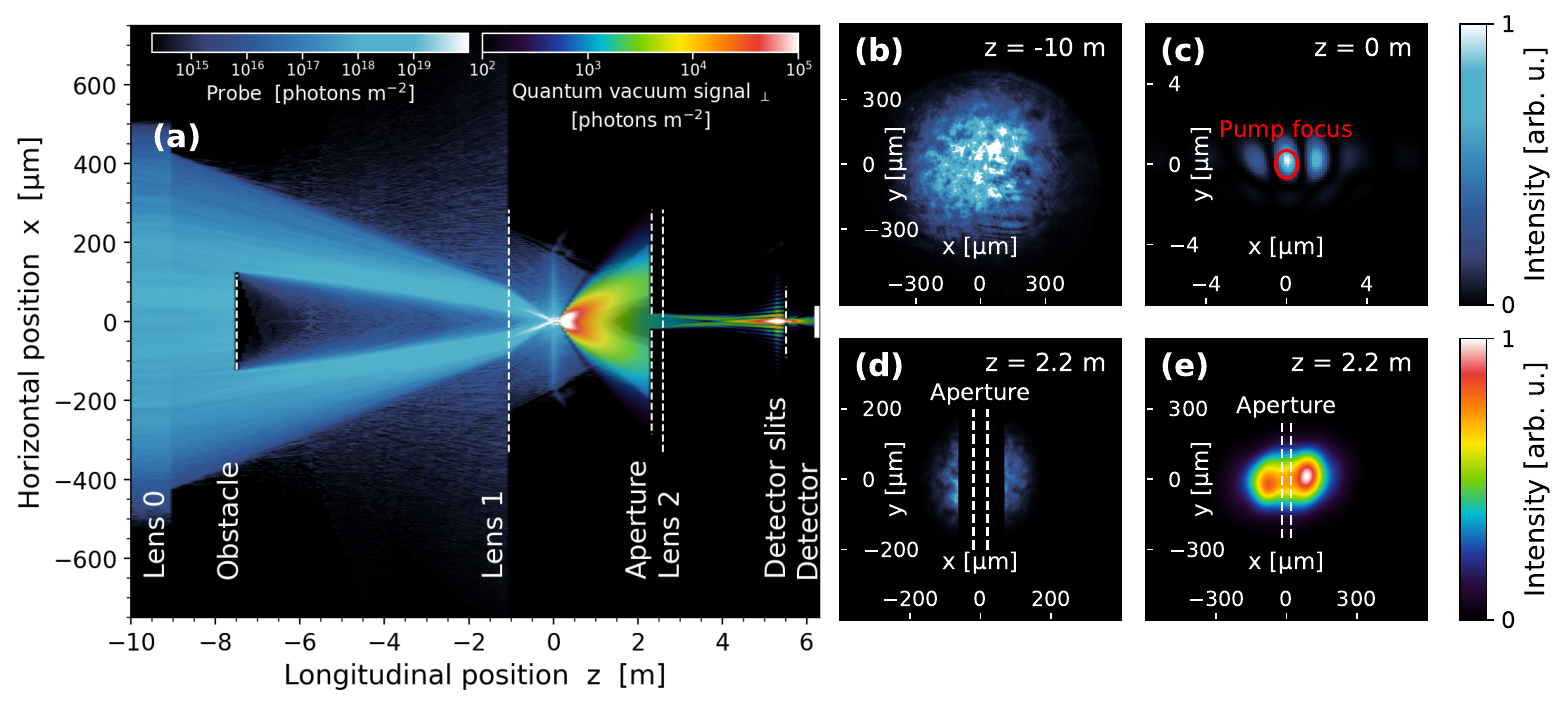}
	\caption{VIBE simulation of the experimental setup. (a) Top view of the simulated setup; the intensities of the probe and of the quantum vacuum signal are displayed on logarithmic color scales (upper color bars). The initial probe propagating from left to right is shown in blue. The obstacle, aperture, detector slits, and lenses are overlaid as white dotted lines. The quantum vacuum signal, emitted at $z=0$, is shown in a semi-transparent multicolor colormap. (b) Measured probe profile used as the initial profile in the VIBE simulation. (c) Simulated probe focus intensity distribution at $z=0$. The red ellipse represents the FWHM of the pump focus [fitted from Fig.~\ref{fig:relax_focus}(a)] for reference. (d) and (e) Probe and quantum vacuum signal intensity distributions just before the aperture. The two white dashed lines represent the aperture opening. Panels (b) to (e) are in linear color scale.}
	\label{fig:VIBE1}
\end{figure*}

Figure~\ref{fig:VIBE1}(a) shows the simulated setup in top view. The input probe profile is taken from a typical experimental shot [Fig.~\ref{fig:VIBE1}(b)] and kept fixed, as the results depend only weakly on the initial profile, which remained stable throughout data taking. The probe is pre-focused by Lens~0, a CRL consisting of a single diamond lens with radius of curvature \SI{200}{\micro m}, geometric aperture \SI{850}{\micro m}, and apex thickness $\sim$\SI{30}{\micro m}, located at $z = \SI{-9.1}{m}$, with focal length $f=\SI{10.54}{m}$ at \SI{8.766}{keV}. For all lenses in the simulation, both the absorption of the parabolic lens profile and the scattering caused by measured lens imperfections are included (Sec.~\ref{sec:lens_scattering}). The central part of the probe is then blocked by the obstacle, a vertical tungsten wire of \SI{250}{\micro m} diameter at $z = \SI{-7.52}{m}$. The probe reaching its focal plane at $z=0$, is focused by Lens~1, a CRL of three diamond lenses (radius of curvature \SI{125}{\micro m}, aperture \SI{475}{\micro m}, apex $\sim$\SI{30}{\micro m}) located at $z = \SI{-1.11}{m}$, with $f \simeq \SI{2.2}{m}$ at \SI{8766}{eV}. At its focus the probe exhibits the interference pattern shown in Fig.~\ref{fig:VIBE1}(c), where it collides with the pump, whose profile is the measured one of Fig.~\ref{fig:relax_focus}(a).

The quantum vacuum signal computed by VIBE is shown as the multicolor colormap in Fig.~\ref{fig:VIBE1}(a). Because the interaction takes place close to the probe's focal plane, the signal acquires an angular distribution distinct from that of the probe: downstream of the focus the signal remains close to maximal on axis, whereas the probe is strongly suppressed on axis. Lens~1 images the obstacle onto the plane $z = \SI{2.27}{m}$, where the aperture is placed. The probe and signal intensity profiles in this plane are shown in Figs.~\ref{fig:VIBE1}(d) and \ref{fig:VIBE1}(e) together with the aperture opening (dashed lines): the aperture blocks the bulk of the probe while transmitting a portion of the quantum vacuum signal. Lens~2, a CRL of five lenses identical to those of Lens~1 with $f \simeq \SI{1.32}{m}$, placed just after the aperture at $z=\SI{2.41}{m}$, refocuses the signal onto the detector aperture at $z \simeq \SI{5.48}{m}$, of size about $\SI{5}{\micro m} \times \SI{5}{\micro m}$.
Before reaching the detectors at $z \simeq \SI{6.28}{m}$, the beam traverses \SI{40}{cm} of air; the resulting absorption and scattering are both included in the simulation using air scattering and transmission maps generated with GEANT4 simulations \cite{Geant4:2003}.

The simulation also reproduces roughly the measured transmissions of the individual elements: $0.74$ simulated versus $0.8$ measured for Lens~0, $0.42$ versus $0.45$ for the obstacle, $0.79$ versus $0.8$ for Lens~1, and $0.66$ simulated for the transmission of the quantum vacuum signal through Lens~2. 

\subsection{Simulation of the spatiotemporal jitter}
\label{sec:vibe_jitter}

\begin{figure*}[hbt]
	\centering
	\includegraphics[width=2.0\columnwidth]{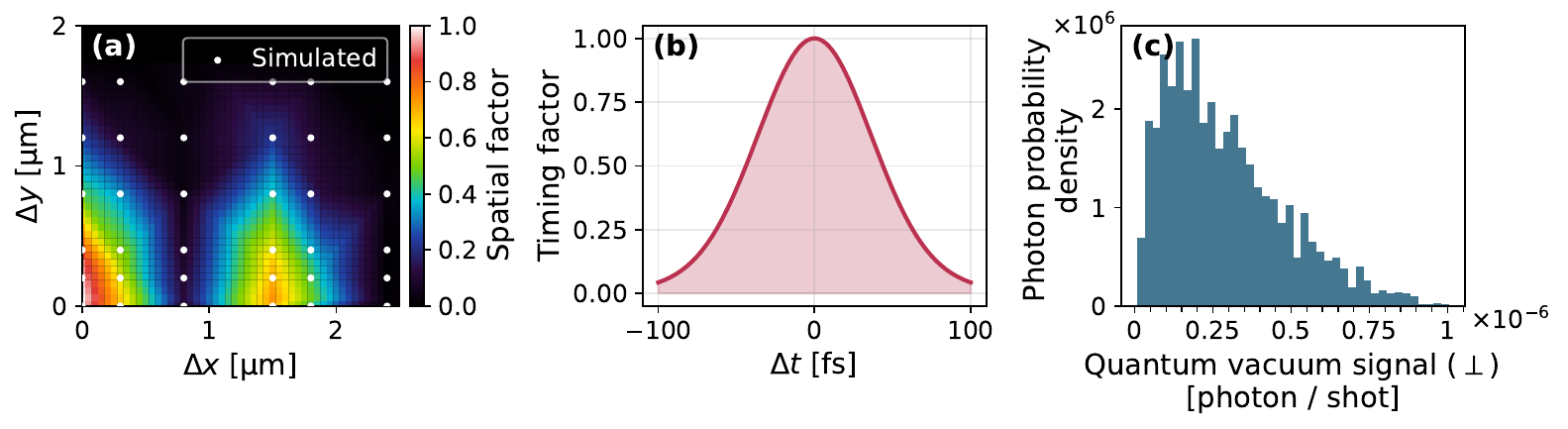}
	\caption{Simulation of the imperfect spatiotemporal pump--probe alignment. (a) Two-dimensional map of the spatial factor as a function of the transverse misalignment $(\Delta x, \Delta y)$. The white dots correspond to simulated values. (b) Timing factor as defined by Eq.~(\ref{eq:timing_factor}). (c) Corresponding simulated quantum vacuum signal distribution in the $\perp$ channel, obtained from the experimental spatiotemporal misalignment information [Fig.~\ref{fig:3Dscatter}].}
	\label{fig:alignement}
\end{figure*}

The spatiotemporal jitter makes the quantum vacuum signal fluctuate from shot to shot, and the analysis must account for it. As described in Sec.~\ref{sec:pershot}, each pump-on shot is characterized by its misalignment $(\Delta x, \Delta y, \Delta t)$. The corresponding signal prediction is obtained in two steps: the spatial offset is treated by dedicated simulations, and the temporal offset by the timing factor (see also \cite{Matheron:2025owt}).

For the spatial part, the full start-to-end simulation of Sec.~\ref{sec:vibe_aligned} is repeated with the pump displaced by prescribed offsets $(\Delta x, \Delta y)$ on a grid, and the resulting quantum vacuum signal is interpolated between grid points to form the map shown in Fig.~\ref{fig:alignement}(a) for the $\perp$ channel (the $\parallel$-channel map has the same shape, scaled up by roughly $121/9$). The signal is maximal at $\Delta x = \Delta y = 0$ and drops sharply away from it, but interestingly rises again near $\Delta x \simeq \SI{1.5}{\micro m}$: this reflects the interference pattern of the probe focus [Fig.~\ref{fig:VIBE1}(c)], whose secondary intensity peak at $\simeq$\SI{1.5}{\micro m} also generates signal when overlapped with the pump. Vertically, the signal falls to half its maximum already at $\Delta y \simeq \SI{0.8}{\micro m}$. Compared with the measured pointing jitter of less than $\pm\SI{1}{\micro m}$ rms in both directions [Fig.~\ref{fig:spatial_alignment}(e)], this ensures that the vast majority of shots contribute substantially to the expected signal. The temporal offset is then accounted for by multiplying the perfectly timed signal with the timing factor of Eq.~\eqref{eq:timing_factor} [Fig.~\ref{fig:alignement}(b)].

This procedure maps the measured per-shot coordinates $(\Delta x, \Delta y, \Delta t)$ shown in Fig.~\ref{fig:3Dscatter} directly onto the expected quantum vacuum signal of that shot in both channels, equivalent to running a dedicated simulation for every shot with its actual parameters. This simulation therefore predicts the expected per-shot quantum vacuum signal distribution shown in Fig.~\ref{fig:alignement}(c) for the $\perp$ channel. This quantity is denoted $N_n^{\rm QED}$ in the statistical analysis of Sec.~\ref{sec:limit}.

\section{Sources of background}
\label{sec:scattering}

The background consists of probe photons that reach the detectors without having scattered off the pump, because they were imperfectly rejected by the dark-field optics and/or the polarizing beam splitter; at the detector they are indistinguishable from the quantum vacuum signal (Sec.~\ref{sec:df_concept}). Understanding and minimizing this background is therefore essential, and this section discusses its dominant sources: scattering in the lenses and scattering off the aperture.

\subsection{Scattering in the lenses}
\label{sec:lens_scattering}

\begin{figure*}[hbt]
	\centering
	\includegraphics[width=2\columnwidth]{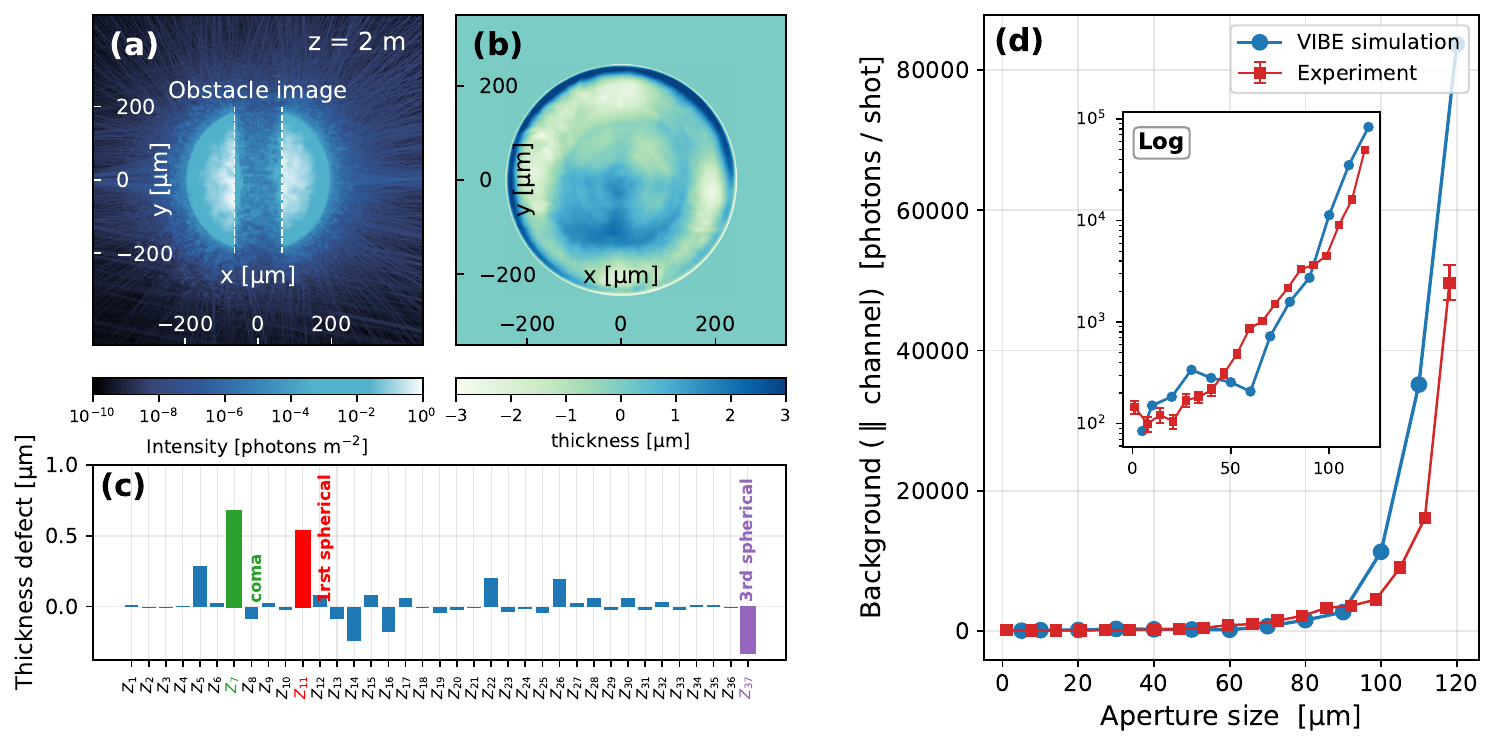}
	\caption{Simulation of the background generated by lens defects. (a) Simulated probe intensity distribution at $z=\SI{2}{m}$, on a logarithmic color scale. The image of the obstacle (dark vertical band) is distorted by the lens imperfections, and scattered light leaks into the dark-field region (dashed lines). (b) Measured thickness-defect map of one of the diamond lenses used in Lens~1, obtained by \mbox{at-wavelength} speckle tracking. (c) Zernike decomposition (Noll convention) of the defect map shown in (b); the dominant aberrations are coma (green), first-order spherical (red), and third-order spherical (purple). (d) Background photon number per shot in the $\parallel$ channel as a function of the aperture size: VIBE simulation (blue) and experiment (red). The inset shows the same data on a logarithmic scale.}
	\label{fig:VIBE2}
\end{figure*}

In an idealized dark-field setup with perfect optics and an infinite aperture, the probe would be fully extinguished in the dark-field region, at the image plane of the object, and the background would vanish. In practice, the CRLs exhibit shape imperfections that scatter probe light into the dark field~\cite{Lengeler:ht2006,Roth2014,Gasilov:mo5165,Celestre:2020}. Figure~\ref{fig:VIBE2}(a) shows the simulated probe intensity at $z=\SI{2}{m}$: scattered light contaminates the region between the dashed lines that the aperture would otherwise keep dark.

We now describe how the CRLs and their defects are modeled in the VIBE code. Each individual lens is represented by its ideal parabolic phase-and-absorption profile, to which a measured thickness-defect map is added as an additional phase screen~\cite{Celestre:2020srw,Celestre:2020,Lyatun:2020}. The lenses used in this experiment are polished diamond lenses~\cite{Celestre:2022} whose defect maps were measured at the ESRF's BM05 beamline~\cite{Ziegler2004} using at-wavelength X-ray speckle vector tracking (XSVT)~\cite{Berujon2020}. A typical map, with defects at the $\pm\SI{3}{\micro m}$ level, is shown in Fig.~\ref{fig:VIBE2}(b). To gain physical insight into these defects, the maps are decomposed on the basis of Zernike polynomials (Noll convention), see Fig.~\ref{fig:VIBE2}(c): the dominant contributions are coma and first- and third-order spherical aberrations, typical signatures of the lens manufacturing and polishing process, which is also observed by other type of 2D x-ray refractive optics~\cite{Seiboth:mo5223}.

The quality of this model is assessed by comparing the measured $\parallel$-channel background with its VIBE prediction for various aperture gap sizes [Fig.~\ref{fig:VIBE2}(d)]. All the lens elements used during the experiment have been characterized and their defect maps have been included in the VIBE simulation. The simulation reproduces the measured background within a factor of a few over the full scan range, without any free parameter: the residual discrepancy is attributed to imperfect knowledge of the setup, other scattering sources and limited spatial resolution of the lens defect maps. Given that the background spans several orders of magnitude over the scan, this level of agreement validates the modeling of the dominant background source. Moreover this agreement confirms the predictive power of the VIBE simulations for future experiments where new geometries can be tested in advance to find the best possible dark-field setup before an experimental implementation.

\subsection{Scattering on the apertures}
\label{sec:aperture_scattering}
\begin{figure}[hbt]
	\centering
	\includegraphics[width=\columnwidth]{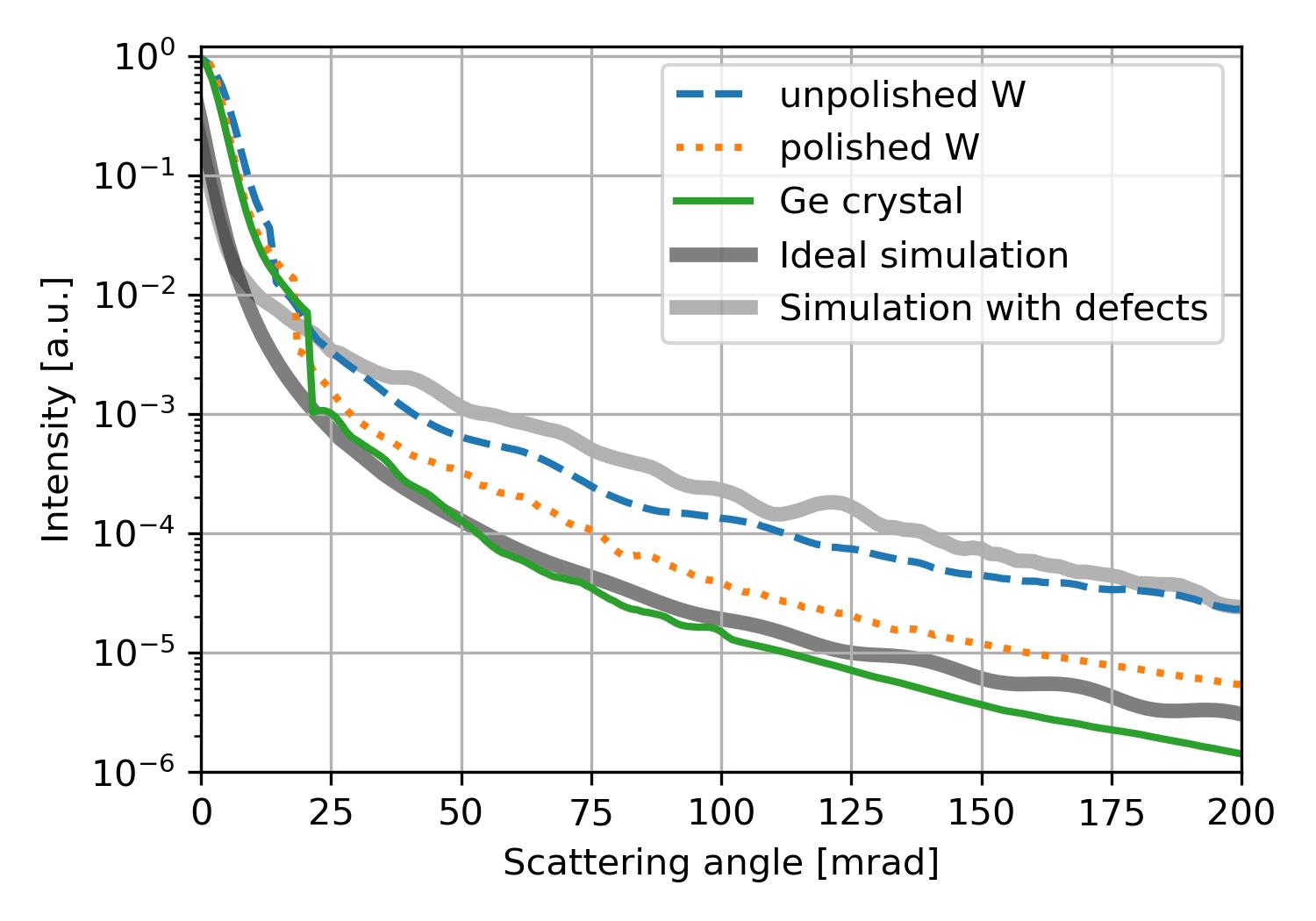}
	\caption{Scattering induced by different aperture systems, and their simulation by VIBE.}
	\label{fig:slits}
\end{figure}

A second source of background is the scattering of the intense probe lobes on the edges of the aperture blades. Even a tiny fraction of light diffusely scattered by its edges into the dark field can compete with the quantum vacuum signal. The scattering properties of several candidate aperture systems with different blades were therefore measured: Germanium (Ge) single-crystal blades fabricated by the Xenocs company, \SI{2}{mm} thick polished Tungsten (W) blades by the JJ X-ray company, and unpolished \SI{2}{mm} thick W blade from Smaract company. Figure~\ref{fig:slits} compares the measured angular distributions of the scattered light. The differences are clear and follow the expected outcome. For the present experiment, the polished W blades were used for the aperture and Ge single-crystal blades for the detector slit system. The Ge crystal blades produce scattering in great agreement with ideal diffractive simulation. If random defects with radii up to 2-\SI{3}{\micro m} are added to the simulation, then the results match well the unpolished blades.

\section{Upper limit on the light-by-light scattering cross section}
\label{sec:limit}

\subsection{Statistical method}
\label{sec:limit_bkg}

Since no excess of photons is observed in the $\perp$ channel (Sec.~\ref{sec:data}), the measurement is used to set an upper limit on the light-by-light scattering cross section, an insightful information to exclude any effect in a given range of the parameter space as it is the case for all rare-process searches \cite{Vacalis:2025}. The measured quantity is the ratio of the measured photon count to the simulated one,
\begin{equation}
	R = \frac{N^{\rm exp}}{N^{\rm QED}}.
	\label{eq:R_def}
\end{equation}

\begin{figure}[hbt]
	\centering
	\includegraphics[width=\columnwidth]{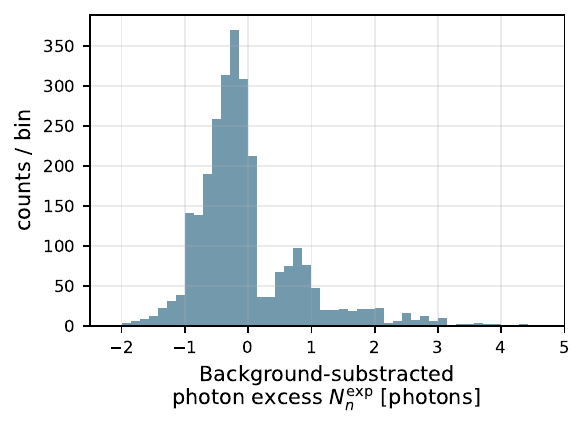}
	\caption{Distribution of the background-subtracted photon excess $N_n^{\rm exp}$ in the $\perp$ channel over the 2751 pump-on shots.}
	\label{fig:delta_N}
\end{figure}

Each pump-on shot contains signal plus background, while the pump-off shots measure the background alone. Because the pump fires on only one X-ray shot out of twenty, every pump-on shot is surrounded by contemporaneous pump-off shots in the acquisition sequence, enabling a local background estimate: to each pump-on shot $n$ we associate the average over the $n^{\rm off}=8$ pump-off shots recorded closest in time (four before and four after),
\begin{equation}
\bar N_n^{\mathrm{off}} = \frac{1}{8} \sum_{\substack{j = n-4 \\ j \neq n}}^{n+4} N^{\mathrm{off}}_{j}\,.
\label{eq:local_bkg}
\end{equation}
where an example of pump-off averaging is shown in Fig.~\ref{fig:perp_parall}(e)]. The background-subtracted photon excess of each pump-on shot is then formed as
\begin{equation}
	N_n^{\rm exp} = N_n^{\rm on} - \bar{N}_n^{\rm off},
	\label{eq:excess}
\end{equation}
where $N_n^{\rm on}$ is the number of photons detected in the $\perp$ channel for a pump-on shot. This subtraction suppresses the slow drifts of the background while preserving per-shot statistical independence. The distribution of $N_n^{\rm exp}$ over the 2751 pump-on shots is shown in Fig.~\ref{fig:delta_N}: it is centered on $-0.0024$ photons with a standard deviation of $\sigma_{N^{\rm exp}} = 0.994$, consistent with zero excess ($\sigma_{N^{\rm exp}}/\sqrt{N_{\rm shots}} \approx 0.019$).

\begin{figure}[hbt]
	\centering
	\includegraphics[width=0.9\columnwidth]{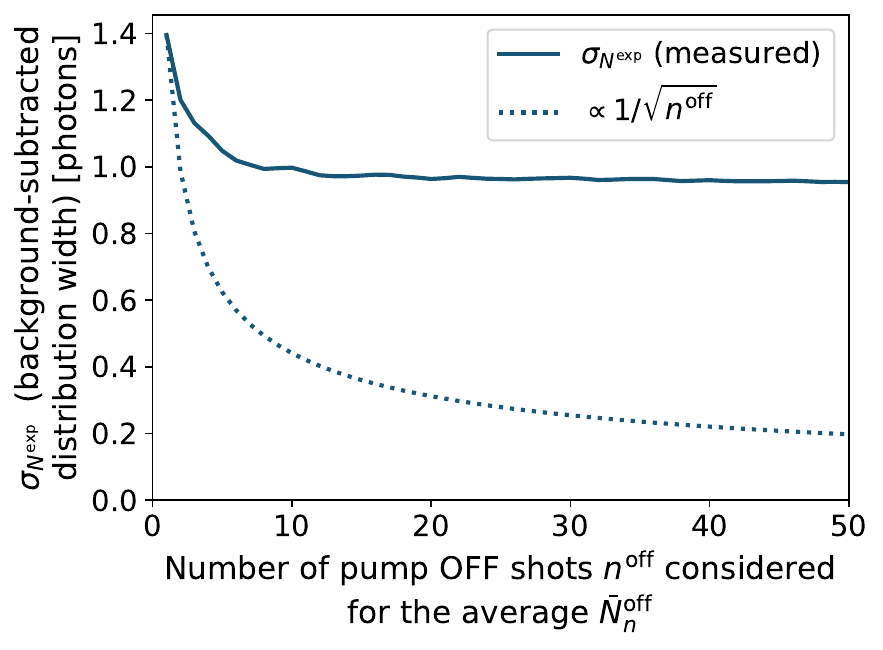}
	\caption{Width $\sigma_{N^{\rm exp}}$ of the background-subtracted photon-excess distribution as a function of the number $n^{\rm off}$ of pump-off shots included in the local background average $\bar{N}_n^{\rm off}$ (solid line), compared with the $\propto 1/\sqrt{n^{\rm off}}$ scaling expected for a purely statistical background estimate (dotted line).}
	\label{fig:Local_bg_average}
\end{figure}

The choice of $n^{\rm off}=8$ results from a trade-off, illustrated in Fig.~\ref{fig:Local_bg_average}, which shows the measured width $\sigma_{N^{\rm exp}}$ of the background-subtracted distribution as a function of the number of pump-off shots included in the local average $\bar{N}_n^{\rm off}$. If the background level were constant in time, the local estimate would benefit from the inverse-square-root scaling of the rms of an average, and including ever more pump-off shots would keep improving the sensitivity (dotted line in Fig.~\ref{fig:Local_bg_average}). In reality, two regimes are clearly visible. Up to $n^{\rm off} \sim 8$, the width decreases markedly: averaging progressively suppresses the statistical shot-to-shot fluctuations of the background estimate. Beyond $n^{\rm off} \sim 8$, corresponding to a time window of about \SI{1}{s}, the width plateaus at $\sigma_{N^{\rm exp}} \sim 1$ photon: on these time scales, the fluctuations of the XFEL machine decorrelate the additional pump-off shots from the local background of the pump-on shot, so that including them no longer improves the estimate. Averaging over $n^{\rm off}=8$ pump-off shots therefore realizes the best compromise between these two regimes and is used throughout the analysis. This study also highlights that, in the background-dominated regime, reducing and precisely tracking the shot-to-shot background fluctuations is as crucial to the sensitivity as lowering the mean background level itself: the mean of the $N_n^{\rm exp}$ distribution must be resolved against a width that is currently dominated by these fluctuations.

Estimating $R$ from the model $\langle N^{\text{exp}}_n\rangle = R N^{\text{QED}}_n$ with a given background, is a linear regression through the origin. The simplest estimator is the ratio of totals \cite{Cowan:1998ji}, $\hat{R} = \sum_n N_n^{\rm{exp}} / \sum_n N^{\mathrm{QED}}_n$ ,
which simply compares the total measured excess to the total predicted signal. This estimator gives every shot the same weight: a poorly aligned shot, whose predicted signal $N^{\mathrm{QED}}_n$ is essentially zero, contributes as much noise to the numerator as a perfectly aligned one, while adding no signal. 

The matched filter linear estimator \cite{Turin:1960,Cowan:1998ji} instead weights each shot in proportion to the signal it is expected to carry :
\begin{equation}
	\hat{R} = \frac{\sum_n N^{\text{QED}}_n N^{\text{exp}}_n}{\sum_n \bigl(N^{\text{QED}}_n\bigr)^2}
	\ \ \text{and} \ \
	\sigma_{\hat{R}} = \frac{\sigma_{N^{\text{exp}}}}{\sqrt{\sum_n \bigl(N^{\text{QED}}_n\bigr)^2}}\,,
	\label{eq:R+sigma}
\end{equation}
with $\sigma_{\hat{R}}^2$ its variance. The data $\{N_n^{\rm{exp}}\}$ are correlated against the per-shot expected quantum vacuum signal calculated in Sec.~\ref{sec:vibe_jitter}, $\{N^{\mathrm{QED}}_n\}$ which serves as a proxy for spatiotemporal alignment, so that well-aligned shots dominate the estimate while badly aligned shots, which carry noise but almost no signal, are suppressed. Importantly this estimator allows for a continuous and signal-based spatiotemporal selection of shots, rather than operating an arbitrary shot selection at a given value of $(\Delta x, \Delta y, \Delta t)$. The evaluation of Eq.~\eqref{eq:R+sigma} gives $\hat{R} \simeq -1.9\times 10^4$ with standard deviation $\sigma_{\hat{R}} = 5.45\times10^4$, compatible with no excess observed.

Following the standard prescription for a one-sided upper bound \cite{Cowan:1998ji,Cowan:2010js,ParticleDataGroup:2024cfk}, the 95\% confidence-level upper limit is
\begin{equation}
	R_{\rm up} = \hat{R} + Z\,\sigma_{\hat{R}} = 7\times 10^{4},
	\label{eq:Rup}
\end{equation}
with $Z=1.645$. Using the expression $\sigma_{\gamma\gamma}^{\rm exp}/\sigma_{\gamma\gamma}^{\rm QED}=N^{\rm exp}/N^{\rm QED}$~\cite{Bernard:2000ovj}, we express a new bound on the cross section: $\sigma_{\gamma\gamma}^{\rm exp}/\sigma_{\gamma\gamma}^{\rm QED}\leq 7\times10^4$.

Equivalently, the limit can be expressed in detected photons rather than as a cross section ratio: the maximum quantum vacuum signal that could remain hidden in the data set is
\begin{equation}
	S_{\mathrm{up}} \;=\; R_{\mathrm{up}} \sum_n N^{\mathrm{QED}}_n \;\simeq\; 56 \ \mathrm{photons} .
\end{equation}
In other words, had light-by-light scattering produced more than about 56 photons integrated over the 2751 pump-on shots, it would have appeared as a significant excess; QED predicts a total of $\sum_n N^{\mathrm{QED}}_n \approx 8\times10^{-4}$ photons for the same shots.

\subsection{Error estimation on the upper limit}
\label{sec:limit_errors}

The upper limit of Eq.~\eqref{eq:Rup} relies on the per-shot signal predictions $N_n^{\rm QED}$, which are themselves computed from measured interaction parameters; their uncertainties therefore propagate into $R_{\rm up}$. The main sources of error are estimated to be (i) the imperfect knowledge of the absolute timing between the two beams, with a systematic error conservatively estimated at roughly \SI{30}{fs}, and (ii) the jitter between the proxy diagnostics and the true values at the interaction point. The latter is quantified from the correlation measurements between the interaction-point diagnostics and their proxies [e.g., Figs.~\ref{fig:spatial_alignment}(c) and \ref{fig:spatial_alignment}(d)], whose correlation coefficients $R^2$ translate into an rms spread between the value reported by the proxy and the true value at the interaction point: \SI{0.2}{\micro m} rms in the horizontal plane and \SI{0.18}{\micro m} rms in the vertical plane for the equivalent plane camera versus the 50$\times$ interaction-point imaging, and \SI{30}{fs} rms between the proxy and interaction-point timing diagnostics. In addition, the X-ray flux at the interaction point varies by 8.5\% rms and the ReLaX energy at focus by 1.8\% rms.

These spreads are fed into a Monte Carlo scheme that recomputes the upper limit 2000 times, each time varying all interaction parameters within their corresponding spreads. The resulting distribution of 2000 effective upper limits allows us to place a one-sigma error on the value $R_{\rm up} = 7\times10^{4}$, yielding a value less than $1.1\times10^{5}$ at one-sigma confidence.

\subsection{Discussion}
\label{sec:limit_discussion}

\begin{figure}[hbt]
	\centering
	\includegraphics[width=\columnwidth]{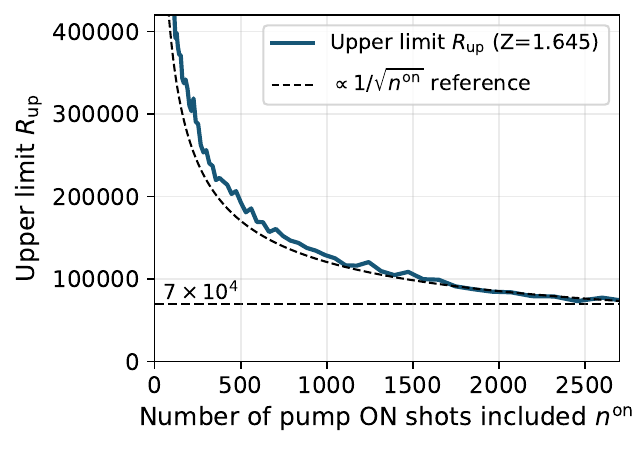}
	\caption{Evolution of the upper limit $R_{\rm up}$ ($Z=1.645$) with the number of pump-on shots included in the analysis. The dashed line shows the $1/\sqrt{n^{\rm{on}}}$ scaling expected for a background-dominated measurement; the horizontal dashed line marks the final value $R_{\rm up} \approx 7\times10^{4}$.}
	\label{fig:upper_limit}
\end{figure}

Figure~\ref{fig:upper_limit} shows how the upper limit improves as pump-on shots are accumulated: it follows the $1/\sqrt{n^{\rm on}}$ scaling characteristic of a background-dominated measurement, converging to $R_{\rm up} \approx 7\times10^{4}$. This scaling confirms that the sensitivity was limited by the background level and the available statistics rather than by systematic effects. Following this conclusion, the background subtraction itself offers room for improvement. Beyond the local 8-shot average of Eq.~\eqref{eq:local_bkg}, a more advanced background estimate can be constructed from the multiple diagnostics available along the XFEL beamline. To this end, several per-shot measurements, the incoming probe flux, the probe spectrum, the horizontal and vertical probe pointing, and the probe size, were correlated with the counts recorded on the $\perp$ and $\parallel$ detectors and fed into a machine-learning regression implemented with the scikit-learn library~\cite{scikit-learn}. The algorithm is trained exclusively on pump-off shots and then predicts the background expected on each pump-on shot from its simultaneous diagnostic readings. This prediction substitutes for $\bar{N}_n^{\rm off}$ in Eq.~\eqref{eq:excess} as a per-shot, multi-diagnostic background estimate. With this multi-diagnostic background estimate, the bound improves from
$R_{\rm up} = 7\times10^{4}$ to $R_{\rm up} = 5.3\times10^{4}$. We nevertheless quote the 8-shot-average result as our main bound, since it relies on a simpler and more transparent analysis than a machine-learning regression. The size of the improvement is set by how well the diagnostics predict the detected counts. In the $\perp$ channel the correlation between predicted and measured counts reaches only $R^2 = 0.25$, and the width of the background-subtracted distribution decreases modestly, from $\sigma_{N^{\rm exp}} = 0.994$ to $0.84$ photons. In the $\parallel$ channel the diagnostics capture the XFEL fluctuations better ($R^2 = 0.7$), and the width is nearly halved, from $6647$ to $3371$ photons. These numbers also indicate the ultimate potential of the method. With perfectly correlated diagnostics ($R^2 \rightarrow 1$), all predictable XFEL fluctuations would be removed by the background estimator, leaving only the Poisson floor set by the mean photon number. In the $\perp$ channel this floor lies at $\sqrt{0.4649} \approx 0.68$ photons, only $\sim 30\%$ below the present width of $0.994$: this channel already operates close to its statistical limit. In the $\parallel$ channel, by contrast, the floor lies at $\sqrt{7264} \approx 85$ photons, roughly two orders of magnitude below the currently measured fluctuations, so that an improved background prediction would yield a dramatic gain there.

The obtained bound improves upon the previous best laboratory limits on elastic photon--photon scattering in the low-energy regime by approximately 10 orders of magnitude~\cite{Bernard:2000ovj}. 
We emphasize that a direct comparison of low-energy bounds on light-by-light scattering with those obtained in the high-energy regime~\cite{ATLAS:2017fur,ATLAS:2019azn,CMS:2018erd} is not possible for three main reasons: first, they probe an entirely different parameter space of the theory, second, they probe the effect via virtual photons, and third, the interpretation of high-energy bounds requires accounting for other SM sectors, going beyond a pure QED framework. 
Moreover, we note that experiments searching for vacuum birefringence using static magnetic fields, such as the Polarizzazione del Vuoto con LASer (PVLAS) experiment~\cite{Ejlli:2020yhk} and others~\cite{Battesti:2018bgc,Fan:2017fnd,Cadene:2013bva,Spector:2025esz}, constrain the difference between the low-energy constants $a-b$.
PVLAS, which provides the most stringent laboratory bound on vacuum birefringence to date, utilizes a single continuous-wave probe photon beam that interacts with the vacuum in a quasi-constant magnetic field.
As there is only a single photon beam, PVLAS results do not allow light-by-light scattering cross-sections to be derived, so cannot be directly compared to the main result of \cite{Matheron:2026} shown in Fig~\ref{fig:upper_limit}. However, the results from the EuXFEL campaign, allow a bound for vacuum birefringence to be obtained,  as the result shown in Fig~\ref{fig:upper_limit} bounds the $\perp$-polarized signal component. The PVLAS data constrains the refractive index difference $\Delta n^{\rm exp}/\Delta n^{\rm QED}$. This can be rescaled to a photon number bound via $N^{\rm exp}/N^{\rm QED}=(\Delta n^{\rm exp}/\Delta n^{\rm QED})^2$~\cite{Karbstein:2021otv}, yielding $N^{\rm exp}/N^{\rm QED}\big\vert_{\rm PVLAS} \lesssim 49$. At this point we emphasize, that the set-up at the EuXFEL scales to positive detection within the nominal operating parameters and geometrical constraints (i.e. using a modified set-up at nominal performance). 
The high sensitivity of an XFEL probe allows precision measurements beyond discovery. For example the integration time of the initial experiment described here requires an integration time of only a few hours, while the PVLAS measurement was achieved after many years of operation and data acquisition.

\section{Conclusion}
\label{sec:conclusion}

We have presented the experimental details searching for quantum vacuum signal using the dark-field approach, performed by colliding the European XFEL beam with the ReLaX high-power laser at the HED scientific instrument. This campaign demonstrated (i) the operation of a dark-field setup suppressing the probe by $\sim 4.1\times 10^{-12}$ while remaining sensitive to both polarization channels of the scattered light, (ii) per-shot tracking of the spatiotemporal overlap of the two beams via calibrated proxy diagnostics, and (iii) a start-to-end simulation pipeline, that predicts both the expected quantum vacuum signal and the dominant background from measured beam profiles and measured optics imperfections. From 2751 pump--probe collisions, an upper limit on the light-by-light scattering cross section of $7\times10^{4}$ times the QED prediction at a center-of-mass photon energy $\omega_\star=116\,{\rm eV}$ was derived at 95\% confidence level, improving on previous low-energy experiments by about 10 orders of magnitude \cite{Bernard:2000ovj}.

The sensitivity of this first implementation was limited by four factors, each of which can be substantially improved in future campaigns. First, the positions and focal lengths of the lenses were constrained by the existing beamline infrastructure; optimized lens positions are expected to reduce the background by about two orders of magnitude while increasing the signal by about one order of magnitude. Second, the pump laser was operated below its published specifications~\cite{LasoGarcia:2021}. At published operation, the on-target energy can reach \SI{11}{J}, the beam quality factor can increase from 0.24 to 0.35, and exploiting the full near-field beam diameter of \SI{15}{cm} would enable stronger focusing, yielding FWHM spot sizes below \SI{1.1}{\upmu m} in both dimensions. Together, these improvements would increase the peak intensity by a factor of $\sim 14$. Third, improved performance of the seeded pulse generation, with increased stability and an average energy of about \SI{600}{\upmu J}, could increase the signal by an additional factor of four. The use of a monochromator upstream of the dark-field scheme would also allow to reach a much better polarisation purity (about two orders of magnitude enhancement). Finally, the background subtraction was limited by the weak correlation between the available beamline diagnostics and the detected counts; a multi-diagnostic machine-learning estimate already lowers the bound to $5.3\times10^{4}$ (Sec.~\ref{sec:limit_discussion}). With improved diagnostics approaching full correlation ($R^2 \rightarrow 1$) with the detector counts, the subtraction would approach the shot-noise Poisson limit, a substantial gain in the $\parallel$ channel which is crucial for accessing the low-energy constants $a$ and $b$. Together, these enhancements bring the first detection of quantum-vacuum signals in the low-energy regime within reach at existing facilities and, beyond detection, open the way to precision measurements of the low-energy constants $a$ and $b$ through the simultaneous two-channel polarimetry unique to the dark-field scheme \cite{Karbstein:2022uwf}.

\begin{acknowledgments}
We acknowledge the European XFEL in Schenefeld, Germany, for provision of X-ray free electron
laser beam time at the Scientific Instrument HED (High Energy Density Science) under proposal
number 10692 and would like to thank the staff of European XFEL and DESY for their assistance.
The authors are indebted to the Helmholtz International Beamline for Extreme Fields (HIBEF)
user consortium for the provision of instrumentation and staff that enabled this experiment.

This work has been funded by the Deutsche Forschungsgemeinschaft (DFG, German Research Foundation) under Grants Nos.\ 392856280, 416611371, 416607684, 416702141, and 416708866 within the Research Unit FOR2783/2.
NM thanks the partial financial support from the Czech Ministry of Education, Youth, and Sports (grant nr. LM2023068, and the European Commission Horizon MSCA-SE Project MAMBA [HORIZON-MSCA-SE-2022 GAN 101131245]). 
\end{acknowledgments}

\bibliographystyle{apsrev4-2}

\begin{thebibliography}{71}%
	\makeatletter
	\providecommand \@ifxundefined [1]{%
		\@ifx{#1\undefined}
	}%
	\providecommand \@ifnum [1]{%
		\ifnum #1\expandafter \@firstoftwo
		\else \expandafter \@secondoftwo
		\fi
	}%
	\providecommand \@ifx [1]{%
		\ifx #1\expandafter \@firstoftwo
		\else \expandafter \@secondoftwo
		\fi
	}%
	\providecommand \natexlab [1]{#1}%
	\providecommand \enquote  [1]{``#1''}%
	\providecommand \bibnamefont  [1]{#1}%
	\providecommand \bibfnamefont [1]{#1}%
	\providecommand \citenamefont [1]{#1}%
	\providecommand \href@noop [0]{\@secondoftwo}%
	\providecommand \href [0]{\begingroup \@sanitize@url \@href}%
	\providecommand \@href[1]{\@@startlink{#1}\@@href}%
	\providecommand \@@href[1]{\endgroup#1\@@endlink}%
	\providecommand \@sanitize@url [0]{\catcode `\\12\catcode `\$12\catcode
		`\&12\catcode `\#12\catcode `\^12\catcode `\_12\catcode `\%12\relax}%
	\providecommand \@@startlink[1]{}%
	\providecommand \@@endlink[0]{}%
	\providecommand \url  [0]{\begingroup\@sanitize@url \@url }%
	\providecommand \@url [1]{\endgroup\@href {#1}{\urlprefix }}%
	\providecommand \urlprefix  [0]{URL }%
	\providecommand \Eprint [0]{\href }%
	\providecommand \doibase [0]{https://doi.org/}%
	\providecommand \selectlanguage [0]{\@gobble}%
	\providecommand \bibinfo  [0]{\@secondoftwo}%
	\providecommand \bibfield  [0]{\@secondoftwo}%
	\providecommand \translation [1]{[#1]}%
	\providecommand \BibitemOpen [0]{}%
	\providecommand \bibitemStop [0]{}%
	\providecommand \bibitemNoStop [0]{.\EOS\space}%
	\providecommand \EOS [0]{\spacefactor3000\relax}%
	\providecommand \BibitemShut  [1]{\csname bibitem#1\endcsname}%
	\let\auto@bib@innerbib\@empty
	\bibitem [{\citenamefont {Euler}\ and\ \citenamefont
		{Kockel}(1935)}]{Euler:1935zz}%
	\BibitemOpen
	\bibfield  {author} {\bibinfo {author} {\bibfnamefont {H.}~\bibnamefont
			{Euler}}\ and\ \bibinfo {author} {\bibfnamefont {B.}~\bibnamefont {Kockel}},\
	}\href {https://doi.org/10.1007/BF01493898} {\bibfield  {journal} {\bibinfo
			{journal} {Naturwiss.}\ }\textbf {\bibinfo {volume} {23}},\ \bibinfo {pages}
		{246} (\bibinfo {year} {1935})}\BibitemShut {NoStop}%
	\bibitem [{\citenamefont {Euler}(1936)}]{Euler:1935qgl}%
	\BibitemOpen
	\bibfield  {author} {\bibinfo {author} {\bibfnamefont {H.}~\bibnamefont
			{Euler}},\ }\href {https://doi.org/10.1002/andp.19364180503} {\bibfield
		{journal} {\bibinfo  {journal} {Annalen Phys.}\ }\textbf {\bibinfo {volume}
			{26}},\ \bibinfo {pages} {398} (\bibinfo {year} {1936})}\BibitemShut
	{NoStop}%
	\bibitem [{\citenamefont {Heisenberg}\ and\ \citenamefont
		{Euler}(1936)}]{Heisenberg:1936nmg}%
	\BibitemOpen
	\bibfield  {author} {\bibinfo {author} {\bibfnamefont {W.}~\bibnamefont
			{Heisenberg}}\ and\ \bibinfo {author} {\bibfnamefont {H.}~\bibnamefont
			{Euler}},\ }\href {https://doi.org/10.1007/BF01343663} {\bibfield  {journal}
		{\bibinfo  {journal} {Z. Phys.}\ }\textbf {\bibinfo {volume} {98}},\ \bibinfo
		{pages} {714} (\bibinfo {year} {1936})},\ \Eprint
	{https://arxiv.org/abs/physics/0605038} {arXiv:physics/0605038} \BibitemShut
	{NoStop}%
	\bibitem [{\citenamefont {Karplus}\ and\ \citenamefont
		{Neuman}(1950)}]{Karplus:1950zza}%
	\BibitemOpen
	\bibfield  {author} {\bibinfo {author} {\bibfnamefont {R.}~\bibnamefont
			{Karplus}}\ and\ \bibinfo {author} {\bibfnamefont {M.}~\bibnamefont
			{Neuman}},\ }\href {https://doi.org/10.1103/PhysRev.80.380} {\bibfield
		{journal} {\bibinfo  {journal} {Phys. Rev.}\ }\textbf {\bibinfo {volume}
			{80}},\ \bibinfo {pages} {380} (\bibinfo {year} {1950})}\BibitemShut
	{NoStop}%
	\bibitem [{\citenamefont {Karplus}\ and\ \citenamefont
		{Neuman}(1951)}]{Karplus:1950zz}%
	\BibitemOpen
	\bibfield  {author} {\bibinfo {author} {\bibfnamefont {R.}~\bibnamefont
			{Karplus}}\ and\ \bibinfo {author} {\bibfnamefont {M.}~\bibnamefont
			{Neuman}},\ }\href {https://doi.org/10.1103/PhysRev.83.776} {\bibfield
		{journal} {\bibinfo  {journal} {Phys. Rev.}\ }\textbf {\bibinfo {volume}
			{83}},\ \bibinfo {pages} {776} (\bibinfo {year} {1951})}\BibitemShut
	{NoStop}%
	\bibitem [{\citenamefont {{ATLAS Collaboration}}(2017)}]{ATLAS:2017fur}%
	\BibitemOpen
	\bibfield  {author} {\bibinfo {author} {\bibnamefont {{ATLAS
					Collaboration}}},\ }\href {https://doi.org/10.1038/nphys4208} {\bibfield
		{journal} {\bibinfo  {journal} {Nature Phys.}\ }\textbf {\bibinfo {volume}
			{13}},\ \bibinfo {pages} {852} (\bibinfo {year} {2017})},\ \Eprint
	{https://arxiv.org/abs/1702.01625} {arXiv:1702.01625} \BibitemShut {NoStop}%
	\bibitem [{\citenamefont {{CMS Collaboration}}(2019)}]{CMS:2018erd}%
	\BibitemOpen
	\bibfield  {author} {\bibinfo {author} {\bibnamefont {{CMS Collaboration}}},\
	}\href {https://doi.org/10.1016/j.physletb.2019.134826} {\bibfield  {journal}
		{\bibinfo  {journal} {Phys. Lett. B}\ }\textbf {\bibinfo {volume} {797}},\
		\bibinfo {pages} {134826} (\bibinfo {year} {2019})},\ \Eprint
	{https://arxiv.org/abs/1810.04602} {arXiv:1810.04602} \BibitemShut {NoStop}%
	\bibitem [{\citenamefont {{ATLAS Collaboration}}(2019)}]{ATLAS:2019azn}%
	\BibitemOpen
	\bibfield  {author} {\bibinfo {author} {\bibnamefont {{ATLAS
					Collaboration}}},\ }\href {https://doi.org/10.1103/PhysRevLett.123.052001}
	{\bibfield  {journal} {\bibinfo  {journal} {Phys. Rev. Lett.}\ }\textbf
		{\bibinfo {volume} {123}},\ \bibinfo {pages} {052001} (\bibinfo {year}
		{2019})},\ \Eprint {https://arxiv.org/abs/1904.03536} {arXiv:1904.03536}
	\BibitemShut {NoStop}%
	\bibitem [{\citenamefont {{BIREF@HIBEF Collaboration, \textit{Letter of
					Intent,}}}(2025)}]{Ahmadiniaz:2024xob}%
	\BibitemOpen
	\bibfield  {author} {\bibinfo {author} {\bibnamefont {{BIREF@HIBEF
					Collaboration, \textit{Letter of Intent,}}}},\ }\href
	{https://doi.org/10.1017/hpl.2024.70} {\bibfield  {journal} {\bibinfo
			{journal} {High Power Laser Sci. Eng.}\ }\textbf {\bibinfo {volume} {13}},\
		\bibinfo {pages} {e7} (\bibinfo {year} {2025})},\ \Eprint
	{https://arxiv.org/abs/2405.18063} {arXiv:2405.18063} \BibitemShut {NoStop}%
	\bibitem [{\citenamefont {Moulin}\ \emph {et~al.}(1996)\citenamefont {Moulin},
		\citenamefont {Bernard},\ and\ \citenamefont {Amiranoff}}]{Moulin:1996vv}%
	\BibitemOpen
	\bibfield  {author} {\bibinfo {author} {\bibfnamefont {F.}~\bibnamefont
			{Moulin}}, \bibinfo {author} {\bibfnamefont {D.}~\bibnamefont {Bernard}},\
		and\ \bibinfo {author} {\bibfnamefont {F.}~\bibnamefont {Amiranoff}},\ }\href
	{https://doi.org/10.1007/s002880050282} {\bibfield  {journal} {\bibinfo
			{journal} {Z. Phys. C}\ }\textbf {\bibinfo {volume} {72}},\ \bibinfo {pages}
		{607} (\bibinfo {year} {1996})}\BibitemShut {NoStop}%
	\bibitem [{\citenamefont {Bernard}\ \emph {et~al.}(2000)\citenamefont
		{Bernard}, \citenamefont {Moulin}, \citenamefont {Amiranoff}, \citenamefont
		{Braun}, \citenamefont {Chambaret}, \citenamefont {Darpentigny},
		\citenamefont {Grillon}, \citenamefont {Ranc},\ and\ \citenamefont
		{Perrone}}]{Bernard:2000ovj}%
	\BibitemOpen
	\bibfield  {author} {\bibinfo {author} {\bibfnamefont {D.}~\bibnamefont
			{Bernard}}, \bibinfo {author} {\bibfnamefont {F.}~\bibnamefont {Moulin}},
		\bibinfo {author} {\bibfnamefont {F.}~\bibnamefont {Amiranoff}}, \bibinfo
		{author} {\bibfnamefont {A.}~\bibnamefont {Braun}}, \bibinfo {author}
		{\bibfnamefont {J.~P.}\ \bibnamefont {Chambaret}}, \bibinfo {author}
		{\bibfnamefont {G.}~\bibnamefont {Darpentigny}}, \bibinfo {author}
		{\bibfnamefont {G.}~\bibnamefont {Grillon}}, \bibinfo {author} {\bibfnamefont
			{S.}~\bibnamefont {Ranc}},\ and\ \bibinfo {author} {\bibfnamefont
			{F.}~\bibnamefont {Perrone}},\ }\href {https://doi.org/10.1007/s100530050535}
	{\bibfield  {journal} {\bibinfo  {journal} {Eur. Phys. J. D}\ }\textbf
		{\bibinfo {volume} {10}},\ \bibinfo {pages} {141} (\bibinfo {year} {2000})},\
	\Eprint {https://arxiv.org/abs/1007.0104} {arXiv:1007.0104} \BibitemShut
	{NoStop}%
	\bibitem [{\citenamefont {Yamaji}\ \emph {et~al.}(2016)\citenamefont {Yamaji},
		\citenamefont {Inada}, \citenamefont {Yamazaki}, \citenamefont {Namba},
		\citenamefont {Asai}, \citenamefont {Kobayashi}, \citenamefont {Tamasaku},
		\citenamefont {Tanaka}, \citenamefont {Inubushi}, \citenamefont {Sawada},
		\citenamefont {Yabashi},\ and\ \citenamefont {Ishikawa}}]{Yamaji:2016xws}%
	\BibitemOpen
	\bibfield  {author} {\bibinfo {author} {\bibfnamefont {T.}~\bibnamefont
			{Yamaji}}, \bibinfo {author} {\bibfnamefont {T.}~\bibnamefont {Inada}},
		\bibinfo {author} {\bibfnamefont {T.}~\bibnamefont {Yamazaki}}, \bibinfo
		{author} {\bibfnamefont {T.}~\bibnamefont {Namba}}, \bibinfo {author}
		{\bibfnamefont {S.}~\bibnamefont {Asai}}, \bibinfo {author} {\bibfnamefont
			{T.}~\bibnamefont {Kobayashi}}, \bibinfo {author} {\bibfnamefont
			{K.}~\bibnamefont {Tamasaku}}, \bibinfo {author} {\bibfnamefont
			{Y.}~\bibnamefont {Tanaka}}, \bibinfo {author} {\bibfnamefont
			{Y.}~\bibnamefont {Inubushi}}, \bibinfo {author} {\bibfnamefont
			{K.}~\bibnamefont {Sawada}}, \bibinfo {author} {\bibfnamefont
			{M.}~\bibnamefont {Yabashi}},\ and\ \bibinfo {author} {\bibfnamefont
			{T.}~\bibnamefont {Ishikawa}},\ }\href
	{https://doi.org/10.1016/j.physletb.2016.11.003} {\bibfield  {journal}
		{\bibinfo  {journal} {Phys. Lett. B}\ }\textbf {\bibinfo {volume} {763}},\
		\bibinfo {pages} {454} (\bibinfo {year} {2016})},\ \Eprint
	{https://arxiv.org/abs/1608.06369} {arXiv:1608.06369} \BibitemShut {NoStop}%
	\bibitem [{\citenamefont {Watt}\ \emph {et~al.}(2025)\citenamefont {Watt} \emph
		{et~al.}}]{Watt:2024brh}%
	\BibitemOpen
	\bibfield  {author} {\bibinfo {author} {\bibfnamefont {R.}~\bibnamefont
			{Watt}} \emph {et~al.},\ }\href
	{https://doi.org/10.1016/j.physletb.2025.139247} {\bibfield  {journal}
		{\bibinfo  {journal} {Phys. Lett. B}\ }\textbf {\bibinfo {volume} {861}},\
		\bibinfo {pages} {139247} (\bibinfo {year} {2025})},\ \Eprint
	{https://arxiv.org/abs/2407.12915} {arXiv:2407.12915} \BibitemShut {NoStop}%
	\bibitem [{\citenamefont {Ritus}(1975)}]{Ritus:1975pcc}%
	\BibitemOpen
	\bibfield  {author} {\bibinfo {author} {\bibfnamefont {V.~I.}\ \bibnamefont
			{Ritus}},\ }\href@noop {} {\bibfield  {journal} {\bibinfo  {journal} {J. Exp.
				Theor. Phys.}\ }\textbf {\bibinfo {volume} {42}},\ \bibinfo {pages} {774}
		(\bibinfo {year} {1975})}\BibitemShut {NoStop}%
	\bibitem [{\citenamefont {Dittrich}\ and\ \citenamefont
		{Reuter}(1985)}]{Dittrich:1985yb}%
	\BibitemOpen
	\bibfield  {author} {\bibinfo {author} {\bibfnamefont {W.}~\bibnamefont
			{Dittrich}}\ and\ \bibinfo {author} {\bibfnamefont {M.}~\bibnamefont
			{Reuter}},\ }\href {https://doi.org/10.1007/3-540-15182-6} {\emph {\bibinfo
			{title} {{Effective Lagrangians in Quantum Electrodynamics}}}},\ \bibinfo
	{series} {Lecture Notes in Physics}, Vol.\ \bibinfo {volume} {220}\ (\bibinfo
	{publisher} {Springer},\ \bibinfo {address} {Berlin, Heidelberg},\ \bibinfo
	{year} {1985})\BibitemShut {NoStop}%
	\bibitem [{\citenamefont {Gies}\ and\ \citenamefont
		{Karbstein}(2017)}]{Gies:2016yaa}%
	\BibitemOpen
	\bibfield  {author} {\bibinfo {author} {\bibfnamefont {H.}~\bibnamefont
			{Gies}}\ and\ \bibinfo {author} {\bibfnamefont {F.}~\bibnamefont
			{Karbstein}},\ }\href {https://doi.org/10.1007/JHEP03(2017)108} {\bibfield
		{journal} {\bibinfo  {journal} {JHEP}\ }\textbf {\bibinfo {volume}
			{03}}\bibfield  {number} {\bibinfo  {number} { (3)},\ \bibinfo {pages}
			{108}},\ }\Eprint {https://arxiv.org/abs/1612.07251} {arXiv:1612.07251}
	\BibitemShut {NoStop}%
	\bibitem [{\citenamefont {Karbstein}\ \emph {et~al.}(2022)\citenamefont
		{Karbstein}, \citenamefont {Ullmann}, \citenamefont {Mosman},\ and\
		\citenamefont {Zepf}}]{Karbstein:2022uwf}%
	\BibitemOpen
	\bibfield  {author} {\bibinfo {author} {\bibfnamefont {F.}~\bibnamefont
			{Karbstein}}, \bibinfo {author} {\bibfnamefont {D.}~\bibnamefont {Ullmann}},
		\bibinfo {author} {\bibfnamefont {E.~A.}\ \bibnamefont {Mosman}},\ and\
		\bibinfo {author} {\bibfnamefont {M.}~\bibnamefont {Zepf}},\ }\href
	{https://doi.org/10.1103/PhysRevLett.129.061802} {\bibfield  {journal}
		{\bibinfo  {journal} {Phys. Rev. Lett.}\ }\textbf {\bibinfo {volume} {129}},\
		\bibinfo {pages} {061802} (\bibinfo {year} {2022})},\ \Eprint
	{https://arxiv.org/abs/2207.09866} {arXiv:2207.09866} \BibitemShut {NoStop}%
	\bibitem [{\citenamefont {Aleksandrov}\ \emph {et~al.}(1985)\citenamefont
		{Aleksandrov}, \citenamefont {Ansel'm},\ and\ \citenamefont
		{Moskalev}}]{Aleksandrov:1985}%
	\BibitemOpen
	\bibfield  {author} {\bibinfo {author} {\bibfnamefont {E.~B.}\ \bibnamefont
			{Aleksandrov}}, \bibinfo {author} {\bibfnamefont {A.~A.}\ \bibnamefont
			{Ansel'm}},\ and\ \bibinfo {author} {\bibfnamefont {A.~N.}\ \bibnamefont
			{Moskalev}},\ }\href@noop {} {\bibfield  {journal} {\bibinfo  {journal} {Sov.
				Phys. JETP}\ }\textbf {\bibinfo {volume} {62}},\ \bibinfo {pages} {680}
		(\bibinfo {year} {1985})},\ \bibinfo {note} {[Zh.\ Eksp.\ Teor.\ Fiz.\ {\bf
			89}, 1181--1189 (1985)]}\BibitemShut {NoStop}%
	\bibitem [{\citenamefont {Marx}\ \emph {et~al.}(2013)\citenamefont {Marx},
		\citenamefont {Sch{\"u}lze}, \citenamefont {Uschmann}, \citenamefont
		{K{\"a}mpfer}, \citenamefont {Lotzsch}, \citenamefont {Wehrhan},
		\citenamefont {Wagner}, \citenamefont {Detlefs}, \citenamefont {Roth},
		\citenamefont {H{\"a}rtwig}, \citenamefont {F{\"o}rster}, \citenamefont
		{St{\"o}hlker},\ and\ \citenamefont {Paulus}}]{Marx:2013}%
	\BibitemOpen
	\bibfield  {author} {\bibinfo {author} {\bibfnamefont {B.}~\bibnamefont
			{Marx}}, \bibinfo {author} {\bibfnamefont {K.~S.}\ \bibnamefont
			{Sch{\"u}lze}}, \bibinfo {author} {\bibfnamefont {I.}~\bibnamefont
			{Uschmann}}, \bibinfo {author} {\bibfnamefont {T.}~\bibnamefont
			{K{\"a}mpfer}}, \bibinfo {author} {\bibfnamefont {R.}~\bibnamefont
			{Lotzsch}}, \bibinfo {author} {\bibfnamefont {O.}~\bibnamefont {Wehrhan}},
		\bibinfo {author} {\bibfnamefont {W.}~\bibnamefont {Wagner}}, \bibinfo
		{author} {\bibfnamefont {C.}~\bibnamefont {Detlefs}}, \bibinfo {author}
		{\bibfnamefont {T.}~\bibnamefont {Roth}}, \bibinfo {author} {\bibfnamefont
			{J.}~\bibnamefont {H{\"a}rtwig}}, \bibinfo {author} {\bibfnamefont
			{E.}~\bibnamefont {F{\"o}rster}}, \bibinfo {author} {\bibfnamefont
			{T.}~\bibnamefont {St{\"o}hlker}},\ and\ \bibinfo {author} {\bibfnamefont
			{G.~G.}\ \bibnamefont {Paulus}},\ }\href
	{https://doi.org/10.1103/PhysRevLett.110.254801} {\bibfield  {journal}
		{\bibinfo  {journal} {Phys. Rev. Lett.}\ }\textbf {\bibinfo {volume} {110}},\
		\bibinfo {pages} {254801} (\bibinfo {year} {2013})}\BibitemShut {NoStop}%
	\bibitem [{\citenamefont {Schulze}\ \emph {et~al.}(2022)\citenamefont
		{Schulze}, \citenamefont {Grabiger}, \citenamefont {Loetzsch}, \citenamefont
		{Marx-Glowna}, \citenamefont {Schmitt}, \citenamefont {Garcia}, \citenamefont
		{Hippler}, \citenamefont {Huang}, \citenamefont {Karbstein}, \citenamefont
		{Konôpková}, \citenamefont {Schlenvoigt}, \citenamefont {Schwinkendorf},
		\citenamefont {Strohm}, \citenamefont {Toncian}, \citenamefont {Uschmann},
		\citenamefont {Wille}, \citenamefont {Zastrau}, \citenamefont {Röhlsberger},
		\citenamefont {Stöhlker}, \citenamefont {Cowan},\ and\ \citenamefont
		{Paulus}}]{Schulze:2022}%
	\BibitemOpen
	\bibfield  {author} {\bibinfo {author} {\bibfnamefont {K.~S.}\ \bibnamefont
			{Schulze}}, \bibinfo {author} {\bibfnamefont {B.}~\bibnamefont {Grabiger}},
		\bibinfo {author} {\bibfnamefont {R.}~\bibnamefont {Loetzsch}}, \bibinfo
		{author} {\bibfnamefont {B.}~\bibnamefont {Marx-Glowna}}, \bibinfo {author}
		{\bibfnamefont {A.~T.}\ \bibnamefont {Schmitt}}, \bibinfo {author}
		{\bibfnamefont {A.~L.}\ \bibnamefont {Garcia}}, \bibinfo {author}
		{\bibfnamefont {W.}~\bibnamefont {Hippler}}, \bibinfo {author} {\bibfnamefont
			{L.}~\bibnamefont {Huang}}, \bibinfo {author} {\bibfnamefont
			{F.}~\bibnamefont {Karbstein}}, \bibinfo {author} {\bibfnamefont
			{Z.}~\bibnamefont {Konôpková}}, \bibinfo {author} {\bibfnamefont {H.-P.}\
			\bibnamefont {Schlenvoigt}}, \bibinfo {author} {\bibfnamefont {J.-P.}\
			\bibnamefont {Schwinkendorf}}, \bibinfo {author} {\bibfnamefont
			{C.}~\bibnamefont {Strohm}}, \bibinfo {author} {\bibfnamefont
			{T.}~\bibnamefont {Toncian}}, \bibinfo {author} {\bibfnamefont
			{I.}~\bibnamefont {Uschmann}}, \bibinfo {author} {\bibfnamefont {H.-C.}\
			\bibnamefont {Wille}}, \bibinfo {author} {\bibfnamefont {U.}~\bibnamefont
			{Zastrau}}, \bibinfo {author} {\bibfnamefont {R.}~\bibnamefont
			{Röhlsberger}}, \bibinfo {author} {\bibfnamefont {T.}~\bibnamefont
			{Stöhlker}}, \bibinfo {author} {\bibfnamefont {T.~E.}\ \bibnamefont
			{Cowan}},\ and\ \bibinfo {author} {\bibfnamefont {G.~G.}\ \bibnamefont
			{Paulus}},\ }\href {https://doi.org/10.1103/PhysRevResearch.4.013222}
	{\bibfield  {journal} {\bibinfo  {journal} {Phys. Rev. Research}\ }\textbf
		{\bibinfo {volume} {4}},\ \bibinfo {pages} {013222} (\bibinfo {year}
		{2022})}\BibitemShut {NoStop}%
	\bibitem [{\citenamefont {Heinzl}\ \emph {et~al.}(2006)\citenamefont {Heinzl},
		\citenamefont {Liesfeld}, \citenamefont {Amthor}, \citenamefont {Schwoerer},
		\citenamefont {Sauerbrey},\ and\ \citenamefont {Wipf}}]{Heinzl:2006xc}%
	\BibitemOpen
	\bibfield  {author} {\bibinfo {author} {\bibfnamefont {T.}~\bibnamefont
			{Heinzl}}, \bibinfo {author} {\bibfnamefont {B.}~\bibnamefont {Liesfeld}},
		\bibinfo {author} {\bibfnamefont {K.-U.}\ \bibnamefont {Amthor}}, \bibinfo
		{author} {\bibfnamefont {H.}~\bibnamefont {Schwoerer}}, \bibinfo {author}
		{\bibfnamefont {R.}~\bibnamefont {Sauerbrey}},\ and\ \bibinfo {author}
		{\bibfnamefont {A.}~\bibnamefont {Wipf}},\ }\href
	{https://doi.org/10.1016/j.optcom.2006.06.053} {\bibfield  {journal}
		{\bibinfo  {journal} {Opt. Commun.}\ }\textbf {\bibinfo {volume} {267}},\
		\bibinfo {pages} {318} (\bibinfo {year} {2006})},\ \Eprint
	{https://arxiv.org/abs/hep-ph/0601076} {arXiv:hep-ph/0601076} \BibitemShut
	{NoStop}%
	\bibitem [{\citenamefont {Di~Piazza}\ \emph {et~al.}(2006)\citenamefont
		{Di~Piazza}, \citenamefont {Hatsagortsyan},\ and\ \citenamefont
		{Keitel}}]{DiPiazza:2006pr}%
	\BibitemOpen
	\bibfield  {author} {\bibinfo {author} {\bibfnamefont {A.}~\bibnamefont
			{Di~Piazza}}, \bibinfo {author} {\bibfnamefont {K.~Z.}\ \bibnamefont
			{Hatsagortsyan}},\ and\ \bibinfo {author} {\bibfnamefont {C.~H.}\
			\bibnamefont {Keitel}},\ }\href
	{https://doi.org/10.1103/PhysRevLett.97.083603} {\bibfield  {journal}
		{\bibinfo  {journal} {Phys. Rev. Lett.}\ }\textbf {\bibinfo {volume} {97}},\
		\bibinfo {pages} {083603} (\bibinfo {year} {2006})},\ \Eprint
	{https://arxiv.org/abs/hep-ph/0602039} {arXiv:hep-ph/0602039} \BibitemShut
	{NoStop}%
	\bibitem [{\citenamefont {Schlenvoigt}\ \emph {et~al.}(2016)\citenamefont
		{Schlenvoigt}, \citenamefont {Heinzl}, \citenamefont {Schramm}, \citenamefont
		{Cowan},\ and\ \citenamefont {Sauerbrey}}]{Schlenvoigt:2016}%
	\BibitemOpen
	\bibfield  {author} {\bibinfo {author} {\bibfnamefont {H.-P.}\ \bibnamefont
			{Schlenvoigt}}, \bibinfo {author} {\bibfnamefont {T.}~\bibnamefont {Heinzl}},
		\bibinfo {author} {\bibfnamefont {U.}~\bibnamefont {Schramm}}, \bibinfo
		{author} {\bibfnamefont {T.~E.}\ \bibnamefont {Cowan}},\ and\ \bibinfo
		{author} {\bibfnamefont {R.}~\bibnamefont {Sauerbrey}},\ }\href
	{https://doi.org/10.1088/0031-8949/91/2/023010} {\bibfield  {journal}
		{\bibinfo  {journal} {Phys. Scr.}\ }\textbf {\bibinfo {volume} {91}},\
		\bibinfo {pages} {023010} (\bibinfo {year} {2016})}\BibitemShut {NoStop}%
	\bibitem [{\citenamefont {Mosman}\ and\ \citenamefont
		{Karbstein}(2021)}]{Mosman:2021vua}%
	\BibitemOpen
	\bibfield  {author} {\bibinfo {author} {\bibfnamefont {E.~A.}\ \bibnamefont
			{Mosman}}\ and\ \bibinfo {author} {\bibfnamefont {F.}~\bibnamefont
			{Karbstein}},\ }\href {https://doi.org/10.1103/PhysRevD.104.013006}
	{\bibfield  {journal} {\bibinfo  {journal} {Phys. Rev. D}\ }\textbf {\bibinfo
			{volume} {104}},\ \bibinfo {pages} {013006} (\bibinfo {year} {2021})},\
	\Eprint {https://arxiv.org/abs/2104.05103} {arXiv:2104.05103} \BibitemShut
	{NoStop}%
	\bibitem [{\citenamefont {Karbstein}\ and\ \citenamefont
		{Mosman}(2020)}]{Karbstein:2020gzg}%
	\BibitemOpen
	\bibfield  {author} {\bibinfo {author} {\bibfnamefont {F.}~\bibnamefont
			{Karbstein}}\ and\ \bibinfo {author} {\bibfnamefont {E.~A.}\ \bibnamefont
			{Mosman}},\ }\href {https://doi.org/10.1103/PhysRevD.101.113002} {\bibfield
		{journal} {\bibinfo  {journal} {Phys. Rev. D}\ }\textbf {\bibinfo {volume}
			{101}},\ \bibinfo {pages} {113002} (\bibinfo {year} {2020})},\ \Eprint
	{https://arxiv.org/abs/2004.04268} {arXiv:2004.04268 [quant-ph]} \BibitemShut
	{NoStop}%
	\bibitem [{\citenamefont {{\v{S}}m{\'i}d}\ \emph {et~al.}(2025)\citenamefont
		{{\v{S}}m{\'i}d} \emph {et~al.}}]{Smid:2025wmj}%
	\BibitemOpen
	\bibfield  {author} {\bibinfo {author} {\bibfnamefont {M.}~\bibnamefont
			{{\v{S}}m{\'i}d}} \emph {et~al.},\ }\href {https://doi.org/10.1103/xpxy-ntwz}
	{\bibfield  {journal} {\bibinfo  {journal} {Phys. Rev. A}\ }\textbf {\bibinfo
			{volume} {112}},\ \bibinfo {pages} {063103} (\bibinfo {year} {2025})},\
	\Eprint {https://arxiv.org/abs/2506.11649} {arXiv:2506.11649} \BibitemShut
	{NoStop}%
	\bibitem [{\citenamefont {Matheron}\ \emph
		{et~al.}(2026{\natexlab{a}})\citenamefont {Matheron}, \citenamefont
		{{\v{S}}m{\'i}d}, \citenamefont {Karbstein}, \citenamefont {Zepf} \emph
		{et~al.}}]{Matheron:2026}%
	\BibitemOpen
	\bibfield  {author} {\bibinfo {author} {\bibfnamefont {A.}~\bibnamefont
			{Matheron}}, \bibinfo {author} {\bibfnamefont {M.}~\bibnamefont
			{{\v{S}}m{\'i}d}}, \bibinfo {author} {\bibfnamefont {F.}~\bibnamefont
			{Karbstein}}, \bibinfo {author} {\bibfnamefont {M.}~\bibnamefont {Zepf}},
		\emph {et~al.},\ }\href@noop {} {\bibfield  {journal} {\bibinfo  {journal}
			{companion Letter, submitted to Phys. Rev. Lett.}\ } (\bibinfo {year}
		{2026}{\natexlab{a}})}\BibitemShut {NoStop}%
	\bibitem [{\citenamefont {Matheron}\ \emph
		{et~al.}(2026{\natexlab{b}})\citenamefont {Matheron}, \citenamefont
		{{\v{S}}m{\'i}d}, \citenamefont {Zepf},\ and\ \citenamefont
		{Karbstein}}]{Matheron:2025owt}%
	\BibitemOpen
	\bibfield  {author} {\bibinfo {author} {\bibfnamefont {A.}~\bibnamefont
			{Matheron}}, \bibinfo {author} {\bibfnamefont {M.}~\bibnamefont
			{{\v{S}}m{\'i}d}}, \bibinfo {author} {\bibfnamefont {M.}~\bibnamefont
			{Zepf}},\ and\ \bibinfo {author} {\bibfnamefont {F.}~\bibnamefont
			{Karbstein}},\ }\href {https://doi.org/10.1103/qqcw-461k} {\bibfield
		{journal} {\bibinfo  {journal} {Phys. Rev. D}\ }\textbf {\bibinfo {volume}
			{113}},\ \bibinfo {pages} {033005} (\bibinfo {year} {2026}{\natexlab{b}})},\
	\Eprint {https://arxiv.org/abs/2512.17641} {arXiv:2512.17641} \BibitemShut
	{NoStop}%
	\bibitem [{\citenamefont {Baronova}\ and\ \citenamefont
		{Stepanenko}(2003)}]{Baronova:2003}%
	\BibitemOpen
	\bibfield  {author} {\bibinfo {author} {\bibfnamefont {E.~O.}\ \bibnamefont
			{Baronova}}\ and\ \bibinfo {author} {\bibfnamefont {M.~M.}\ \bibnamefont
			{Stepanenko}},\ }\href {https://doi.org/10.1088/0741-3335/45/7/305}
	{\bibfield  {journal} {\bibinfo  {journal} {Plasma Phys. Control. Fusion}\
		}\textbf {\bibinfo {volume} {45}},\ \bibinfo {pages} {1113} (\bibinfo {year}
		{2003})}\BibitemShut {NoStop}%
	\bibitem [{\citenamefont {Wallace}\ \emph {et~al.}(2020)\citenamefont
		{Wallace}, \citenamefont {Haefner},\ and\ \citenamefont
		{et~al.}}]{Wallace:2020}%
	\BibitemOpen
	\bibfield  {author} {\bibinfo {author} {\bibfnamefont {M.~S.}\ \bibnamefont
			{Wallace}}, \bibinfo {author} {\bibfnamefont {C.}~\bibnamefont {Haefner}},\
		and\ \bibinfo {author} {\bibnamefont {et~al.}},\ }\href
	{https://doi.org/10.1063/1.5120530} {\bibfield  {journal} {\bibinfo
			{journal} {Rev. Sci. Instrum.}\ }\textbf {\bibinfo {volume} {91}},\ \bibinfo
		{pages} {023105} (\bibinfo {year} {2020})}\BibitemShut {NoStop}%
	\bibitem [{\citenamefont {Presura}(2021)}]{Presura:2021}%
	\BibitemOpen
	\bibfield  {author} {\bibinfo {author} {\bibfnamefont {R.}~\bibnamefont
			{Presura}},\ }\href {https://doi.org/10.1063/5.0047035} {\bibfield  {journal}
		{\bibinfo  {journal} {Rev. Sci. Instrum.}\ }\textbf {\bibinfo {volume}
			{92}},\ \bibinfo {pages} {073102} (\bibinfo {year} {2021})}\BibitemShut
	{NoStop}%
	\bibitem [{\citenamefont {Zastrau}\ \emph {et~al.}(2021)\citenamefont {Zastrau}
		\emph {et~al.}}]{Zastrau:2021}%
	\BibitemOpen
	\bibfield  {author} {\bibinfo {author} {\bibfnamefont {U.}~\bibnamefont
			{Zastrau}} \emph {et~al.},\ }\href
	{https://doi.org/10.1107/S1600577521007335} {\bibfield  {journal} {\bibinfo
			{journal} {J. Synchrotron Radiat.}\ }\textbf {\bibinfo {volume} {28}},\
		\bibinfo {pages} {1393} (\bibinfo {year} {2021})}\BibitemShut {NoStop}%
	\bibitem [{\citenamefont {Decking}\ \emph {et~al.}(2020)\citenamefont {Decking}
		\emph {et~al.}}]{Decking:2020}%
	\BibitemOpen
	\bibfield  {author} {\bibinfo {author} {\bibfnamefont {W.}~\bibnamefont
			{Decking}} \emph {et~al.},\ }\href
	{https://doi.org/10.1038/s41566-020-0607-z} {\bibfield  {journal} {\bibinfo
			{journal} {Nature Photon.}\ }\textbf {\bibinfo {volume} {14}},\ \bibinfo
		{pages} {391} (\bibinfo {year} {2020})}\BibitemShut {NoStop}%
	\bibitem [{\citenamefont {Snigirev}\ \emph {et~al.}(1996)\citenamefont
		{Snigirev}, \citenamefont {Kohn}, \citenamefont {Snigireva},\ and\
		\citenamefont {Lengeler}}]{Snigirev1996}%
	\BibitemOpen
	\bibfield  {author} {\bibinfo {author} {\bibfnamefont {A.}~\bibnamefont
			{Snigirev}}, \bibinfo {author} {\bibfnamefont {V.}~\bibnamefont {Kohn}},
		\bibinfo {author} {\bibfnamefont {I.}~\bibnamefont {Snigireva}},\ and\
		\bibinfo {author} {\bibfnamefont {B.}~\bibnamefont {Lengeler}},\ }\href
	{https://doi.org/10.1038/384049a0} {\bibfield  {journal} {\bibinfo  {journal}
			{Nature}\ }\textbf {\bibinfo {volume} {384}},\ \bibinfo {pages} {49–51}
		(\bibinfo {year} {1996})}\BibitemShut {NoStop}%
	\bibitem [{\citenamefont {Tomie}(1997)}]{Tomie1997}%
	\BibitemOpen
	\bibfield  {author} {\bibinfo {author} {\bibfnamefont {T.}~\bibnamefont
			{Tomie}},\ }\href {https://patents.google.com/patent/US5594773A/en} {\bibinfo
		{title} {{X-ray lens}}},\ \bibinfo {howpublished} {U.S. patent} (\bibinfo
	{year} {1997})\BibitemShut {NoStop}%
	\bibitem [{\citenamefont {Celestre}\ \emph
		{et~al.}(2020{\natexlab{a}})\citenamefont {Celestre}, \citenamefont
		{Berujon}, \citenamefont {Roth}, \citenamefont {Sanchez~del Rio},\ and\
		\citenamefont {Barrett}}]{Celestre:2020}%
	\BibitemOpen
	\bibfield  {author} {\bibinfo {author} {\bibfnamefont {R.}~\bibnamefont
			{Celestre}}, \bibinfo {author} {\bibfnamefont {S.}~\bibnamefont {Berujon}},
		\bibinfo {author} {\bibfnamefont {T.}~\bibnamefont {Roth}}, \bibinfo {author}
		{\bibfnamefont {M.}~\bibnamefont {Sanchez~del Rio}},\ and\ \bibinfo {author}
		{\bibfnamefont {R.}~\bibnamefont {Barrett}},\ }\href
	{https://doi.org/10.1107/S1600577519017235} {\bibfield  {journal} {\bibinfo
			{journal} {J. Synchrotron Radiat.}\ }\textbf {\bibinfo {volume} {27}},\
		\bibinfo {pages} {305} (\bibinfo {year} {2020}{\natexlab{a}})}\BibitemShut
	{NoStop}%
	\bibitem [{\citenamefont {Sikorski}\ \emph {et~al.}(2023)\citenamefont
		{Sikorski} \emph {et~al.}}]{Sikorski:2023}%
	\BibitemOpen
	\bibfield  {author} {\bibinfo {author} {\bibfnamefont {M.}~\bibnamefont
			{Sikorski}} \emph {et~al.},\ }\href
	{https://doi.org/10.3389/fphy.2023.1303247} {\bibfield  {journal} {\bibinfo
			{journal} {Front. Phys.}\ }\textbf {\bibinfo {volume} {11}},\ \bibinfo
		{pages} {1303247} (\bibinfo {year} {2023})}\BibitemShut {NoStop}%
	\bibitem [{\citenamefont {Chalupsky}\ \emph {et~al.}(2010)\citenamefont
		{Chalupsky}, \citenamefont {Krzywinski}, \citenamefont {Juha}, \citenamefont
		{Hajkova}, \citenamefont {Cihelka}, \citenamefont {Burian}, \citenamefont
		{Vysin}, \citenamefont {Gaudin}, \citenamefont {Gleeson}, \citenamefont
		{Jurek}, \citenamefont {Khorsand}, \citenamefont {Klinger}, \citenamefont
		{Wabnitz}, \citenamefont {Sobierajski}, \citenamefont {St\"{o}rmer},
		\citenamefont {Tiedtke},\ and\ \citenamefont {Toleikis}}]{Chalupsk2010}%
	\BibitemOpen
	\bibfield  {author} {\bibinfo {author} {\bibfnamefont {J.}~\bibnamefont
			{Chalupsky}}, \bibinfo {author} {\bibfnamefont {J.}~\bibnamefont
			{Krzywinski}}, \bibinfo {author} {\bibfnamefont {L.}~\bibnamefont {Juha}},
		\bibinfo {author} {\bibfnamefont {V.}~\bibnamefont {Hajkova}}, \bibinfo
		{author} {\bibfnamefont {J.}~\bibnamefont {Cihelka}}, \bibinfo {author}
		{\bibfnamefont {T.}~\bibnamefont {Burian}}, \bibinfo {author} {\bibfnamefont
			{L.}~\bibnamefont {Vysin}}, \bibinfo {author} {\bibfnamefont
			{J.}~\bibnamefont {Gaudin}}, \bibinfo {author} {\bibfnamefont
			{A.}~\bibnamefont {Gleeson}}, \bibinfo {author} {\bibfnamefont
			{M.}~\bibnamefont {Jurek}}, \bibinfo {author} {\bibfnamefont {A.~R.}\
			\bibnamefont {Khorsand}}, \bibinfo {author} {\bibfnamefont {D.}~\bibnamefont
			{Klinger}}, \bibinfo {author} {\bibfnamefont {H.}~\bibnamefont {Wabnitz}},
		\bibinfo {author} {\bibfnamefont {R.}~\bibnamefont {Sobierajski}}, \bibinfo
		{author} {\bibfnamefont {M.}~\bibnamefont {St\"{o}rmer}}, \bibinfo {author}
		{\bibfnamefont {K.}~\bibnamefont {Tiedtke}},\ and\ \bibinfo {author}
		{\bibfnamefont {S.}~\bibnamefont {Toleikis}},\ }\href
	{https://doi.org/10.1364/oe.18.027836} {\bibfield  {journal} {\bibinfo
			{journal} {Optics Express}\ }\textbf {\bibinfo {volume} {18}},\ \bibinfo
		{pages} {27836} (\bibinfo {year} {2010})}\BibitemShut {NoStop}%
	\bibitem [{\citenamefont {Chalupsky}\ \emph {et~al.}(2023)\citenamefont
		{Chalupsky}, \citenamefont {Vozda}, \citenamefont {Hering}, \citenamefont
		{Kybic}, \citenamefont {Burian}, \citenamefont {Dziarzhytski}, \citenamefont
		{Frantalova}, \citenamefont {Hajkova}, \citenamefont {Jelinek}, \citenamefont
		{Juha}, \citenamefont {Keitel}, \citenamefont {Kuglerova}, \citenamefont
		{Kuhlmann}, \citenamefont {Petryshak}, \citenamefont {Ruiz-Lopez},
		\citenamefont {Vyain}, \citenamefont {Wodzinski},\ and\ \citenamefont
		{Pl\"{o}njes}}]{Chalupsk2023}%
	\BibitemOpen
	\bibfield  {author} {\bibinfo {author} {\bibfnamefont {J.}~\bibnamefont
			{Chalupsky}}, \bibinfo {author} {\bibfnamefont {V.}~\bibnamefont {Vozda}},
		\bibinfo {author} {\bibfnamefont {J.}~\bibnamefont {Hering}}, \bibinfo
		{author} {\bibfnamefont {J.}~\bibnamefont {Kybic}}, \bibinfo {author}
		{\bibfnamefont {T.}~\bibnamefont {Burian}}, \bibinfo {author} {\bibfnamefont
			{S.}~\bibnamefont {Dziarzhytski}}, \bibinfo {author} {\bibfnamefont
			{K.}~\bibnamefont {Frantalova}}, \bibinfo {author} {\bibfnamefont
			{V.}~\bibnamefont {Hajkova}}, \bibinfo {author} {\bibfnamefont
			{S.}~\bibnamefont {Jelinek}}, \bibinfo {author} {\bibfnamefont
			{L.}~\bibnamefont {Juha}}, \bibinfo {author} {\bibfnamefont {B.}~\bibnamefont
			{Keitel}}, \bibinfo {author} {\bibfnamefont {Z.}~\bibnamefont {Kuglerova}},
		\bibinfo {author} {\bibfnamefont {M.}~\bibnamefont {Kuhlmann}}, \bibinfo
		{author} {\bibfnamefont {B.}~\bibnamefont {Petryshak}}, \bibinfo {author}
		{\bibfnamefont {M.}~\bibnamefont {Ruiz-Lopez}}, \bibinfo {author}
		{\bibfnamefont {L.}~\bibnamefont {Vyain}}, \bibinfo {author} {\bibfnamefont
			{T.}~\bibnamefont {Wodzinski}},\ and\ \bibinfo {author} {\bibfnamefont
			{E.}~\bibnamefont {Pl\"{o}njes}},\ }\href {https://doi.org/10.1364/oe.481776}
	{\bibfield  {journal} {\bibinfo  {journal} {Optics Express}\ }\textbf
		{\bibinfo {volume} {31}},\ \bibinfo {pages} {19703} (\bibinfo {year}
		{2023})}\BibitemShut {NoStop}%
	\bibitem [{\citenamefont {Duchon}(1976)}]{Duchon:1976}%
	\BibitemOpen
	\bibfield  {author} {\bibinfo {author} {\bibfnamefont {J.}~\bibnamefont
			{Duchon}},\ }\href {https://doi.org/10.1051/m2an/197610R300051} {\bibfield
		{journal} {\bibinfo  {journal} {Revue fran{\c c}aise d'automatique,
				informatique, recherche op{\'e}rationnelle. Analyse num{\'e}rique}\ }\textbf
		{\bibinfo {volume} {10}},\ \bibinfo {pages} {5} (\bibinfo {year}
		{1976})}\BibitemShut {NoStop}%
	\bibitem [{\citenamefont {Laso~Garcia}\ \emph {et~al.}(2021)\citenamefont
		{Laso~Garcia} \emph {et~al.}}]{LasoGarcia:2021}%
	\BibitemOpen
	\bibfield  {author} {\bibinfo {author} {\bibfnamefont {A.}~\bibnamefont
			{Laso~Garcia}} \emph {et~al.},\ }\href {https://doi.org/10.1017/hpl.2021.47}
	{\bibfield  {journal} {\bibinfo  {journal} {High Power Laser Sci. Eng.}\
		}\textbf {\bibinfo {volume} {9}},\ \bibinfo {pages} {e59} (\bibinfo {year}
		{2021})}\BibitemShut {NoStop}%
	\bibitem [{\citenamefont {Gerchberg}\ and\ \citenamefont
		{Saxton}(1972)}]{GerchbergSaxton:1972}%
	\BibitemOpen
	\bibfield  {author} {\bibinfo {author} {\bibfnamefont {R.~W.}\ \bibnamefont
			{Gerchberg}}\ and\ \bibinfo {author} {\bibfnamefont {W.~O.}\ \bibnamefont
			{Saxton}},\ }\href@noop {} {\bibfield  {journal} {\bibinfo  {journal}
			{Optik}\ }\textbf {\bibinfo {volume} {35}},\ \bibinfo {pages} {237} (\bibinfo
		{year} {1972})}\BibitemShut {NoStop}%
	\bibitem [{\citenamefont {Allen}\ and\ \citenamefont
		{Oxley}(2001)}]{Allen_2001_Phase_retrieval_OpC}%
	\BibitemOpen
	\bibfield  {author} {\bibinfo {author} {\bibfnamefont {L.}~\bibnamefont
			{Allen}}\ and\ \bibinfo {author} {\bibfnamefont {M.}~\bibnamefont {Oxley}},\
	}\href {https://doi.org/10.1016/S0030-4018(01)01556-5} {\bibfield  {journal}
		{\bibinfo  {journal} {Optics Communications}\ }\textbf {\bibinfo {volume}
			{199}},\ \bibinfo {pages} {65} (\bibinfo {year} {2001})}\BibitemShut
	{NoStop}%
	\bibitem [{\citenamefont {Chalupsky}\ \emph {et~al.}(2015)\citenamefont
		{Chalupsky}, \citenamefont {Bohacek}, \citenamefont {Burian}, \citenamefont
		{Hajkova}, \citenamefont {Hau-Riege}, \citenamefont {Heimann}, \citenamefont
		{Juha}, \citenamefont {Messerschmidt}, \citenamefont {Moeller}, \citenamefont
		{Nagler}, \citenamefont {Rowen}, \citenamefont {Schlotter}, \citenamefont
		{Swiggers}, \citenamefont {Turner},\ and\ \citenamefont
		{Krzywinski}}]{Chalupsky_2015_Imprinting_PRA}%
	\BibitemOpen
	\bibfield  {author} {\bibinfo {author} {\bibfnamefont {J.}~\bibnamefont
			{Chalupsky}}, \bibinfo {author} {\bibfnamefont {P.}~\bibnamefont {Bohacek}},
		\bibinfo {author} {\bibfnamefont {T.}~\bibnamefont {Burian}}, \bibinfo
		{author} {\bibfnamefont {V.}~\bibnamefont {Hajkova}}, \bibinfo {author}
		{\bibfnamefont {S.~P.}\ \bibnamefont {Hau-Riege}}, \bibinfo {author}
		{\bibfnamefont {P.~A.}\ \bibnamefont {Heimann}}, \bibinfo {author}
		{\bibfnamefont {L.}~\bibnamefont {Juha}}, \bibinfo {author} {\bibfnamefont
			{M.}~\bibnamefont {Messerschmidt}}, \bibinfo {author} {\bibfnamefont {S.~P.}\
			\bibnamefont {Moeller}}, \bibinfo {author} {\bibfnamefont {B.}~\bibnamefont
			{Nagler}}, \bibinfo {author} {\bibfnamefont {M.}~\bibnamefont {Rowen}},
		\bibinfo {author} {\bibfnamefont {W.~F.}\ \bibnamefont {Schlotter}}, \bibinfo
		{author} {\bibfnamefont {M.~L.}\ \bibnamefont {Swiggers}}, \bibinfo {author}
		{\bibfnamefont {J.~J.}\ \bibnamefont {Turner}},\ and\ \bibinfo {author}
		{\bibfnamefont {J.}~\bibnamefont {Krzywinski}},\ }\href
	{https://doi.org/10.1103/PhysRevApplied.4.014004} {\bibfield  {journal}
		{\bibinfo  {journal} {Physical Review Applied}\ }\textbf {\bibinfo {volume}
			{4}},\ \bibinfo {pages} {014004} (\bibinfo {year} {2015})}\BibitemShut
	{NoStop}%
	\bibitem [{\citenamefont {Jelinek}\ \emph {et~al.}(2026)\citenamefont
		{Jelinek}, \citenamefont {Dudzak}, \citenamefont {Dostal}, \citenamefont
		{Hajkova}, \citenamefont {Juha}, \citenamefont {Krupka}, \citenamefont
		{Morace}, \citenamefont {Valdova}, \citenamefont {Vozda},\ and\ \citenamefont
		{Chalupsky}}]{Jelinek_2026_HPLSE}%
	\BibitemOpen
	\bibfield  {author} {\bibinfo {author} {\bibfnamefont {S.}~\bibnamefont
			{Jelinek}}, \bibinfo {author} {\bibfnamefont {R.}~\bibnamefont {Dudzak}},
		\bibinfo {author} {\bibfnamefont {J.}~\bibnamefont {Dostal}}, \bibinfo
		{author} {\bibfnamefont {V.}~\bibnamefont {Hajkova}}, \bibinfo {author}
		{\bibfnamefont {L.}~\bibnamefont {Juha}}, \bibinfo {author} {\bibfnamefont
			{M.}~\bibnamefont {Krupka}}, \bibinfo {author} {\bibfnamefont
			{A.}~\bibnamefont {Morace}}, \bibinfo {author} {\bibfnamefont
			{Z.}~\bibnamefont {Valdova}}, \bibinfo {author} {\bibfnamefont
			{V.}~\bibnamefont {Vozda}},\ and\ \bibinfo {author} {\bibfnamefont
			{J.}~\bibnamefont {Chalupsky}},\ }\href
	{https://doi.org/10.1017/hpl.2026.10125} {\bibfield  {journal} {\bibinfo
			{journal} {High Power Laser Sci. Eng.}\ }\textbf {\bibinfo {volume} {14}},\
		\bibinfo {pages} {e39} (\bibinfo {year} {2026})}\BibitemShut {NoStop}%
	\bibitem [{\citenamefont {Matheron}\ \emph {et~al.}(2024)\citenamefont
		{Matheron} \emph {et~al.}}]{Matheron:2024gev}%
	\BibitemOpen
	\bibfield  {author} {\bibinfo {author} {\bibfnamefont {A.}~\bibnamefont
			{Matheron}} \emph {et~al.},\ }\href@noop {} {\bibinfo {title} {{Compton
				photons at the GeV scale from self-aligned collisions with a plasma mirror}}}
	(\bibinfo {year} {2024}),\ \Eprint {https://arxiv.org/abs/2412.19337}
	{arXiv:2412.19337} \BibitemShut {NoStop}%
	\bibitem [{\citenamefont {Karbstein}(2018)}]{Karbstein:2018omb}%
	\BibitemOpen
	\bibfield  {author} {\bibinfo {author} {\bibfnamefont {F.}~\bibnamefont
			{Karbstein}},\ }\href {https://doi.org/10.1103/PhysRevD.98.056010} {\bibfield
		{journal} {\bibinfo  {journal} {Phys. Rev. D}\ }\textbf {\bibinfo {volume}
			{98}},\ \bibinfo {pages} {056010} (\bibinfo {year} {2018})},\ \Eprint
	{https://arxiv.org/abs/1807.03302} {arXiv:1807.03302 [quant-ph]} \BibitemShut
	{NoStop}%
	\bibitem [{\citenamefont {Cowan}(1998)}]{Cowan:1998ji}%
	\BibitemOpen
	\bibfield  {author} {\bibinfo {author} {\bibfnamefont {G.}~\bibnamefont
			{Cowan}},\ }\href@noop {} {\emph {\bibinfo {title} {{Statistical Data
					Analysis}}}}\ (\bibinfo  {publisher} {Oxford University Press},\ \bibinfo
	{year} {1998})\BibitemShut {NoStop}%
	\bibitem [{\citenamefont {Vdovin}\ \emph {et~al.}(1997)\citenamefont {Vdovin},
		\citenamefont {van Brug},\ and\ \citenamefont {van Goor}}]{LightPipes:1997}%
	\BibitemOpen
	\bibfield  {author} {\bibinfo {author} {\bibfnamefont {G.}~\bibnamefont
			{Vdovin}}, \bibinfo {author} {\bibfnamefont {H.}~\bibnamefont {van Brug}},\
		and\ \bibinfo {author} {\bibfnamefont {F.}~\bibnamefont {van Goor}},\ }in\
	\href {https://doi.org/10.1117/12.294366} {\emph {\bibinfo {booktitle} {Proc.
				5th Int. Top. Meet. Educ. Train. Opt.}}},\ \bibinfo {series} {Proc. SPIE},
	Vol.\ \bibinfo {volume} {3190}\ (\bibinfo {year} {1997})\ pp.\ \bibinfo
	{pages} {82--93}\BibitemShut {NoStop}%
	\bibitem [{\citenamefont {Vdovin}\ \emph {et~al.}(2023)\citenamefont {Vdovin},
		\citenamefont {van Goor} \emph {et~al.}}]{LightPipes:2.1.5}%
	\BibitemOpen
	\bibfield  {author} {\bibinfo {author} {\bibfnamefont {G.}~\bibnamefont
			{Vdovin}}, \bibinfo {author} {\bibfnamefont {F.}~\bibnamefont {van Goor}},
		\emph {et~al.},\ }\href {https://opticspy.github.io/lightpipes/} {\bibinfo
		{title} {{LightPipes for Python 2.1.5}}} (\bibinfo {year} {2023})\BibitemShut
	{NoStop}%
	\bibitem [{\citenamefont {Agostinelli}\ \emph {et~al.}(2003)\citenamefont
		{Agostinelli} \emph {et~al.}}]{Geant4:2003}%
	\BibitemOpen
	\bibfield  {author} {\bibinfo {author} {\bibfnamefont {S.}~\bibnamefont
			{Agostinelli}} \emph {et~al.},\ }\href
	{https://doi.org/10.1016/S0168-9002(03)01368-8} {\bibfield  {journal}
		{\bibinfo  {journal} {Nucl. Instrum. Meth. A}\ }\textbf {\bibinfo {volume}
			{506}},\ \bibinfo {pages} {250} (\bibinfo {year} {2003})}\BibitemShut
	{NoStop}%
	\bibitem [{\citenamefont {Lengeler}\ \emph {et~al.}(1999)\citenamefont
		{Lengeler}, \citenamefont {Schroer}, \citenamefont {T{\"{u}}mmler},
		\citenamefont {Benner}, \citenamefont {Richwin}, \citenamefont {Snigirev},
		\citenamefont {Snigireva},\ and\ \citenamefont
		{Drakopoulos}}]{Lengeler:ht2006}%
	\BibitemOpen
	\bibfield  {author} {\bibinfo {author} {\bibfnamefont {B.}~\bibnamefont
			{Lengeler}}, \bibinfo {author} {\bibfnamefont {C.}~\bibnamefont {Schroer}},
		\bibinfo {author} {\bibfnamefont {J.}~\bibnamefont {T{\"{u}}mmler}}, \bibinfo
		{author} {\bibfnamefont {B.}~\bibnamefont {Benner}}, \bibinfo {author}
		{\bibfnamefont {M.}~\bibnamefont {Richwin}}, \bibinfo {author} {\bibfnamefont
			{A.}~\bibnamefont {Snigirev}}, \bibinfo {author} {\bibfnamefont
			{I.}~\bibnamefont {Snigireva}},\ and\ \bibinfo {author} {\bibfnamefont
			{M.}~\bibnamefont {Drakopoulos}},\ }\href
	{https://doi.org/10.1107/S0909049599009747} {\bibfield  {journal} {\bibinfo
			{journal} {Journal of Synchrotron Radiation}\ }\textbf {\bibinfo {volume}
			{6}},\ \bibinfo {pages} {1153} (\bibinfo {year} {1999})}\BibitemShut
	{NoStop}%
	\bibitem [{\citenamefont {Roth}\ \emph {et~al.}(2014)\citenamefont {Roth},
		\citenamefont {Helfen}, \citenamefont {Hallmann}, \citenamefont {Samoylova},
		\citenamefont {Kwaśniewski}, \citenamefont {Lengeler},\ and\ \citenamefont
		{Madsen}}]{Roth2014}%
	\BibitemOpen
	\bibfield  {author} {\bibinfo {author} {\bibfnamefont {T.}~\bibnamefont
			{Roth}}, \bibinfo {author} {\bibfnamefont {L.}~\bibnamefont {Helfen}},
		\bibinfo {author} {\bibfnamefont {J.}~\bibnamefont {Hallmann}}, \bibinfo
		{author} {\bibfnamefont {L.}~\bibnamefont {Samoylova}}, \bibinfo {author}
		{\bibfnamefont {P.}~\bibnamefont {Kwaśniewski}}, \bibinfo {author}
		{\bibfnamefont {B.}~\bibnamefont {Lengeler}},\ and\ \bibinfo {author}
		{\bibfnamefont {A.}~\bibnamefont {Madsen}},\ }in\ \href
	{https://doi.org/10.1117/12.2061127} {\emph {\bibinfo {booktitle} {Advances
				in X-Ray/EUV Optics and Components IX}}},\ Vol.\ \bibinfo {volume} {9207},\
	\bibinfo {editor} {edited by\ \bibinfo {editor} {\bibfnamefont
			{C.}~\bibnamefont {Morawe}}, \bibinfo {editor} {\bibfnamefont {A.~M.}\
			\bibnamefont {Khounsary}},\ and\ \bibinfo {editor} {\bibfnamefont
			{S.}~\bibnamefont {Goto}}}\ (\bibinfo  {publisher} {SPIE},\ \bibinfo {year}
	{2014})\ p.\ \bibinfo {pages} {920702}\BibitemShut {NoStop}%
	\bibitem [{\citenamefont {Gasilov}\ \emph {et~al.}(2017)\citenamefont
		{Gasilov}, \citenamefont {Mittone}, \citenamefont {dos Santos~Rolo},
		\citenamefont {Polyakov}, \citenamefont {Zholudev}, \citenamefont
		{Terentyev}, \citenamefont {Blank}, \citenamefont {Bravin},\ and\
		\citenamefont {Baumbach}}]{Gasilov:mo5165}%
	\BibitemOpen
	\bibfield  {author} {\bibinfo {author} {\bibfnamefont {S.}~\bibnamefont
			{Gasilov}}, \bibinfo {author} {\bibfnamefont {A.}~\bibnamefont {Mittone}},
		\bibinfo {author} {\bibfnamefont {T.}~\bibnamefont {dos Santos~Rolo}},
		\bibinfo {author} {\bibfnamefont {S.}~\bibnamefont {Polyakov}}, \bibinfo
		{author} {\bibfnamefont {S.}~\bibnamefont {Zholudev}}, \bibinfo {author}
		{\bibfnamefont {S.}~\bibnamefont {Terentyev}}, \bibinfo {author}
		{\bibfnamefont {V.}~\bibnamefont {Blank}}, \bibinfo {author} {\bibfnamefont
			{A.}~\bibnamefont {Bravin}},\ and\ \bibinfo {author} {\bibfnamefont
			{T.}~\bibnamefont {Baumbach}},\ }\href
	{https://doi.org/10.1107/S1600577517012772} {\bibfield  {journal} {\bibinfo
			{journal} {Journal of Synchrotron Radiation}\ }\textbf {\bibinfo {volume}
			{24}},\ \bibinfo {pages} {1137} (\bibinfo {year} {2017})}\BibitemShut
	{NoStop}%
	\bibitem [{\citenamefont {Celestre}\ \emph
		{et~al.}(2020{\natexlab{b}})\citenamefont {Celestre}, \citenamefont {Chubar},
		\citenamefont {Roth}, \citenamefont {Sanchez~del Rio},\ and\ \citenamefont
		{Barrett}}]{Celestre:2020srw}%
	\BibitemOpen
	\bibfield  {author} {\bibinfo {author} {\bibfnamefont {R.}~\bibnamefont
			{Celestre}}, \bibinfo {author} {\bibfnamefont {O.}~\bibnamefont {Chubar}},
		\bibinfo {author} {\bibfnamefont {T.}~\bibnamefont {Roth}}, \bibinfo {author}
		{\bibfnamefont {M.}~\bibnamefont {Sanchez~del Rio}},\ and\ \bibinfo {author}
		{\bibfnamefont {R.}~\bibnamefont {Barrett}},\ }in\ \href
	{https://doi.org/10.1117/12.2567947} {\emph {\bibinfo {booktitle} {Adv.
				Comput. Methods X-Ray Opt. V}}},\ \bibinfo {series} {Proc. SPIE}, Vol.\
	\bibinfo {volume} {11493}\ (\bibinfo {year} {2020})\ p.\ \bibinfo {pages}
	{114930J}\BibitemShut {NoStop}%
	\bibitem [{\citenamefont {Lyatun}\ \emph {et~al.}(2020)\citenamefont {Lyatun},
		\citenamefont {Ershov}, \citenamefont {Snigireva},\ and\ \citenamefont
		{Snigirev}}]{Lyatun:2020}%
	\BibitemOpen
	\bibfield  {author} {\bibinfo {author} {\bibfnamefont {I.}~\bibnamefont
			{Lyatun}}, \bibinfo {author} {\bibfnamefont {P.}~\bibnamefont {Ershov}},
		\bibinfo {author} {\bibfnamefont {I.}~\bibnamefont {Snigireva}},\ and\
		\bibinfo {author} {\bibfnamefont {A.}~\bibnamefont {Snigirev}},\ }\href
	{https://doi.org/10.1107/S1600577519015625} {\bibfield  {journal} {\bibinfo
			{journal} {J. Synchrotron Radiat.}\ }\textbf {\bibinfo {volume} {27}},\
		\bibinfo {pages} {44} (\bibinfo {year} {2020})}\BibitemShut {NoStop}%
	\bibitem [{\citenamefont {Celestre}\ \emph {et~al.}(2022)\citenamefont
		{Celestre}, \citenamefont {Antipov}, \citenamefont {Gomez}, \citenamefont
		{Zinn}, \citenamefont {Barrett},\ and\ \citenamefont {Roth}}]{Celestre:2022}%
	\BibitemOpen
	\bibfield  {author} {\bibinfo {author} {\bibfnamefont {R.}~\bibnamefont
			{Celestre}}, \bibinfo {author} {\bibfnamefont {S.}~\bibnamefont {Antipov}},
		\bibinfo {author} {\bibfnamefont {E.}~\bibnamefont {Gomez}}, \bibinfo
		{author} {\bibfnamefont {T.}~\bibnamefont {Zinn}}, \bibinfo {author}
		{\bibfnamefont {R.}~\bibnamefont {Barrett}},\ and\ \bibinfo {author}
		{\bibfnamefont {T.}~\bibnamefont {Roth}},\ }\href
	{https://doi.org/10.1107/S1600577522001795} {\bibfield  {journal} {\bibinfo
			{journal} {J. Synchrotron Radiat.}\ }\textbf {\bibinfo {volume} {29}},\
		\bibinfo {pages} {629} (\bibinfo {year} {2022})}\BibitemShut {NoStop}%
	\bibitem [{\citenamefont {Ziegler}(2004)}]{Ziegler2004}%
	\BibitemOpen
	\bibfield  {author} {\bibinfo {author} {\bibfnamefont {E.}~\bibnamefont
			{Ziegler}},\ }in\ \href {https://doi.org/10.1063/1.1757827} {\emph {\bibinfo
			{booktitle} {AIP Conference Proceedings}}},\ Vol.\ \bibinfo {volume} {705}\
	(\bibinfo  {publisher} {AIP},\ \bibinfo {year} {2004})\ p.\ \bibinfo {pages}
	{436–439}\BibitemShut {NoStop}%
	\bibitem [{\citenamefont {Berujon}\ \emph {et~al.}(2020)\citenamefont
		{Berujon}, \citenamefont {Cojocaru}, \citenamefont {Piault}, \citenamefont
		{Celestre}, \citenamefont {Roth}, \citenamefont {Barrett},\ and\
		\citenamefont {Ziegler}}]{Berujon2020}%
	\BibitemOpen
	\bibfield  {author} {\bibinfo {author} {\bibfnamefont {S.}~\bibnamefont
			{Berujon}}, \bibinfo {author} {\bibfnamefont {R.}~\bibnamefont {Cojocaru}},
		\bibinfo {author} {\bibfnamefont {P.}~\bibnamefont {Piault}}, \bibinfo
		{author} {\bibfnamefont {R.}~\bibnamefont {Celestre}}, \bibinfo {author}
		{\bibfnamefont {T.}~\bibnamefont {Roth}}, \bibinfo {author} {\bibfnamefont
			{R.}~\bibnamefont {Barrett}},\ and\ \bibinfo {author} {\bibfnamefont
			{E.}~\bibnamefont {Ziegler}},\ }\href
	{https://doi.org/10.1107/s1600577520000508} {\bibfield  {journal} {\bibinfo
			{journal} {Journal of Synchrotron Radiation}\ }\textbf {\bibinfo {volume}
			{27}},\ \bibinfo {pages} {293–304} (\bibinfo {year} {2020})}\BibitemShut
	{NoStop}%
	\bibitem [{\citenamefont {Seiboth}\ \emph {et~al.}(2020)\citenamefont
		{Seiboth}, \citenamefont {Br{\"{u}}ckner}, \citenamefont {Kahnt},
		\citenamefont {Lyubomirskiy}, \citenamefont {Wittwer}, \citenamefont
		{Dzhigaev}, \citenamefont {Ullsperger}, \citenamefont {Nolte}, \citenamefont
		{Koch}, \citenamefont {David}, \citenamefont {Garrevoet}, \citenamefont
		{Falkenberg},\ and\ \citenamefont {Schroer}}]{Seiboth:mo5223}%
	\BibitemOpen
	\bibfield  {author} {\bibinfo {author} {\bibfnamefont {F.}~\bibnamefont
			{Seiboth}}, \bibinfo {author} {\bibfnamefont {D.}~\bibnamefont
			{Br{\"{u}}ckner}}, \bibinfo {author} {\bibfnamefont {M.}~\bibnamefont
			{Kahnt}}, \bibinfo {author} {\bibfnamefont {M.}~\bibnamefont {Lyubomirskiy}},
		\bibinfo {author} {\bibfnamefont {F.}~\bibnamefont {Wittwer}}, \bibinfo
		{author} {\bibfnamefont {D.}~\bibnamefont {Dzhigaev}}, \bibinfo {author}
		{\bibfnamefont {T.}~\bibnamefont {Ullsperger}}, \bibinfo {author}
		{\bibfnamefont {S.}~\bibnamefont {Nolte}}, \bibinfo {author} {\bibfnamefont
			{F.}~\bibnamefont {Koch}}, \bibinfo {author} {\bibfnamefont {C.}~\bibnamefont
			{David}}, \bibinfo {author} {\bibfnamefont {J.}~\bibnamefont {Garrevoet}},
		\bibinfo {author} {\bibfnamefont {G.}~\bibnamefont {Falkenberg}},\ and\
		\bibinfo {author} {\bibfnamefont {C.~G.}\ \bibnamefont {Schroer}},\ }\href
	{https://doi.org/10.1107/S1600577520007900} {\bibfield  {journal} {\bibinfo
			{journal} {Journal of Synchrotron Radiation}\ }\textbf {\bibinfo {volume}
			{27}},\ \bibinfo {pages} {1121} (\bibinfo {year} {2020})}\BibitemShut
	{NoStop}%
	\bibitem [{\citenamefont {Vacalis}\ \emph {et~al.}(2025)\citenamefont {Vacalis}
		\emph {et~al.}}]{Vacalis:2025}%
	\BibitemOpen
	\bibfield  {author} {\bibinfo {author} {\bibfnamefont {G.}~\bibnamefont
			{Vacalis}} \emph {et~al.},\ }\href {https://doi.org/10.1103/vgvg-hcbr}
	{\bibfield  {journal} {\bibinfo  {journal} {Phys. Rev. Lett.}\ }\textbf
		{\bibinfo {volume} {135}},\ \bibinfo {pages} {195003} (\bibinfo {year}
		{2025})}\BibitemShut {NoStop}%
	\bibitem [{\citenamefont {Turin}(1960)}]{Turin:1960}%
	\BibitemOpen
	\bibfield  {author} {\bibinfo {author} {\bibfnamefont {G.~L.}\ \bibnamefont
			{Turin}},\ }\href {https://doi.org/10.1109/TIT.1960.1057571} {\bibfield
		{journal} {\bibinfo  {journal} {IRE Trans. Inf. Theory}\ }\textbf {\bibinfo
			{volume} {6}},\ \bibinfo {pages} {311} (\bibinfo {year} {1960})}\BibitemShut
	{NoStop}%
	\bibitem [{\citenamefont {Cowan}\ \emph {et~al.}(2011)\citenamefont {Cowan},
		\citenamefont {Cranmer}, \citenamefont {Gross},\ and\ \citenamefont
		{Vitells}}]{Cowan:2010js}%
	\BibitemOpen
	\bibfield  {author} {\bibinfo {author} {\bibfnamefont {G.}~\bibnamefont
			{Cowan}}, \bibinfo {author} {\bibfnamefont {K.}~\bibnamefont {Cranmer}},
		\bibinfo {author} {\bibfnamefont {E.}~\bibnamefont {Gross}},\ and\ \bibinfo
		{author} {\bibfnamefont {O.}~\bibnamefont {Vitells}},\ }\href
	{https://doi.org/10.1140/epjc/s10052-011-1554-0} {\bibfield  {journal}
		{\bibinfo  {journal} {Eur. Phys. J. C}\ }\textbf {\bibinfo {volume} {71}},\
		\bibinfo {pages} {1554} (\bibinfo {year} {2011})},\ \bibinfo {note}
	{[Erratum: Eur. Phys. J. C 73, 2501 (2013)]},\ \Eprint
	{https://arxiv.org/abs/1007.1727} {arXiv:1007.1727} \BibitemShut {NoStop}%
	\bibitem [{\citenamefont {Navas}\ \emph {et~al.}(2024)\citenamefont {Navas}
		\emph {et~al.}}]{ParticleDataGroup:2024cfk}%
	\BibitemOpen
	\bibfield  {author} {\bibinfo {author} {\bibfnamefont {S.}~\bibnamefont
			{Navas}} \emph {et~al.} (\bibinfo {collaboration} {Particle Data Group}),\
	}\href {https://doi.org/10.1103/PhysRevD.110.030001} {\bibfield  {journal}
		{\bibinfo  {journal} {Phys. Rev. D}\ }\textbf {\bibinfo {volume} {110}},\
		\bibinfo {pages} {030001} (\bibinfo {year} {2024})}\BibitemShut {NoStop}%
	\bibitem [{\citenamefont {Pedregosa}\ \emph {et~al.}(2011)\citenamefont
		{Pedregosa}, \citenamefont {Varoquaux}, \citenamefont {Gramfort},
		\citenamefont {Michel}, \citenamefont {Thirion}, \citenamefont {Grisel},
		\citenamefont {Blondel}, \citenamefont {Prettenhofer}, \citenamefont {Weiss},
		\citenamefont {Dubourg}, \citenamefont {Vanderplas}, \citenamefont {Passos},
		\citenamefont {Cournapeau}, \citenamefont {Brucher}, \citenamefont {Perrot},\
		and\ \citenamefont {Duchesnay}}]{scikit-learn}%
	\BibitemOpen
	\bibfield  {author} {\bibinfo {author} {\bibfnamefont {F.}~\bibnamefont
			{Pedregosa}}, \bibinfo {author} {\bibfnamefont {G.}~\bibnamefont
			{Varoquaux}}, \bibinfo {author} {\bibfnamefont {A.}~\bibnamefont {Gramfort}},
		\bibinfo {author} {\bibfnamefont {V.}~\bibnamefont {Michel}}, \bibinfo
		{author} {\bibfnamefont {B.}~\bibnamefont {Thirion}}, \bibinfo {author}
		{\bibfnamefont {O.}~\bibnamefont {Grisel}}, \bibinfo {author} {\bibfnamefont
			{M.}~\bibnamefont {Blondel}}, \bibinfo {author} {\bibfnamefont
			{P.}~\bibnamefont {Prettenhofer}}, \bibinfo {author} {\bibfnamefont
			{R.}~\bibnamefont {Weiss}}, \bibinfo {author} {\bibfnamefont
			{V.}~\bibnamefont {Dubourg}}, \bibinfo {author} {\bibfnamefont
			{J.}~\bibnamefont {Vanderplas}}, \bibinfo {author} {\bibfnamefont
			{A.}~\bibnamefont {Passos}}, \bibinfo {author} {\bibfnamefont
			{D.}~\bibnamefont {Cournapeau}}, \bibinfo {author} {\bibfnamefont
			{M.}~\bibnamefont {Brucher}}, \bibinfo {author} {\bibfnamefont
			{M.}~\bibnamefont {Perrot}},\ and\ \bibinfo {author} {\bibfnamefont
			{E.}~\bibnamefont {Duchesnay}},\ }\href@noop {} {\bibfield  {journal}
		{\bibinfo  {journal} {Journal of Machine Learning Research}\ }\textbf
		{\bibinfo {volume} {12}},\ \bibinfo {pages} {2825} (\bibinfo {year}
		{2011})}\BibitemShut {NoStop}%
	\bibitem [{\citenamefont {Ejlli}\ \emph {et~al.}(2020)\citenamefont {Ejlli},
		\citenamefont {Della~Valle}, \citenamefont {Gastaldi}, \citenamefont
		{Messineo}, \citenamefont {Pengo}, \citenamefont {Ruoso},\ and\ \citenamefont
		{Zavattini}}]{Ejlli:2020yhk}%
	\BibitemOpen
	\bibfield  {author} {\bibinfo {author} {\bibfnamefont {A.}~\bibnamefont
			{Ejlli}}, \bibinfo {author} {\bibfnamefont {F.}~\bibnamefont {Della~Valle}},
		\bibinfo {author} {\bibfnamefont {U.}~\bibnamefont {Gastaldi}}, \bibinfo
		{author} {\bibfnamefont {G.}~\bibnamefont {Messineo}}, \bibinfo {author}
		{\bibfnamefont {R.}~\bibnamefont {Pengo}}, \bibinfo {author} {\bibfnamefont
			{G.}~\bibnamefont {Ruoso}},\ and\ \bibinfo {author} {\bibfnamefont
			{G.}~\bibnamefont {Zavattini}},\ }\href
	{https://doi.org/10.1016/j.physrep.2020.06.001} {\bibfield  {journal}
		{\bibinfo  {journal} {Phys. Rept.}\ }\textbf {\bibinfo {volume} {871}},\
		\bibinfo {pages} {1} (\bibinfo {year} {2020})},\ \Eprint
	{https://arxiv.org/abs/2005.12913} {arXiv:2005.12913} \BibitemShut {NoStop}%
	\bibitem [{\citenamefont {Battesti}\ \emph {et~al.}(2018)\citenamefont
		{Battesti} \emph {et~al.}}]{Battesti:2018bgc}%
	\BibitemOpen
	\bibfield  {author} {\bibinfo {author} {\bibfnamefont {R.}~\bibnamefont
			{Battesti}} \emph {et~al.},\ }\href
	{https://doi.org/10.1016/j.physrep.2018.07.005} {\bibfield  {journal}
		{\bibinfo  {journal} {Phys. Rept.}\ }\textbf {\bibinfo {volume} {765-766}},\
		\bibinfo {pages} {1} (\bibinfo {year} {2018})},\ \Eprint
	{https://arxiv.org/abs/1803.07547} {arXiv:1803.07547 [physics.ins-det]}
	\BibitemShut {NoStop}%
	\bibitem [{\citenamefont {Fan}\ \emph {et~al.}(2017)\citenamefont {Fan} \emph
		{et~al.}}]{Fan:2017fnd}%
	\BibitemOpen
	\bibfield  {author} {\bibinfo {author} {\bibfnamefont {X.}~\bibnamefont
			{Fan}} \emph {et~al.},\ }\href {https://doi.org/10.1140/epjd/e2017-80290-7}
	{\bibfield  {journal} {\bibinfo  {journal} {Eur. Phys. J. D}\ }\textbf
		{\bibinfo {volume} {71}},\ \bibinfo {pages} {308} (\bibinfo {year} {2017})},\
	\Eprint {https://arxiv.org/abs/1705.00495} {arXiv:1705.00495
		[physics.ins-det]} \BibitemShut {NoStop}%
	\bibitem [{\citenamefont {Cad{\`e}ne}\ \emph {et~al.}(2014)\citenamefont
		{Cad{\`e}ne}, \citenamefont {Berceau}, \citenamefont {Fouch{\'e}},
		\citenamefont {Battesti},\ and\ \citenamefont {Rizzo}}]{Cadene:2013bva}%
	\BibitemOpen
	\bibfield  {author} {\bibinfo {author} {\bibfnamefont {A.}~\bibnamefont
			{Cad{\`e}ne}}, \bibinfo {author} {\bibfnamefont {P.}~\bibnamefont {Berceau}},
		\bibinfo {author} {\bibfnamefont {M.}~\bibnamefont {Fouch{\'e}}}, \bibinfo
		{author} {\bibfnamefont {R.}~\bibnamefont {Battesti}},\ and\ \bibinfo
		{author} {\bibfnamefont {C.}~\bibnamefont {Rizzo}},\ }\href
	{https://doi.org/10.1140/epjd/e2013-40725-9} {\bibfield  {journal} {\bibinfo
			{journal} {Eur. Phys. J. D}\ }\textbf {\bibinfo {volume} {68}},\ \bibinfo
		{pages} {16} (\bibinfo {year} {2014})},\ \Eprint
	{https://arxiv.org/abs/1302.5389} {arXiv:1302.5389 [physics.optics]}
	\BibitemShut {NoStop}%
	\bibitem [{\citenamefont {Spector}\ \emph {et~al.}(2026)\citenamefont
		{Spector}, \citenamefont {Kozlowski},\ and\ \citenamefont
		{Roberts}}]{Spector:2025esz}%
	\BibitemOpen
	\bibfield  {author} {\bibinfo {author} {\bibfnamefont {A.~D.}\ \bibnamefont
			{Spector}}, \bibinfo {author} {\bibfnamefont {T.}~\bibnamefont {Kozlowski}},\
		and\ \bibinfo {author} {\bibfnamefont {L.}~\bibnamefont {Roberts}},\ }\href
	{https://doi.org/10.1103/r33m-v1kn} {\bibfield  {journal} {\bibinfo
			{journal} {Phys. Rev. A}\ }\textbf {\bibinfo {volume} {113}},\ \bibinfo
		{pages} {063518} (\bibinfo {year} {2026})},\ \Eprint
	{https://arxiv.org/abs/2510.14064} {arXiv:2510.14064 [physics.optics]}
	\BibitemShut {NoStop}%
	\bibitem [{\citenamefont {Karbstein}(2022)}]{Karbstein:2021otv}%
	\BibitemOpen
	\bibfield  {author} {\bibinfo {author} {\bibfnamefont {F.}~\bibnamefont
			{Karbstein}},\ }\href {https://doi.org/10.1002/andp.202100137} {\bibfield
		{journal} {\bibinfo  {journal} {Annalen Phys.}\ }\textbf {\bibinfo {volume}
			{534}},\ \bibinfo {pages} {2100137} (\bibinfo {year} {2022})},\ \Eprint
	{https://arxiv.org/abs/2106.06359} {arXiv:2106.06359} \BibitemShut {NoStop}%
\end{thebibliography}

\end{document}